\documentclass[times]{article}

\usepackage{moreverb}
\usepackage{graphicx}

\usepackage{amsmath}

\usepackage{diagbox}

\usepackage{subcaption}
\usepackage{tikz,pgfplots}
\usepackage{xcolor, colortbl}
\usepackage{array}
\usetikzlibrary{shapes,snakes}

\usepackage{tabularx}
\usepackage{bm}
\usepackage{multirow}
\usepackage{float}
\usepackage{authblk}
\usepackage{amsmath, amssymb}
\usepackage[margin=1in]{geometry}

\usepackage[driverfallback=dvipdfm,colorlinks,bookmarksopen,bookmarksnumbered,citecolor=red,urlcolor=red]{hyperref}

\newcommand\BibTeX{{\rmfamily B\kern-.05em \textsc{i\kern-.025em b}\kern-.08em
T\kern-.1667em\lower.7ex\hbox{E}\kern-.125emX}}

\newcommand{\markerfive}{\raisebox{0.6pt}{\tikz{\node[draw,scale=0.42, circle,fill=red!100!, color=red](){};}}}
\newcommand{\markersix}{\raisebox{0.6pt}{\tikz{\node[draw,scale=0.42,thick, circle,color=black](){};}}}

\begin{document}

\title{Design and analysis adaptivity in multi-resolution topology optimization}

\author{Deepak~K.~Gupta, Fred van Keulen and Matthijs Langelaar}

\affil{Department of Precision and Microsystems Engineering, Faculty of 3mE, \\ Delft University of Technology, Mekelweg 2, 2628 CD, Delft, The Netherlands}
\date{Correponding author: M. Langelaar, Email: M.Langelaar@tudelft.nl}

\maketitle

\begin{abstract}
Multiresolution topology optimization (MTO) methods involve decoupling of the design and analysis discretizations, such that a high-resolution design can be obtained at relatively low analysis costs. Recent studies have shown that the MTO method can be approximately 3 and 30 times faster than the traditional topology optimization method for 2D and 3D problems, respectively. To further exploit the potential of decoupling analysis and design, we propose a $dp$-adaptive MTO method, which involves locally increasing/decreasing the shape function orders ($p$) and design resolution ($d$). The adaptive refinement/coarsening is performed using a composite refinement indicator which includes criteria based on analysis error, presence of intermediate densities as well as the occurrence of design artefacts referred to as QR-patterns. While standard MTO must rely on filtering to suppress QR-patterns, the proposed adaptive method ensures efficiently that these artefacts are suppressed in the final design, without sacrificing the design resolution. The applicability of the $dp$-adaptive MTO method is demonstrated on several 2D mechanical design problems. For all the cases, significant speed-ups in computational time are obtained. In particular for design problems involving low material volume fractions, speed-ups of up to a factor of 10 can be obtained over the conventional MTO method.
\end{abstract}


\let\thefootnote\relax\footnotetext{Submitted to \emph{International Journal for  Numerical Methods in Engineering}}

\section{Introduction}
\label{intro}

Topology optimization (TO) can be described as an approach that optimally distributes material in a specified domain under  a set of constraints, such that the performance function of the structure achieves a maximum \cite{Bendsoe1989}. In the past two decades, TO has widely been used in various academic and industrial disciplines. For a survey on the latest developments in TO as well as its recent applications, see the review papers by \mbox{Sigmund and Maute \cite{Sigmund2013}}, van Dijk \emph{et al.} \cite{vanDijk2013}, and Deaton and Grandhi \cite{Deaton2014}.

Typically, in popular density-based TO, the domain is discretized into a finite set of elements and a density value is associated with every finite element \cite{Bendsoe1989}. The density of an element indicates the volume fraction of that element filled with a certain amount of material, and can vary from 0 (void) to 1 (solid). These density values are optimized during the course of optimization. Since in traditional approaches, density is assumed to be constant inside an element, a large number of finite elements as well as associated design variables are needed to obtain a well defined design with the desired structural features and boundary resolution, especially for three-dimensional (3D) problems \cite{Aage2013}. The computational costs associated with TO are mainly determined by the used finite element analysis (FEA) and associated sensitivity analysis, which limits the number of elements and consequently the design resolution.

With the growing popularity of TO, a clear need exists for improved methods that can deliver high quality results at the  lowest computational cost. Various approaches have been proposed in the past to reduce the computational costs associated with solving large-scale TO problems \cite{Wang2007, Suresh2013, Amir2014, Xia2017, Amir2009, Amir2015, Frutos2015}. These focused mainly on improving the efficiency of solving the FEA systems of equations. Another possibility that has been explored in the existing literature is to modify the way the FEA system is defined in the first place through the use of adaptive FEA formulations. Popular adaptive FEA approaches are $h$-refinement and $p$-refinement \cite{Babuska1992}. However, the standard formulations for these methods use FEA based error criteria for adaptation of the mesh. These by themselves are not well suited for TO, as they do not take the need for refinement based on design considerations into account \cite{Bruggi2011}. In the final designs obtained from TO, it is desirable that the material distribution is clearly defined. Thus, the refinement criterion used in TO should depend on the material distribution as well.

Maute and Ramm \cite{Maute1995} proposed an adaptive mesh refinement (AMR) approach which involved optimizing the topology of the design followed by approximating the boundaries using cubic or Be\'{z}ier splines. After every cycle of TO, shape optimization was performed followed by remeshing of the solid domain. The whole process was repeated over a series of cycles and the new mesh generated at the end of each cycle was used as the domain for the TO problem of the next cycle. Van Keulen and Hinton \cite{vanKeulen1996} for the first time combined the TO with an FEA error based refinement strategy. The recovery of material, in their approach, was controlled by the stress level in the adjacent elements and mesh densities were determined using (a) the standard Zienkiewicz-Zhu error estimator and (b) the shortest distance to the material-void boundary. Both these approaches involved remeshing the whole domain at the end of each cycle, which was computationally expensive.

Costa and Alves \cite{Costa2003} presented an AMR strategy which involved refining only the solid material region. For TO problems, intermediate densities are found to be prevalent near the boundaries. On the assumption that refinement of these regions can reduce the intermediate densities, Stainko \cite{Stainko2006} proposed to refine the region only around the material-void boundary. Bruggi and Verani \cite{Bruggi2011} progressed in the direction of the work proposed by \cite{vanKeulen1996}, and proposed a goal-based AMR strategy that properly guides the progression of refinement and coarsening in TO. For refinement/coarsening, a dual-weighted residual based FEA indicator as well as a heuristic density-gradient based indicator were used. While most of these methods helped to achieve the required $h$-adaptivity in TO, the fixed choice of density values for refinement at every cycle of TO led to excessive numbers of elements getting refined, thereby leading to undesired increase in computational costs. Gupta \emph{et al.} \cite{Gupta2016} proposed a heuristic scheme to control the refinement/coarsening bounds at every cycle of TO. The proposed scheme was combined with $h$-refinement and very clear material descriptions with low gray regions were obtained. Other adaptive formulations involving $h$-refinement or a similar approach include adaptive refinement of polygonal elements \cite{Xuan2017, Hoshina2018}, combining a continuous density field representation with adaptive mesh refinement \cite{Lambe2018} and efficient TO based on adaptive quadtree structures \cite{Wu2018}. 

Another possible way to reduce FEA costs is the adaptive $p$-refinement, as stated earlier, where the mesh topology remains the same. Additionally, for smooth problems, the accuracy of $p$-refinement is dramatically higher than that of $h$-refinement for the same computational costs \cite{Babuska1992}. Increasing the polynomial order of the shape functions gives an exponential rate of convergence. Other advantages of $p$-refinement are its robustness against locking effects and high aspect ratios \cite{Parvizian2007}. However, due to the fact that the conventional TO approaches assume an elementwise-constant density distribution, using higher-order shape functions inside a finite element is not an efficient approach. Although it reduces the FEA error to some extent, it cannot improve the material definition  within the element.

The recently proposed Finite Cell Method (FCM) offers new perspectives to overcome this limitation \cite{Parvizian2012}. FCM is an FE-based modeling approach where the analysis mesh is decoupled from the material distribution domain and higher order shape functions are used \cite{Parvizian2007}. This approach can handle a material-void boundary within an element through the use of appropriate integration schemes. Recently, a similar approach was proposed by Nguyen \emph{et al.} \cite{Nguyen2010} for TO, termed as multiresolution topology optimization (MTO), where the analysis and design meshes are decoupled. Here, design mesh denotes the distribution of the design points which are used to generate the material distribution. The density values associated with these points serve as optimization parameters for TO. In MTO, a coarse analysis mesh was used and inside every finite element, a large number of design points were positioned. This allowed a high resolution density distribution inside every finite element, unlike an elementwise-constant density distribution as in standard TO approaches. In spite of using low order shape functions and coarse elements, the method is still capable of generating high resolution structures, albeit with reduced analysis accuracy. To increase this accuracy, recently a $p$-version of MTO has been proposed, where the potential of higher order polynomial shape functions has been investigated in the context of MTO \cite{Nguyen2017}. Other approaches based on a similar concept were further presented in \cite{Nguyen2012, Yiqiang2013a}. Note that in \cite{Nguyen2010} and other research papers thereafter, the term `multi-resolution' refers to allowing the possibility for multiple different design resolutions for the same choice of analysis resolution. In line with these works, we also refer to our formulation as an MTO approach.

It is important to note that although the design and analysis meshes can be decoupled, the iterative updates of the design variables in TO are based on the analysis results. In a recent study, we showed that for a given finite element mesh and polynomial order of FE shape functions, there exists an upper bound on the number of design variables that can be used in TO \cite{Gupta2016a}. A density resolution beyond this threshold cannot be fully captured by the FEA and can lead to issues such as nonuniqueness. For certain cases, it can also lead to poorly performing designs. Thus, when using large numbers of design points inside an element, both for analysis accuracy as well as well-posedness of the TO problem, higher order shape functions and corresponding numerical integration schemes need to be chosen.

Parvizian \emph{et al.	}\cite{Parvizian2012} proposed a TO strategy based on FCM where a coarse analysis mesh with high order shape functions as well as a high order numerical integration scheme is used. Although expected to give more reliable results, FCM-based TO may not necessarily satisfy the bounds proposed in \cite{Gupta2016a}, which implies it might still be prone to numerical issues. Groen \emph{et al.} \cite{Groen2016} presented results related to rigorous numerical investigations of FCM-based TO. Their observations show close resemblance with those in \cite{Gupta2016a}. Also, the authors showed that using FCM-based TO, remarkable speed-ups of more than 3- and 60-folds for 2D and 3D problems, respectively, could be obtained over the traditional TO approach. However, for certain configurations of FCM-based TO, it is possible that the design consists of `QR-patterns', comprising disconnected or loosely connected material parts which cannot be correctly modeled by the employed modeling scheme \cite{Gupta2018}. Use of density filtering with a sufficient filter radius was found to suppress the QR-pattern artifacts \cite{Nguyen2017, Gupta2016a, Groen2016}, but has the undesired consequence of reducing the design resolution. Applying $p$-refinement was also found to reduce the issue, but rapidly raises the computational cost.

Hereafter, we use the term MTO to refer to all the TO approaches (including FCM-based TO) where the design and analysis discretizations are decoupled. The goal of MTO approaches is to obtain high resolution, high quality designs at low analysis costs. Possible ways to increase resolution versus cost could include using a finely discretized density mesh, reducing the filter size, using shape functions of low polynomial order to describe the state field, \emph{etc}. However, each of these approaches has certain limitations which can adversely affect the analysis accuracy. Using too many density cells and low polynomial order shape functions can lead to nonuniqueness in the design field and result in numerical instability \cite{Gupta2016a}. Reducing the filter size can lead to the formation of QR-patterns, which are numerical artefacts and can affect the model accuracy \cite{Groen2016, Gupta2018}. Using higher order shape functions can circumvent these problems, however, the analysis related costs are significantly increased. Due to this, the advantage of MTO over the traditional TO approach could be lost. In an MTO setting, this requires considering adaptivity both of the analysis and the design, which thus far has not been explored.

\indent In this work, we present an adaptive MTO approach that enables a better balance between resolution and computational costs. Local adaptation is applied to both  the analysis and the design description, which allows computational effort to be concentrated in regions of interest. Moreover, the adaptivity allows rigorous prevention of QR-pattern artefacts. We coin the term `$dp$-adaptivity', an adaptive multiresolution TO scheme where both the design resolution $d$ and FE polynomial order $p$ can be locally adapted based on certain refinement/coarsening criteria. Here, the symbol `$d$' should not be confused with the one in $hp$-$d$ adaptivity, where it refers to domain decomposition and mesh overlaying \cite{Rank1992}. It is assumed that computational costs are the limiting factor, and that the manufacturing-imposed length scale is at or below the smallest lengthscale that can be reached by the adaptive TO process. Our approach can obtain high resolution representations of the material field at significantly lower computational costs compared to non-adaptive MTO approaches. At the same time, by jointly adapting design and FE discretization, we ensure that the bounds proposed in \cite{Gupta2016a} are satisfied and instability issues are avoided. For refinement/coarsening after every TO cycle, analysis error, correctness of the design as well as the error associated with QR-patterns are used. For this purpose, we also propose a novel indicator. Various numerical tests are conducted to analyze the capabilities of the method as well as its robustness. The scope of this paper is restricted to  linear elastostatic problems and the material is assumed to be isotropic, however, the method is expected to be applicable to a wider range of problems.

In the following section, theory of multiresolution TO is presented followed by discussions related to choice of design distribution, polynomial orders and numerical integration schemes. Section \ref{dpadap} subsequently presents the theory and formulation for the proposed $dp$-adaptivity approach. The applicability of this method is presented on a set of numerical examples (Section \ref{examples}), and discussion and related conclusions are stated in Section \ref{discuss} and \ref{conclude}, respectively.  

\section{Multiresolution Topology Optimization}
\label{mtop}

\subsection{Domain and variable definitions}
\label{mtop_desc}

\begin{figure*}
\centering
\begin{tikzpicture}
    \node[anchor=south west,inner sep=0] at (0,0){
\includegraphics[scale=0.75]{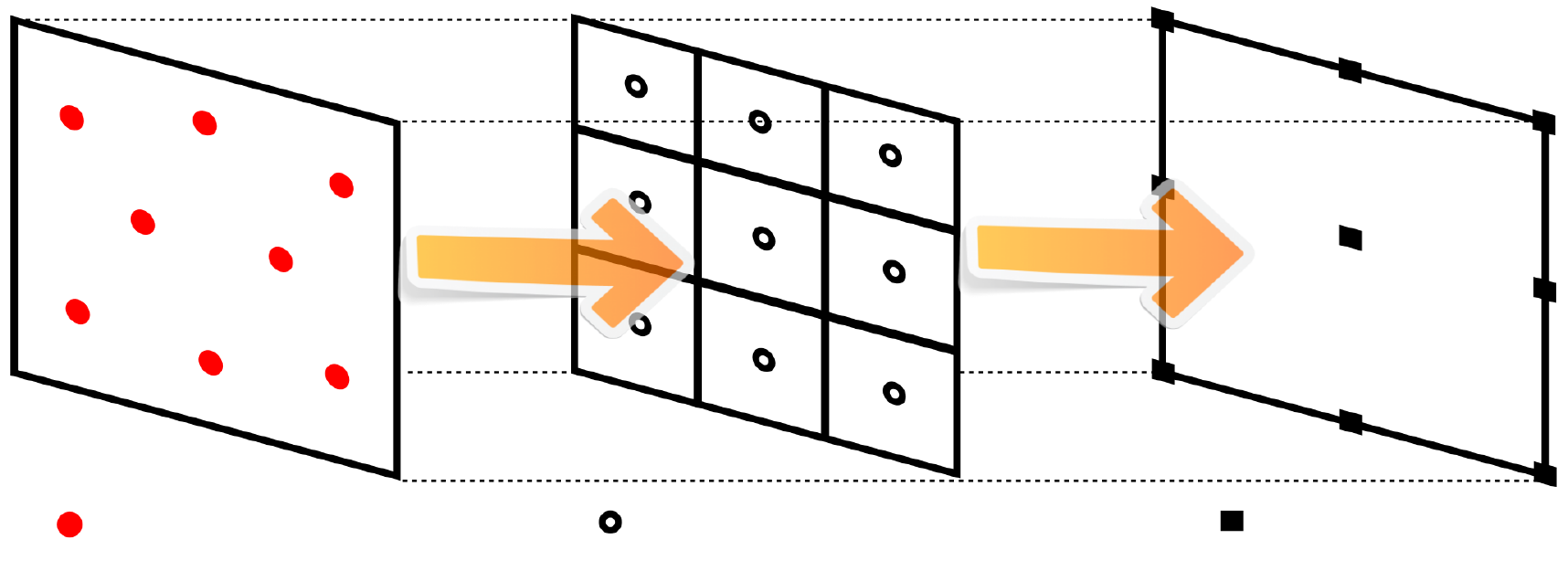}};
\node[text width=2cm] at (1.8, 0.408) 
    {\small design point};
\node[text width=3cm] at (6.8, 0.402) 
    {\small density cell-center};
\node[text width=2cm] at (11.5, 0.41) 
    {\small analysis node};
\node[text width=3cm] at (2.1, 4.9) 
    {\normalsize design domain};
\node[text width=3.2cm] at (6.4, 4.9) 
    {\normalsize background domain};
\node[text width=3cm] at (11.6, 4.9) 
    {\normalsize Q2 finite element};
	\node[text width=1cm] at (4.5, 3) 
    {\small $P_1$};
	\node[text width=1cm] at (9.2, 3.1) 
    {\small $P_2$};
\end{tikzpicture}
\caption{Schematic representation of a Q2/d8 MTO element comprising 3 linked overlapping domains. These domains represent a design domain with 8 design points (left) and a Q2 finite element (right), with a background distribution of $3 \times 3$ density cells (middle). Here, $P_1$ and $P_2$ denote the projections from the design to background domain and from background domain to the finite element, respectively. The design points are distributed in the domain using a variant of the $k$-means clustering approach (Appendix \ref{sec_kmeans}).}
\label{dp_element_1}
\end{figure*}
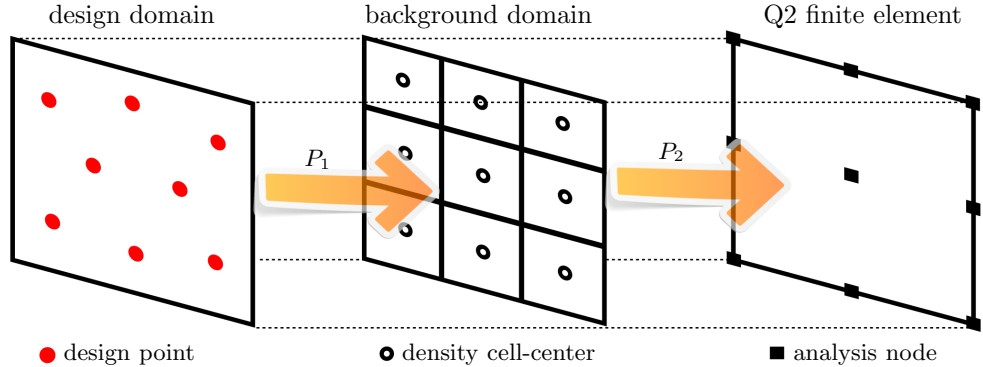

\begin{figure}
\centering
\begin{tikzpicture}
    \node[anchor=south west,inner sep=0] at (0,0){
\includegraphics[scale=0.7]{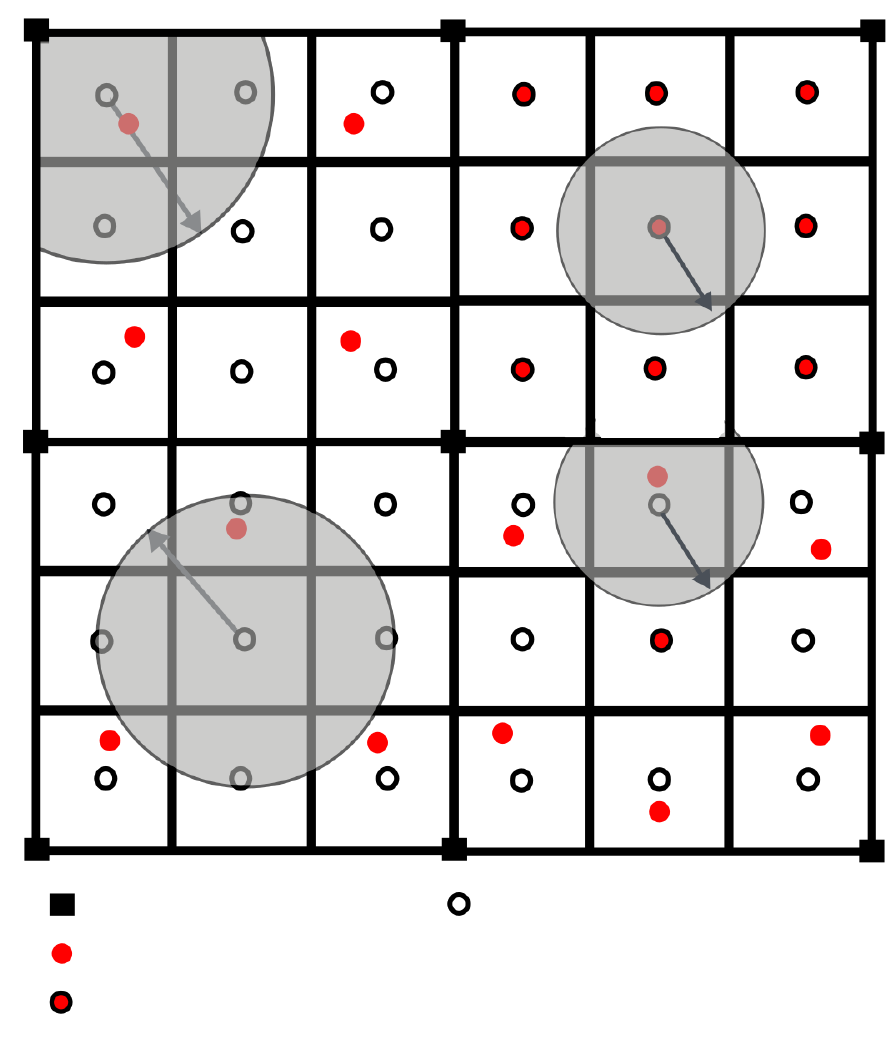}};
	\node[text width=2.5cm] at (2.0, 0.97) 
    {\small analysis node};
    \node[text width=3cm] at (5, 0.99) 
    {\small center of density cell};
    \node[text width=2.5cm] at (2.0, 0.62) 
    {\small design point};
	\node[text width=6cm] at (3.7, 0.05) 
    {\small overlap of design point and center of density cell};
\end{tikzpicture}
\caption{Schematic representation of projection $P_1$ illustrating the projection of density values from the design points in the design mesh to the centers of density cells of the background domain. Four projection regions are indicated in gray. Note that these projections are localized and operate on the design points and the density cell-centers of the same element. Here, the four MTO elements from top-left to bottom-right consist of 4, 9, 3 and 7 design points, respectively. The densities at the centers of the gray projection domains (denoted by \protect\markersix) shown in each MTO element are computed from contributions of all design points (denoted by \protect\markerfive) of the same MTO element within its projection domain.}
\label{dp_element_2a}
\end{figure}
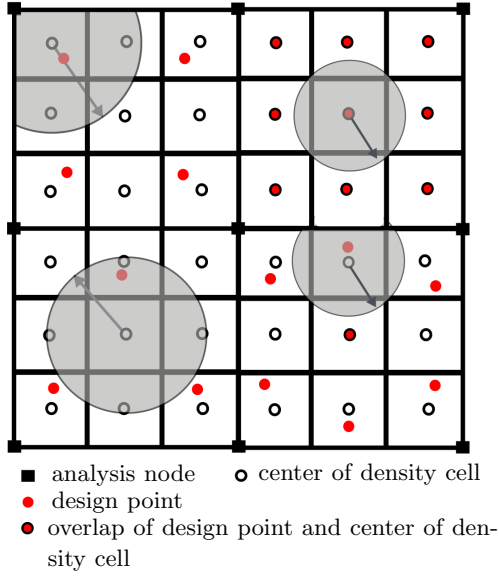

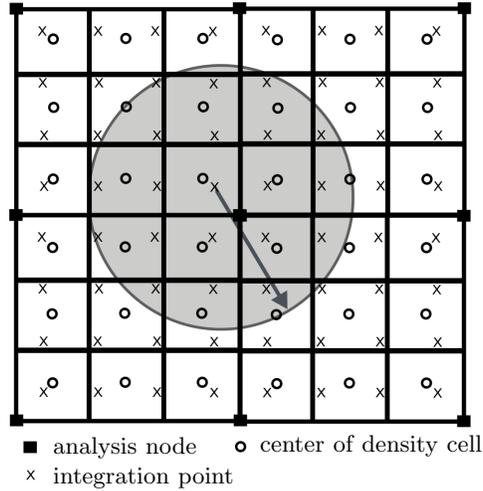
\begin{figure}
\centering
\begin{tikzpicture}
    \node[anchor=south west,inner sep=0] at (0,0){
\includegraphics[scale=0.7, trim = 0 0 1cm 0, clip=true]{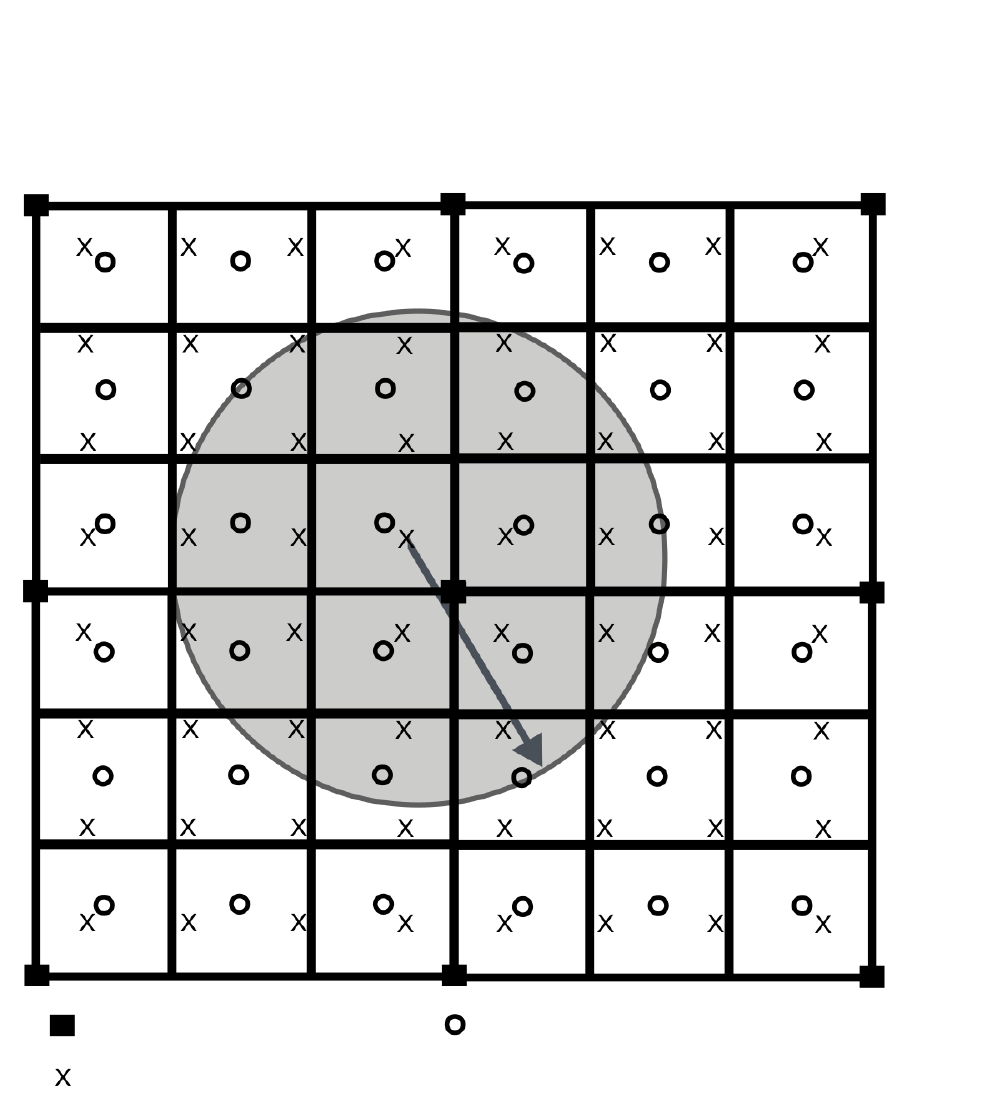}};
	\node[text width=2.5cm] at (2.0, 0.59) 
    {\small analysis node};
    \node[text width=3cm] at (5, 0.61) 
    {\small center of density cell};
	\node[text width=2.5cm] at (2.0, 0.19) 
    {\small integration point};
\end{tikzpicture}
\caption{Schematic representation of projection $P_2$ which projects density values from the background mesh to the integration points of the analysis cell. The problem domain is discretized using $2 \times 2$ MTO elements composed of Q1 finite elements and $3 \times 3$ density cells per element. For numerical integration, a $4 \times 4$ Gaussian quadrature rule is used. The density at the integration point at the center of the gray projection domain (denoted by $\times$) is computed from contributions of all background density cell center points (denoted by \protect\markersix) within its projection domain.}
\label{dp_element_2}
\end{figure}

\indent In this work, we propose an adaptive MTO formulation based on selective refinement/coarsening of the design as well as analysis domains. First a conceptual description is provided, whereas the mathematical formulation follows in Section \ref{mto_math}. The proposed approach uses three meshes: design mesh, background mesh (comprising density cells) and analysis mesh. The analysis mesh is used to determine the solution of the physics at hand (\emph{e.g.} displacement field) and the design mesh represents the distribution of design points in the domain. For simplicity, we use a structured mesh setting, as often used in topology optimization. In an adaptive setting, the analysis resolution and distribution of design points in the domain can be non-uniform. The background mesh is added to provide a convenient link between the analysis and design meshes. More details related to the role of the background mesh follow later in this section.

For practical implementation, we introduce the notion of MTO elements. An MTO element comprises a finite element, a set of design points and an overlapping background element comprising a regular grid of density cells. They all occupy the same spatial domain, and this ordered arrangement is chosen to simplify implementation in an existing FE framework. For example, \mbox{Fig. \ref{dp_element_1}} shows the schematic representation of a Q2/d8 MTO element using a Q2 (bilinear quadrilateral) finite element and consisting of 8 design points distributed non-uniformly in the domain. The overlapping background element comprises $3 \times 3$ density cells. A density design variable is associated with each design point. During optimization, these density variables are updated at every iteration based on the response functions and the corresponding design sensitivities. 

To generate suitably uniform distributions of design points within an element for any number of design variables, a variant of the $k$-means clustering method is used \cite{Mackay2003, Arthur2007}. This approach divides the design domain into $k$ segments (clusters) with roughly equal areas. The design points are assumed to be located at the centroids of these clusters. For self-containment, the details of the method are discussed in Appendix \ref{sec_kmeans}. We use this approach to obtain an approximately uniform distribution of any given number of design points in the MTO element domain. The achievable resolution limit of the design depends on the spacing between the design points. For a given number of design points and without a priori knowledge of the optimal design, a uniform distribution allows the best possible resolution. Note  here that the proposed adaptive MTO approach is independent of the choice of methodology for the distribution of design points, and any other method to distribute points in a domain can be applied, including a set of predefined patterns. 

The aligned background mesh consists of a uniform grid of equally-sized density cells in the whole domain, such that a certain number of these cells overlap with every finite element. For these density cells, the respective finite element is referred as the parent analysis cell. For example, in Fig. \ref{dp_element_1}, $3 \times 3$ density cells overlap with the parent Q2 finite element (analysis cell). The density is defined at the centroid of every density cell and is assumed to be constant inside it. This density is obtained from the design mesh through a localized projection operation.

The density inside any density cell of the background mesh is calculated using projection $P_1$ (as shown in Fig. \ref{dp_element_1}, defined in detail in Section \ref{mto_math}), and only those design points are used which lie within the same MTO element. The role of the localized projection is to define density values in all the density cells of the respective MTO element. The projection is restricted to the considered MTO element for two reasons: (i) to minimize the associated loss in design resolution of MTO elements adjacent to other MTO elements with fewer design points and (ii) to enable element-level implementation. While choosing the local projection radius $P_1$, it needs to be ensured that the density inside each density cell can be defined. The mathematical details related to choosing this projection radius are provided in Section \ref{mto_math}. An example is presented in Fig. \ref{dp_element_2a}, which shows a domain of $2 \times 2$ MTO elements, each comprising a Q1 finite element and $3 \times 3$ density cells. As can be seen, the distribution of design points can be non-uniform. The four MTO elements from top-left to bottom-right consist of 4, 9, 3, and 7 design points, respectively. In the bottom-right MTO element shown in Fig. \ref{dp_element_2a}, a partial projection circle can be seen, which is due to the fact that the projection is restricted to within this MTO element. Mathematical details related to projection $P_2$ are provided in Section \ref{mto_math}.

The stiffness matrix for every MTO element is obtained by numerical integration using a Gaussian quadrature scheme. For this purpose, the stiffness matrix contribution at the integration point needs to be known, which in turn requires knowing the density value at that point. This density value, referred further as `projected density', is obtained through a projection on the background mesh, denoted by $P_2$ (Fig. \ref{dp_element_1}). Fig. \ref{dp_element_2} illustrates how these density values are computed. It shows a mesh of $2 \times 2$ MTO elements, comprising Q1 finite elements and the corresponding background domain with $3 \times 3$ density cells per element. Here, `Q1' refers to quadrilateral finite elements with shape functions of polynomial order 1. Similar to the approach described in \cite{Nguyen2010, Nguyen2017, Groen2016}, the projected densities are computed using a distance-weighted projection of design densities found in the neighborhood of a certain radius $R$ over the background mesh. In this work, density filtering is used for the projection \cite{Bruns2001}.

The use of the background mesh facilitates $d$-adaptivity, \emph{i.e.} the use of different numbers of design points in adjacent elements. In the absence of the background mesh, the non-uniform design field when directly projected on the analysis mesh, can lead to irregular boundary features which are not desired. The design variables are not directly linked to the density cells of the background mesh, because it would not allow an adaptive formulation anymore. Moreover, such a formulation would significantly increase the number of design variables and would lead to nonuniqueness related \mbox{issues \cite{Gupta2016a}}. The background mesh provides the flexibility of having a reference discretization  independent of the number of design variables. Moreover, it simplifies the numerical integration required for the stiffness matrix.

\subsection{Mathematical formulation}
\label{mto_math}
In this paper, the applicability of a $dp$-adaptive MTO approach is demonstrated on mechanical problems of two different types: minimum compliance and compliant mechanism.

For the chosen problems, the problem statement for TO can be expressed as
\begin{align}
& \min_{\boldsymbol \rho} \mathcal{J}(\mathbf{u}, \boldsymbol \rho) = \mathbf{z}^{\intercal}\mathbf{u},	\nonumber \\
& \text{s.t.} \enskip \mathbf{Ku = f},	\nonumber \label{eq_comp_1} \\ 
& V(\boldsymbol \rho) \leq V_0,	\\ 
& \mathbf{0} \leq \boldsymbol \rho \leq \mathbf{1} \nonumber,
\end{align}
where, $\mathcal{J}(\mathcal{\cdot})$ denotes the objective functional, and $\mathbf{K}$, $\mathbf{u}$ and $\mathbf{f}$ denote the global stiffness matrix, displacement vector and load vector, respectively. The vector $\mathbf{z}$ is chosen based on the type of problem and will be discussed in Section \ref{define_test}. The volume constraint restricts the total volume fraction of the given material to be less than certain predefined volume $V_0$.

Next, the details related to various steps  associated with the proposed multiresolution modeling approach are described. The matrix $\mathbf{K}$ in Eq. \ref{eq_comp_1} is obtained from the global assembly of the element stiffness matrices $\mathbf{K}_e$, which can be expressed as
\begin{equation}
\mathbf{K}_e = \int_{\mathrm{\Omega_e}}\mathbf{B}^\intercal\mathbf{DB}d\mathrm{\Omega} = \sum_{i=1}^{N_g}\mathbf{B}_i^\intercal\mathbf{D}_i\mathbf{B}_iw_i,
\end{equation}
where $\mathbf{B}$ and $\mathbf{D}$ denote the strain-displacement matrix and constitutive matrix, respectively, and $N_g$ is the number of integration points. More details related to the choice of numerical integration are discussed in Appendix \ref{num_int}. The subscript $i$ refers to the $i^{\text{th}}$ integration point and $w_i$ denotes the respective integration weight. The construction of the $\mathbf{D}$ matrix depends on the choice of the material interpolation model as well as the material itself. In this work, solid isotropic material interpolation (SIMP) model \cite{Bendsoe1989} is used such that
\begin{equation}
\mathbf{D}_i = \left(E_{\text{min}} + \tilde{\rho}_i^q(E_0 - E_{\text{min}})\right)\mathbf{D}_0,
\end{equation}
where $E_0$ is the Young's modulus of the solid material and $E_{\text{min}}$ is a very small value (typically $10^{-9}E_0$) used to avoid singularity of the system stiffness matrix. Also, $\tilde{\rho}_i$ denotes the density calculated at the $i^{\text{th}}$ integration point, $q$ is the penalization power and $\mathbf{D}_0$ denotes constitutive matrix normalized by the Young's modulus.

The densities at the integration points are calculated by projecting density values from the density cells in the background mesh (Fig \ref{dp_element_2}). For this purpose, we employ a linear projection approach for $P_2$ based on the density filtering method which is widely used in TO \cite{Bruns2001}. Mathematically, it can be stated as
\begin{equation}
\tilde{\rho}_i = \frac{1}{\sum_{j=1}^{n_{\hat{\rho}}}H_{ij}} \sum_{j=1}^{n_{\hat{\rho}}}H_{ij}\hat{\rho}_j,
\end{equation} 
where $\hat{\rho}$ refers to density values for the cells contained in the background mesh with their centers lying within a distance $R$ from the corresponding integration point  (Fig. \ref{dp_element_2}), and their number is denoted by $n_{\hat{\rho}}$. Here, terms $H_{ij}$ reduce linearly with distance from the integration point, \emph{i.e.}, 
\begin{equation}
H_{ij} = R - \text{dist}(i, j),
\end{equation}
where $\text{dist}(\cdot)$ denotes the Euclidean distance operator.

As stated in Section \ref{mtop_desc}, the background mesh densities are calculated using the $P_2$ projection from the design mesh to the background mesh. For the $p^{\text{th}}$ MTO element, the density of the $q^{\text{th}}$ density cell is given as
\begin{equation}
\hat{\rho}_q^{(p)} = \frac{1}{\sum_{s=1}^{n_{\rho}}h_{qs}} \sum_{s=1}^{n_{\rho}}h_{qs}\rho_s,
\end{equation} 
where, $\rho_s$ refers to the density value associated with the $s^{\text{th}}$ design point in the design domain contained within the $p^{\text{th}}$ MTO element, and lying within a distance $r_p$ from the centroid of its $q^{\text{th}}$ density cell. The number of such design points is denoted by $n_{\rho}$, and $r_p$ is the radius of the projection for the $p^{\text{th}}$ element (Fig. \ref{dp_element_2a}). Here, $h_{qs}$ is defined as
\begin{equation}
h_{qs} = r_{p} - \text{dist}(q, s).
\label{eq_projr}
\end{equation}
As stated earlier, the projection radius $r_p$ needs to be chosen such that it is as small as possible, however, large enough to define densities for all the density cells that correspond to the respective element. Here, we define it as
\begin{equation}
r_{p} = 1.04 (dim)^{0.5} \frac{L_p}{\lceil d^{1/dim} \rceil},
\label{eq_rp_emp}
\end{equation}
where $dim$ denotes problem dimension, and $L_p$ is the edge-length of the $p^{\text{th}}$ MTO element. The operator $\lceil \cdot \rceil$ denotes ceiling function which rounds the contained floating-point number to the nearest greater integer value. The term $ \frac{L_p}{\lceil d^{1/dim} \rceil}$ refers to edge-length of the density cells. Next, to obtain a projection length slightly larger than the diagonal, we multiply by $1.04(dim)^{0.5}$. Note that Eq. \ref{eq_rp_emp} has been obtained empirically through observations based on various design distributions obtained using the $k$-means clustering approach. For other approaches of choosing the locations of design points, where for any value of $d$, the distance between the design points can be provided mathematically, it is possible that even lower values of $r_p$ work. Lower values of $r_p$ can help to further reduce the loss in design resolution caused due to the choice of localized projection $P_1$, and this could be a potential direction for future research. 

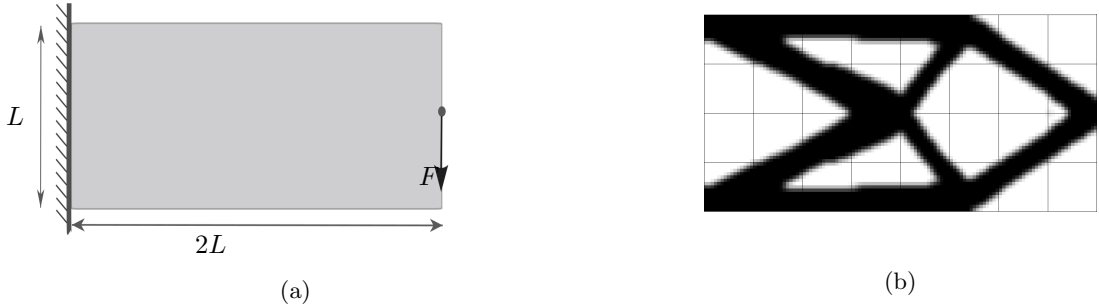
\begin{figure}
\centering
\begin{subfigure}{0.48\textwidth}
	\begin{tikzpicture}
\node[inner sep=0pt](design1) at (0, 0)
{\includegraphics[width=5.8cm, height=3.2cm, trim = 0 0 2cm 1.6cm, clip=true
]{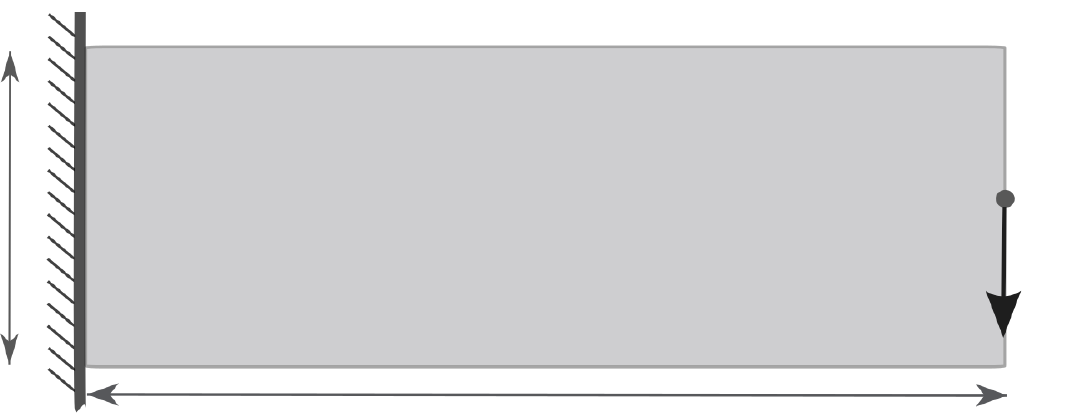}};
\node[] at (2.3, -0.8) {$F$};
\node[] at (-0.6, -1.7) {$2L$};
\node[] at (-3.2, -0.0) {$L$};
\end{tikzpicture}
\caption{}
	\label{cant_point_load_fig1}
\end{subfigure}
\begin{subfigure}{0.48\textwidth}
\centering
	\includegraphics[scale=0.13, trim = 0cm -4.2cm 0cm -1cm, clip=true]{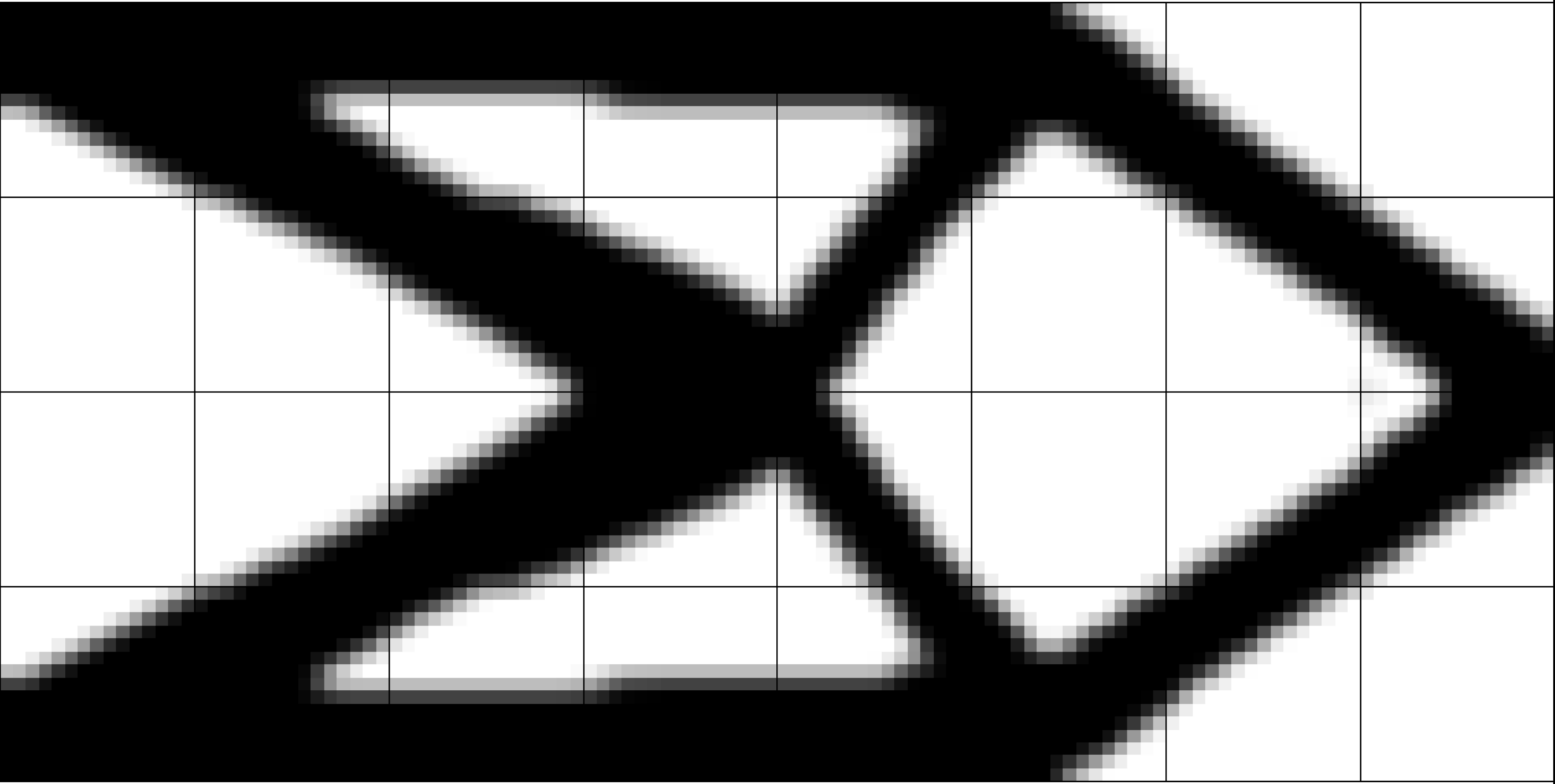}
	\caption{}
	\label{cant_point_load_fig2} 
\end{subfigure}
\caption{(a) Example of a cantilever beam subjected to a point load, and (b) the optimized design obtained using MTO for compliance minimization using $8 \times 4$ Q10/d225 elements and $R = 0.13h$. Here, Q10/d225 refers to a quadrilateral finite element with shape functions of polynomial order 10 and 225 design points.}
\end{figure}
Fig. \ref{cant_point_load_fig1} shows an example of a cantilever beam subjected to a point load, which we will use to illustrate the MTO concept. The domain is discretized using $8 \times 4$ finite elements. For each MTO element, 225 design points, distributed in a square grid of $15 \times 15$, are used to represent the design field. The polynomial order of the shape functions is chosen to be 10.  The choice of shape functions is made in a way that the element-level uniqueness bounds defined in \cite{Gupta2016a} are not violated. As per the uniqueness bound, the number of design points influencing any finite element cannot be greater than the number of deformation modes of that element, With $p$ equal to 10, the number of deformation modes is 239, which is greater than 225. With $p$ and $d$ equal to 10 and 225, respectively, the MTO elements are referred as Q10/d225 type elements. For this example, the projection radius $R$ is set to 0.13 times the element-length, which is equivalent to the size of 2 density cells.

Fig. \ref{cant_point_load_fig2} shows the optimized design obtained using the outlined MTO configuration. Clearly, the employed MTO approach allows the definition of higher resolution material features on relatively coarser MTO elements. However, in Fig. \ref{cant_point_load_fig2}, there are parts of the domain where even lower-order elements and lower design resolution are sufficient. For example, there are totally void MTO elements, where even linear shape functions with only one design point can be used. Clearly, the computational time of the MTO approach can be reduced by exploiting this fact in an efficient way, and in the next section, we propose an approach to do this.

\section{\emph{dp}-adaptivity}
\label{dpadap}
\subsection{General description of the method}
\label{dp_general}
We present here a $dp$-adaptive version of the MTO method which is capable of enhancing further the ratio between the design resolution and analysis cost compared to non-adaptive MTO. The proposed MTO method efficiently distributes the design variables and locally adapts (increases/decreases) the polynomial order of the shape functions. A three-part refinement criterion is defined to select the cells to be refined/coarsened. Note that although the term `refinement' is more commonly used throughout this paper, we implicitly refer to coarsening (reducing the values of $p$ and $d$) as well. Here, `refined' cells are those where additional design points are inserted, or the polynomial order of the shape functions is increased, or both. Similarly, `coarsened' cells are the ones where the design resolution (number of design points) is reduced, or the analysis resolution (shape function order) is reduced, or both. With an adaptive formulation, fewer design variables as well as analysis nodes are used, which provides a computational advantage over the conventional MTO method.

At the start of $dp$-adaptive MTO, a cycle of TO is performed, using a certain initial uniform design- and FE-discretization. A `TO cycle' refers to the entire process from starting with an initial design and optimizing it over a number of iterations (or up to a certain stopping threshold) to reaching an improved design. During a TO cycle, the shape function order and design points of all elements remain fixed. In the optimized design, refinement and coarsening zones are subsequently identified based on an integrated criterion comprising an analysis error-based indicator, a density-based indicator, and a QR-based indicator. Here, QR-error refers to the error due to the incapability of the chosen shape function in modeling the displacement field arising from a high-resolution density representation allowed within that element \cite{Gupta2018}. More details related to these indicators are discussed in \mbox{Section \ref{sec_meshref}}.

All steps from analyzing the design for refinement to updating the $d$ and $p$ values for the whole domain, constitute one cycle of $dp$-adaptivity. The general structure of a $dp$-adaptive MTO cycle is as follows:
\begin{enumerate}
\item Perform optimization of an MTO problem with fixed $p$ and $d$ values.
\item Adapt $p$ values based on analysis error indicator.
\item Adapt $p$ and $d$ values based on density-based criterion.
\item Update $p$ values to reduce QR-errors in every element.
\end{enumerate}
With the new $dp$-adapted mesh, the next cycle of TO is performed. Section \ref{sec_dpalgo} below describes each of the above steps in detail.

\subsection{Refinement criteria}
\label{sec_meshref}

In this section, the details related to the three indicators used in our refinement criterion are provided. As stated earlier, although the term `refinement' is frequently used, we implicitly refer to `coarsening' as well in our adaptive approach. Note that although here certain choices have been made for the refinement indicators, the $dp$-adaptive scheme in itself is not dependent on the choice of refinement indicator, and can be coupled with other appropriate indicators as well.
\subsubsection{Analysis-based refinement indicator}
\label{sec_ana_indic}

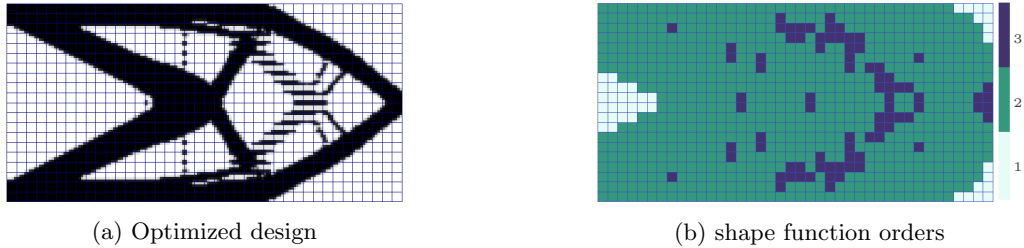
\begin{figure*}
\centering
\begin{subfigure}{0.48\textwidth}
\centering
\begin{tikzpicture}
    \node[anchor=south west,inner sep=0] at (0,0){
\includegraphics[scale=0.2]{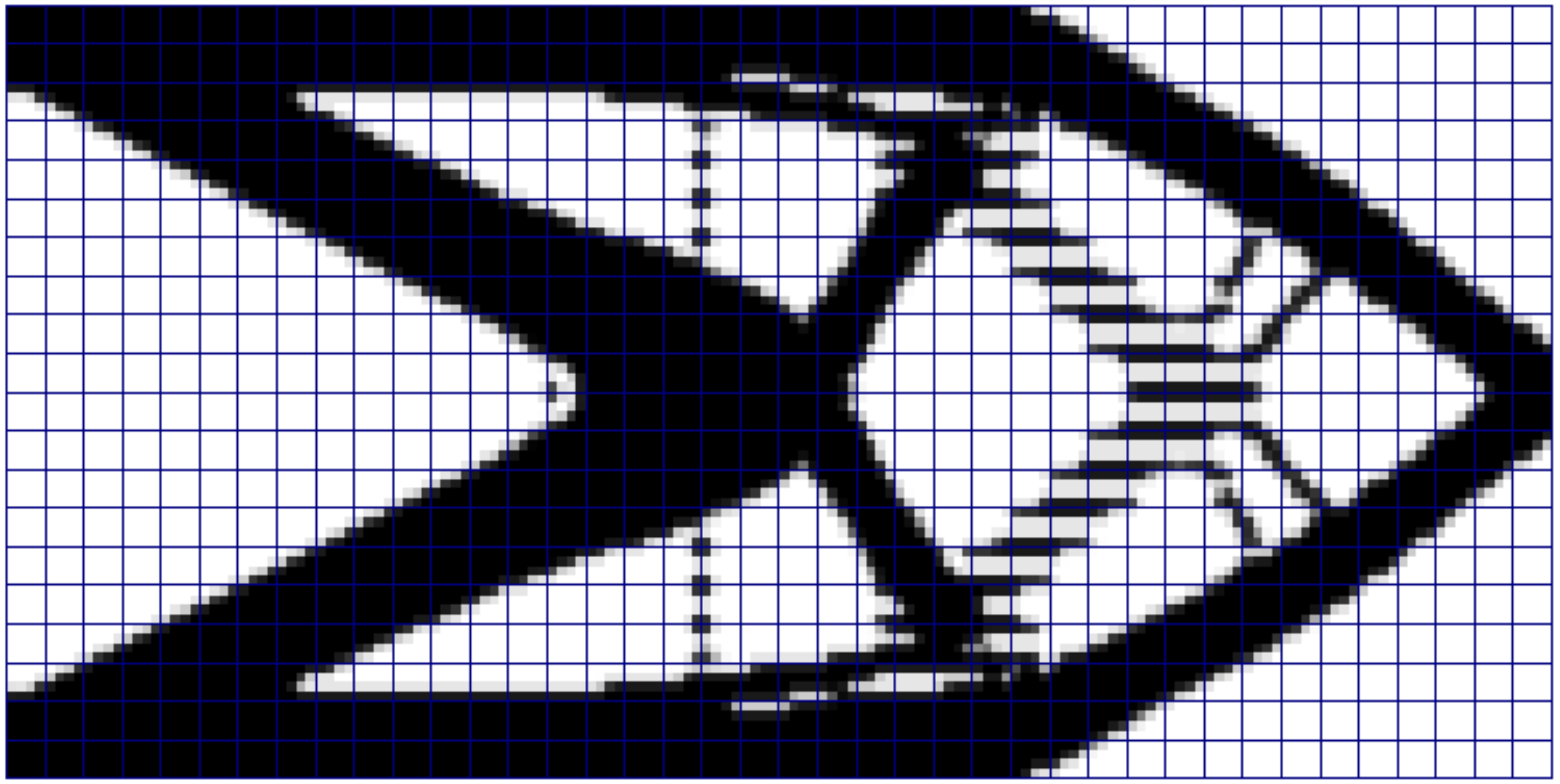}};
\end{tikzpicture}
\caption{Optimized design}
\label{fig_only_anal_a}
\end{subfigure}
\begin{subfigure}{0.48\textwidth}
\centering
\begin{tikzpicture}
    \node[anchor=south west,inner sep=0] at (0,0){
\includegraphics[scale=0.2]{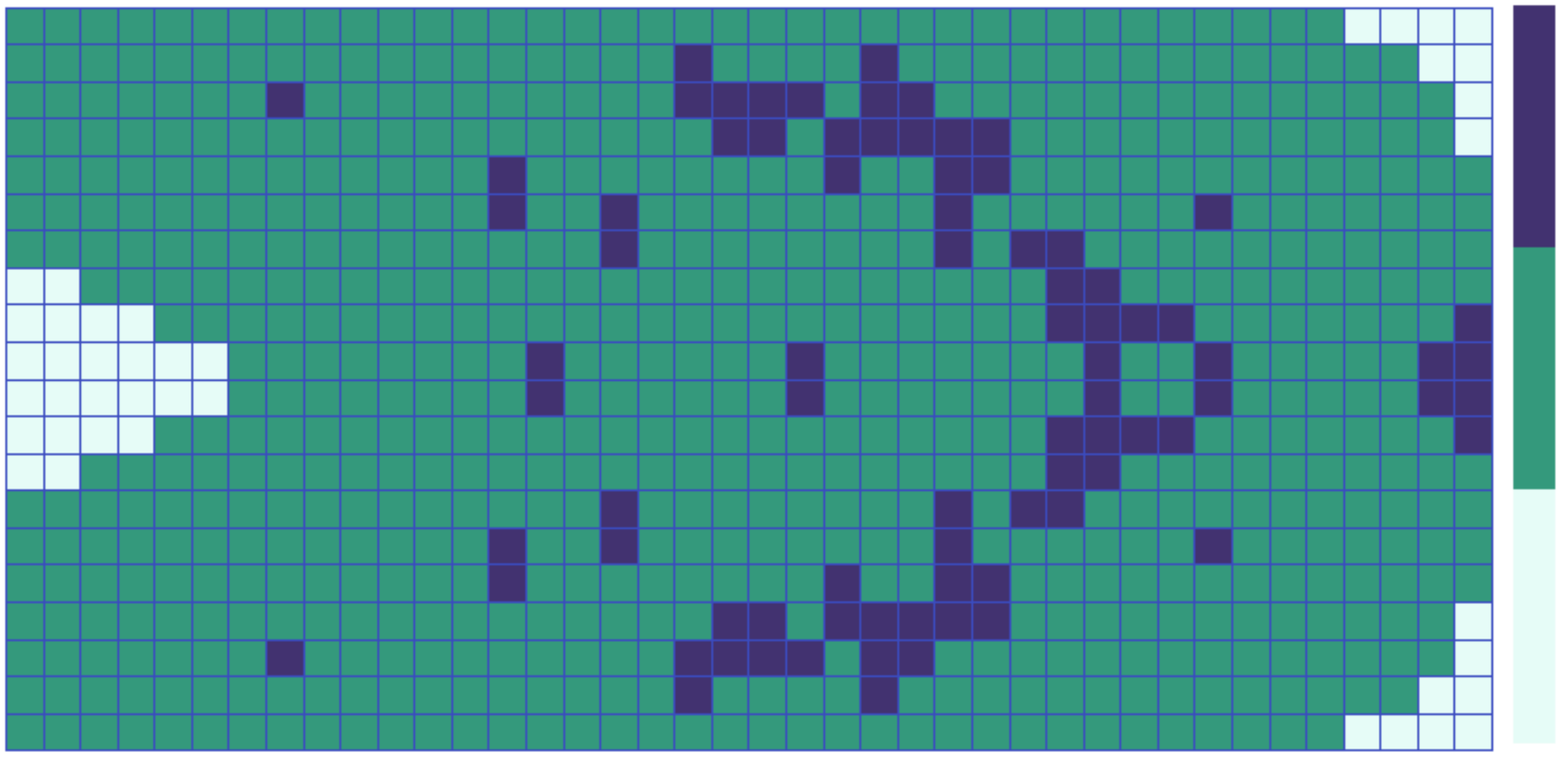}};
\node[text width=0cm] at (5.55, 2.2) 
    {\tiny \color{darkgray}{3}};
    \node[text width=0cm] at (5.55, 1.35) 
    {\tiny \color{darkgray}{2}};
\node[text width=0cm] at (5.55, 0.5) 
    {\tiny \color{darkgray}{1}};
\end{tikzpicture}
\caption{shape function orders}
\label{fig_only_anal_b}
\end{subfigure}
\caption{Optimized design (left), and the distribution of shape function orders (right) obtained from adaptive refinement controlled by only analysis-based refinement indicator for a cantilever subjected to point load, as shown in Fig. \ref{cant_point_load_fig1}. The optimized design clearly shows typical artefacts (QR-patterns) of disconnected structural features. The initial mesh comprises $40 \times 20$ Q2 finite elements with 16 design points and $4 \times 4$ density cells per element. Based on the ranking of analysis-based refinement indicator values, top 10\% and bottom 5\% of the MTO elements have been chosen for refinement and coarsening, respectively.}
\label{fig_only_analysis}
\end{figure*}

For the purpose of analyzing the modeling related error, the Kelly error estimator has been used \cite{Kelly1983}. This error indicator analyzes the jump in the gradient of the solution $\mathbf{u}$ across any face (edge in 2D) of adjacent elements. The error for any element is calculated in a relative sense by integrating the error in the gradient jump across all faces of the respective element. Based on the relative error estimate, only a certain fraction of the MTO elements is selected for updating the orders of the polynomials ($p$). This error estimator can also be understood as a gradient recovery estimator, for details on this aspect, see \cite{Ainsworth2000}. 

There are two reasons to choose the Kelly error estimator instead of more sophisticated recent approaches, \emph{e.g.}, goal-oriented error estimators \cite{ Bruggi2011, Gratsch2005}. The analysis error comprises primarily of two components: \emph{element residual} and \emph{edge residual} \cite{Gratsch2005}. Element residual refers to the error in approximating the gradient field within the element, and edge residual denotes the jumps in gradient across the element edges. The element residual is being taken into account through the QR-error analysis. Thus, the analysis indicator needs to only look at the edge residual term. Moreover, our approach requires only a relative error estimate and not the exact error itself. The use of Kelly error estimator suffices both these requirements. Also, this error estimator is simple to implement and the associated computational costs are negligible.   

For the purpose of ranking the elements for $p$-adaptivity based on the Kelly error estimator, the analysis residual error vector $\boldsymbol\Gamma^a$ needs to be defined. For the $i^\text{th}$ MTO element, $\Gamma^a_i$ can be computed as: 
\begin{equation}
\Gamma^a_i = \sum_{F \in \partial i} c_F \int_{\partial i_F} \left[ \frac{\partial \mathbf{u}}{\partial \mathbf{n}}\right]^2 d\mathbf{s},
\end{equation}
where, $F$ refers to a face (edge in 2D) of the element and operator $[\cdot]$ denotes the jump in the argument across face $F$. Also, $\partial i$ denotes the set of all faces of the element. The constant term $c_F$ is set to $\frac{h_F}{2p_F}$, where $h_F$ is the element diagonal and $p_F$ denotes the maximum among the polynomial degrees of the adjacent elements \cite{Zienkiewicz1981}. The residual errors $\boldsymbol\Gamma^a$ are ranked, and the top 10\% and bottom 5\% of the elements are selected for increasing and decreasing the $p$ values, respectively.

For illustration purposes, we perform a partial adaptive MTO run on the problem shown in \mbox{Fig. \ref{cant_point_load_fig1}}. Fig. \ref{fig_only_anal_a} shows the optimized cantilever beam design obtained for this problem after one TO cycle. The design has been obtained on a mesh of $40 \times 20$ Q2 finite elements with $4 \times 4$ design points per element.  The optimized design clearly shows typical artefacts (QR-patterns) of disconnected structural features. Fig. \ref{fig_only_anal_b} shows the distribution of polynomial shape function orders obtained from $p$-adaptivity controlled by only the analysis-based refinement indicator. It is observed that coarsening (reduction in $p$) has mainly occurred in the void cells which are far from material-void boundaries. This is because the jumps in displacement gradients across the edges for these elements are zero. For refinement (increase in $p$), the elements at the boundary have been preferred.

\subsubsection{Density-based refinement indicator}
\label{sec_des_indic}
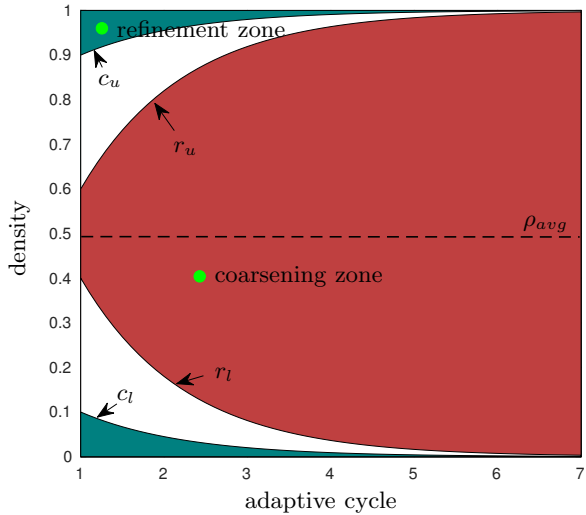
\begin{figure}
\centering
\begin{tikzpicture}
    \node[anchor=south west,inner sep=0] at (0,0){
\includegraphics[scale= 0.6]{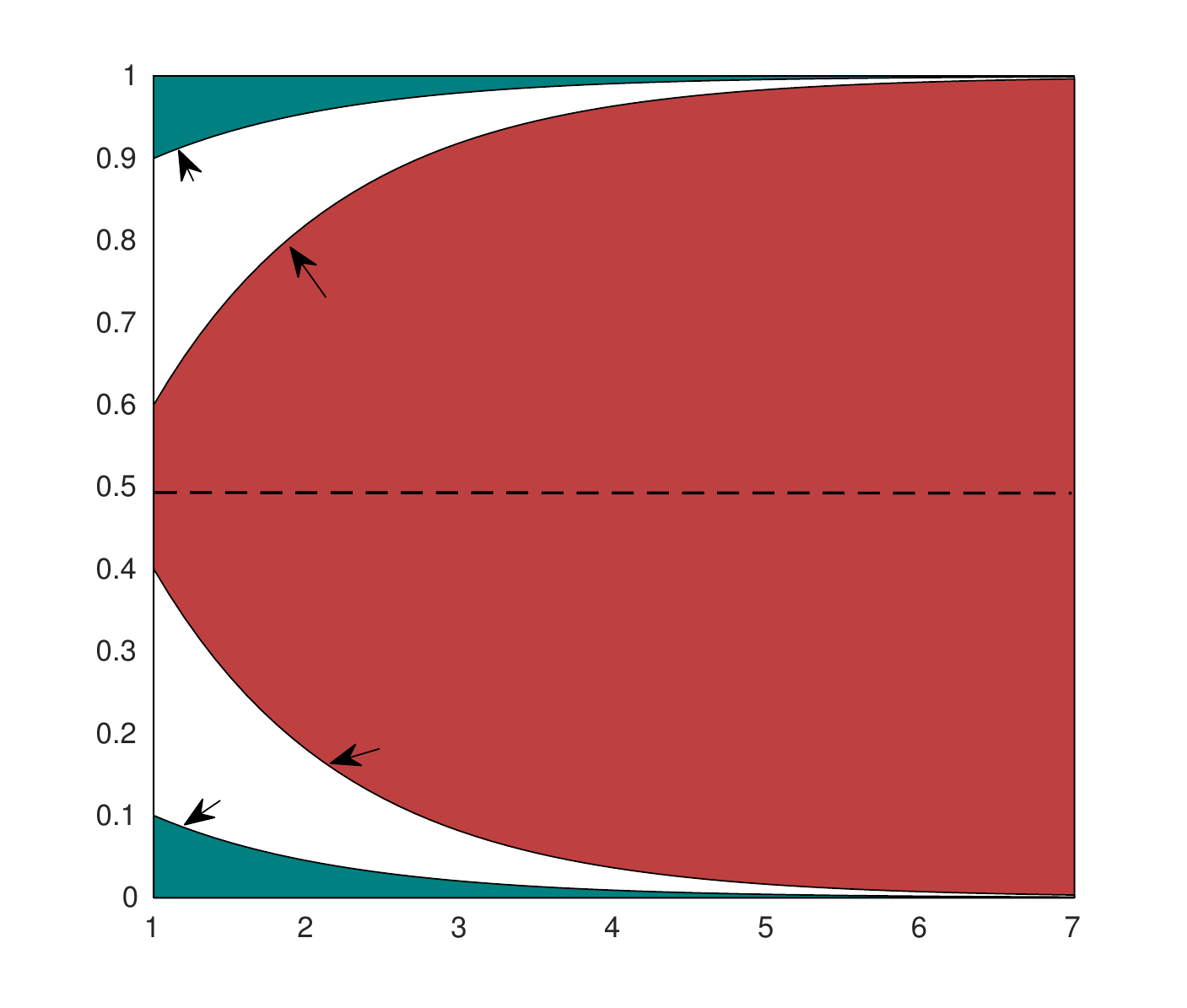}};
\node[rotate=90, text width=4cm] at (0.3, 5.2) 
    {\small density};
\node[text width=4cm] at (5.3, 0.2) 
    {\small adaptive cycle};
\draw[green,fill= green] (2.7, 3.2) circle (.5ex);
\node[text width=4cm] at (4.9, 3.2) 
    {\small coarsening zone};
\draw[green,fill= green] (1.4, 6.5) circle (.5ex);
\node[text width=4cm] at (3.6, 6.5) 
    {\small refinement zone};
\node[text width=4cm] at (3.35, 5.8) 
    {\small $c_u$};
\node[text width=4cm] at (3.6, 1.6) 
    {\small $c_l$};
\node[text width=4cm] at (4.35, 4.9) 
    {\small $r_u$};
\node[text width=4cm] at (4.9, 1.9) 
    {\small $r_l$};
\node[text width=4cm] at (9, 3.9) 
    {\small $\rho_{avg}$};
\end{tikzpicture}
\caption{Bounds for the design refinement indicator as a function of the adaptive cycle \cite{Gupta2016}.}
\label{fig_density_bounds}
\end{figure}

\begin{figure*}[!htb]
\centering
\begin{subfigure}{0.48\textwidth}
\centering
\begin{tikzpicture}
    \node[anchor=south west,inner sep=0] at (0,0){
\includegraphics[scale=0.2]{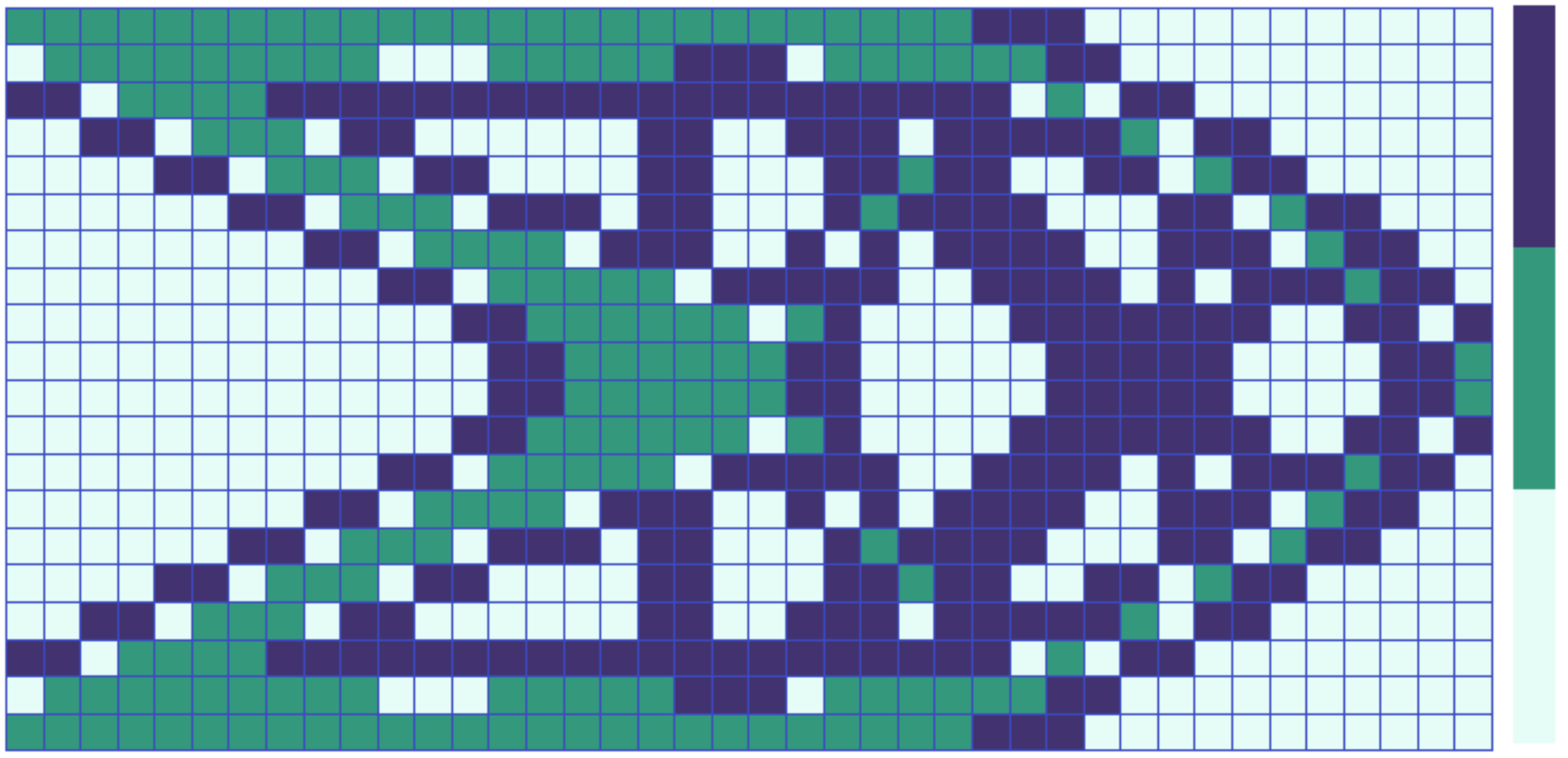}};
\node[text width=0cm] at (5.55, 2.2) 
    {\tiny \color{darkgray}{3}};
    \node[text width=0cm] at (5.55, 1.35) 
    {\tiny \color{darkgray}{2}};
\node[text width=0cm] at (5.55, 0.5) 
    {\tiny \color{darkgray}{1}};
\end{tikzpicture}
\caption{shape function orders}
\label{fig_den_a}
\end{subfigure}
\begin{subfigure}{0.48\textwidth}
\centering
\begin{tikzpicture}
    \node[anchor=south west,inner sep=0] at (0,0){
\includegraphics[scale=0.2]{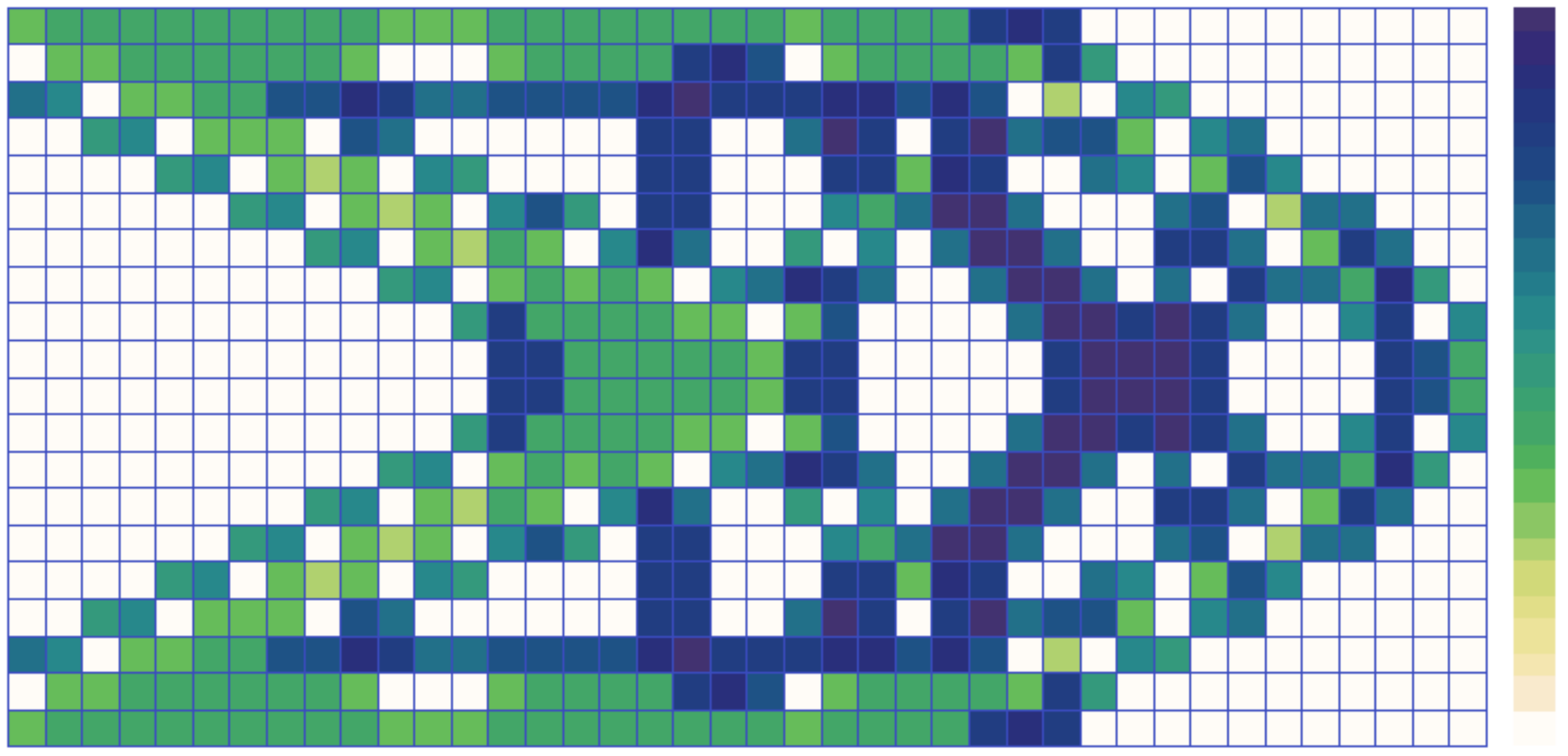}};
\node[text width=0cm] at (5.6, 2.6) 
    {\tiny \color{darkgray}{29}};
    \node[text width=0cm] at (5.6, 1.975) 
    {\tiny \color{darkgray}{23}};
    \node[text width=0cm] at (5.6, 1.35) 
    {\tiny \color{darkgray}{17}};
    \node[text width=0cm] at (5.6, 0.725) 
    {\tiny \color{darkgray}{11}};
\node[text width=0cm] at (5.6, 0.1) 
    {\tiny \color{darkgray}{5}};
\end{tikzpicture}
\caption{design field}
\label{fig_den_b}
\end{subfigure}
\caption{Distribution of polynomial orders of the shape functions (left) and the design field (right), showing the number of design points per MTO element, obtained from adaptive refinement (cycle 1) controlled by only density-based refinement indicator for a cantilever subjected to point load, as shown in Fig. \ref{cant_point_load_fig1}. The initial mesh comprises $40 \times 20$ Q2 finite elements with 16 design points and $4 \times 4$ density cells per element. The optimized design used for adaptive refinement is shown is shown in Fig. \ref{fig_only_analysis}.}
\label{fig_only_density}
\end{figure*}
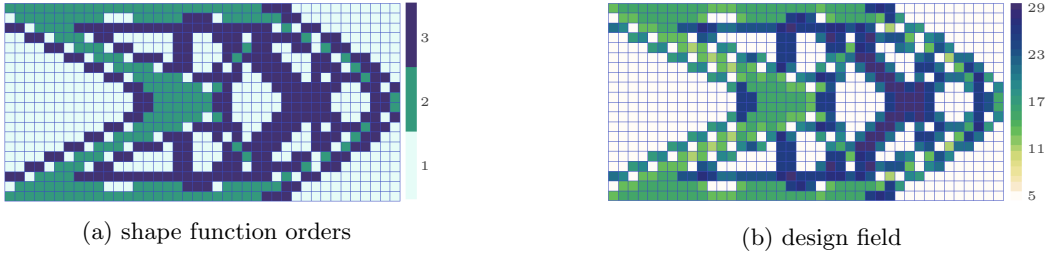

\indent The density-based refinement indicator aims at adaptively choosing MTO elements for refinement/coarsening in way that over a number of cycles, the intermediate densities are reduced, and a crisp and high-resolution boundary representation is obtained. For this purpose, the refinement indicator proposed in \cite{Gupta2016} is adapted for our problem and discussed here. This indicator chooses a certain element for refinement/coarsening based on the density value inside that element. For every cycle of adaptivity, refinement (coarsening) density intervals are defined and associated elements are flagged. We adopt this indicator to regulate the number of design points in each MTO element, based on spatial design information specified by the density values of the voxels of the background mesh. The way this indicator affects the number of design variables is discussed in Section \ref{sec_dpalgo}, here we focus on the definition of the indicator itself.

\indent Fig. \ref{fig_density_bounds} shows the refinement ($r_l \leq \rho \leq r_u$) and coarsening ($\rho < c_l$ or $\rho > c_u$) intervals as a function of adaptive cycle. Unlike the other refinement indicators, here the refinement (coarsening) bounds are chosen not to remain constant. Rather, following \cite{Gupta2016}, the range of density values to be chosen for every adaptive cycle increases. Based on the chosen stopping criterion used for every cycle of TO, it is possible that significant parts of the designs obtained during initial cycles consist of intermediate density values. In such scenarios, selecting all gray (intermediate density) elements for refinement can lead to excessive refinement during the initial cycles, which in turn leads to undesired increase in computational burden. Due to the adaptive nature of the refinement indicator proposed in \cite{Gupta2016}, such problems can be avoided. 

\indent To start, the density-based refinement indicator $\Gamma^d_k$ for the $k^\text{th}$ MTO element is set to 0. To update $\Gamma^d_k$, we iterate over all the density cells of the $k^\text{th}$ MTO element and consider the sum of individual refinement or coarsening contributions of these cells. Let $n_{d,k}$ denote the number of density cells contained within the background mesh associated with the $k^{\text{th}}$ MTO element. Then $\Gamma^d_k$ is updated as follows:
\begin{itemize}
\item Iterate over $j$ from 1 to $n_{d,k}$:
	\begin{enumerate}
	\item Let the density of the $j^{\text{th}}$ voxel be denoted by $\rho_{kj}$.
	\item if $r_l \leq \rho_{kj} \leq \rho_{avg}$, \\ \indent set $\Gamma^d_k = \Gamma^d_k + \frac{1}{n_{d,k}}(\rho_{kj} - r_l)$.
	\item if $r_{avg} < \rho_{kj} \leq \rho_{u}$, \\ \indent set $\Gamma^d_k = \Gamma^d_k + \frac{1}{n_{d,k}}(r_u - \rho_{kj})$.
	\item if $\rho_{kj} \leq c_l$, \\\indent set $\Gamma^d_k = \Gamma^d_k - \frac{1}{n_{d,k}}(c_l - \rho_{kj})$.
	\item if $\rho_{kj} \geq c_u$, \\ \indent set $\Gamma^d_k = \Gamma^d_k - \frac{1}{n_{d,k}}(\rho_{kj} - c_u)$.
	\end{enumerate}
\end{itemize}
Here, the average density $\rho_{avg}$ is defined using the expression $\rho_{avg} = (\rho_{max} + \rho_{min})/2$. The variables $r_l$, $r_u$, $c_l$ and $c_u$ are the bounds used to characterize the refinement and coarsening zones as shown in Fig. \ref{fig_density_bounds}, and are defined as follows:
\begin{align}
&r_l = \rho_{min} + (1-\alpha)\rho_{avg} e^{-\beta (\tilde{k}-1)},\\
&r_u = \rho_{max} - (1 - \alpha)\rho_{avg} e^{-\beta (\tilde{k}-1)},\\
&c_l = \rho_{min} + \alpha\rho_{avg} e^{-\beta (\tilde{k}-1)},\\
&c_u = \rho_{max} - \alpha \rho_{avg} e^{-\beta (\tilde{k}-1)}.
\end{align}
Here, $\tilde{k}$ denotes the adaptive cycle index, and $\alpha$ and $\beta$ are tuning parameters chosen here to be 0.2 and 0.8, respectively.

The tuning parameters $\alpha$ and $\beta$ are independent of the index of the adaptive cycle. However, $\beta$ is sensitive to the rate at which the design converges. As stated earlier, our method assumes that the design has sufficiently converged at the end of every optimization cycle. For different problems as well as different mesh resolutions, the amount of gray region may vary at this point. For problems where the designs of initial cycles of the $dp$-adaptive MTO process are significantly gray, lower values of $\beta$ are recommended. This allows the density range for refinement to expand slowly over a span of cycles. Similarly, for rapidly converging designs, a larger value of $\beta$ is more efficient. Automated adjustment of these parameters could be considered, however, it has not been used in this study.

Fig. \ref{fig_only_density} shows the shape function field and the design field obtained for the optimized cantilever beam design shown in Fig. \ref{cant_point_load_fig1}. The shape function field (Fig. \ref{fig_den_a}) denotes the polynomial order of the shape functions used in every finite element. The design field (Fig. \ref{fig_den_b}) denotes the number of design points used in every analysis element. These distributions have been obtained based on adaptive refinement and coarsening controlled by only the density-based refinement indicator. From Fig. \ref{fig_only_density}, it is seen that the material-void boundaries where the intermediate densities are prominent, have primarily been refined. Coarsening occurs in void parts of the domain.  

\subsubsection{QR-error indicator}
\label{sec_qr_error}

In an MTO scheme, it is possible that the employed shape functions cannot accurately model the displacement field arising due to the allowed high order density representations. As stated earlier, this error arising in an MTO setting due to inappropriate modeling is referred to as QR-error. A closed-form condition to predict this QR-error is currently not known. Groen \emph{et al.} \cite{Groen2016} proposed a method to estimate the average error for the whole domain by determining a reference solution using a refined uniform mesh, and evaluating the obtained MTO solution against it. In the context of $dp$-adaptivity, QR-errors must be quantified at element level. We have proposed a method in \cite{Gupta2018}, where an approximation to the QR-error can be obtained for any element through a comparison with a reference solution obtained by local $p$-refinement. In this work, we use this cost-effective local QR-error indicator proposed in \cite{Gupta2018}. Once a sufficiently converged design has been obtained from a TO cycle, the QR-error is determined by evaluating the effect of local $p$-refinement, as follows. 

Let $\mathbf{K}_k^{(p)}$, $\mathbf{u}_k^{(p)}$ and $\mathbf{f}_k^{(p)}$ denote the element stiffness matrix, displacement solution and internal load vector for the $k^{\text{th}}$ MTO element. Here, $p$ denotes the polynomial degree of the shape functions used in this element. Let  $\mathbf{u}_k^{(p+1)}$ denote the displacement solution obtained for the $k^{\text{th}}$ element using shape functions of polynomial order $p+1$.  Note that $\mathbf{u}_k^{(p+1)}$ will be obtained by solving the element-level system $\mathbf{K}_k^{(p+1)}\mathbf{u}_k^{(p+1)} = \mathbf{f}_k^{(p+1)}$. Here, nodal load $\mathbf{f}^{(p+1)}_k$ is formed by integrating the product of the interpolated original load field $\mathbf{f}^{(p)}_k$ and the refined shape functions.

To obtain a unique solution for $\mathbf{u}_k^{(p+1)}$, sufficient boundary conditions need to be imposed. Thus, degrees of freedom (DOFs) equal to the number of rigid body modes (3 for 2D) need to be fixed. For this purpose, the displacement solution at 3 DOFs of $\mathbf{u}_k^{(p+1)}$ is copied directly from $\mathbf{u}_k^{(p)}$ for the DOFs which overlap, and the solution at the rest of the DOFs is obtained through solving the finite element system. Once $\mathbf{u}_k^{(p+1)}$ has been obtained, the QR-error $\epsilon_k^{QR}$ can be computed as
\begin{equation}
\epsilon_k^{QR} = 1 - \frac{\mathcal{J}_k^{(p)}}{\mathcal{J}_k^{(p+1)}},
\label{eq_qrobj}
\end{equation} 
where $\mathcal{J}^p_k$ refers to element-level strain energy for the $k^{\text{th}}$ finite element using shape functions of order $p$. Thus, $\mathcal{J}^{(p+1)} = \frac{1}{2}\mathbf{u}_k^{(p+1)}\mathbf{K}_k^{(p+1)\intercal}\mathbf{u}_k^{(p+1)}$ and \mbox{$\mathcal{J}^{(p)} = \frac{1}{2}\mathbf{u}_k^{(p)}\mathbf{K}_k^{(p)\intercal}\mathbf{u}_k^{(p)}$} have been used. This strain-energy-based criterion (Eq. \ref{eq_qrobj}) has been found to work well for the cases shown in this paper. 

\begin{figure*}
\centering
\begin{subfigure}{0.32\textwidth}
\centering
\begin{tikzpicture}
    \node[anchor=south west,inner sep=0] at (0,0){
\includegraphics[scale=0.15]{images/cant_design_cycle1.pdf}};
\end{tikzpicture}
\caption{Optimized design}
\label{fig_only_qr_a}
\end{subfigure}
\begin{subfigure}{0.32\textwidth}
\centering
\begin{tikzpicture}
    \node[anchor=south west,inner sep=0] at (0,0){
\includegraphics[scale=0.15]{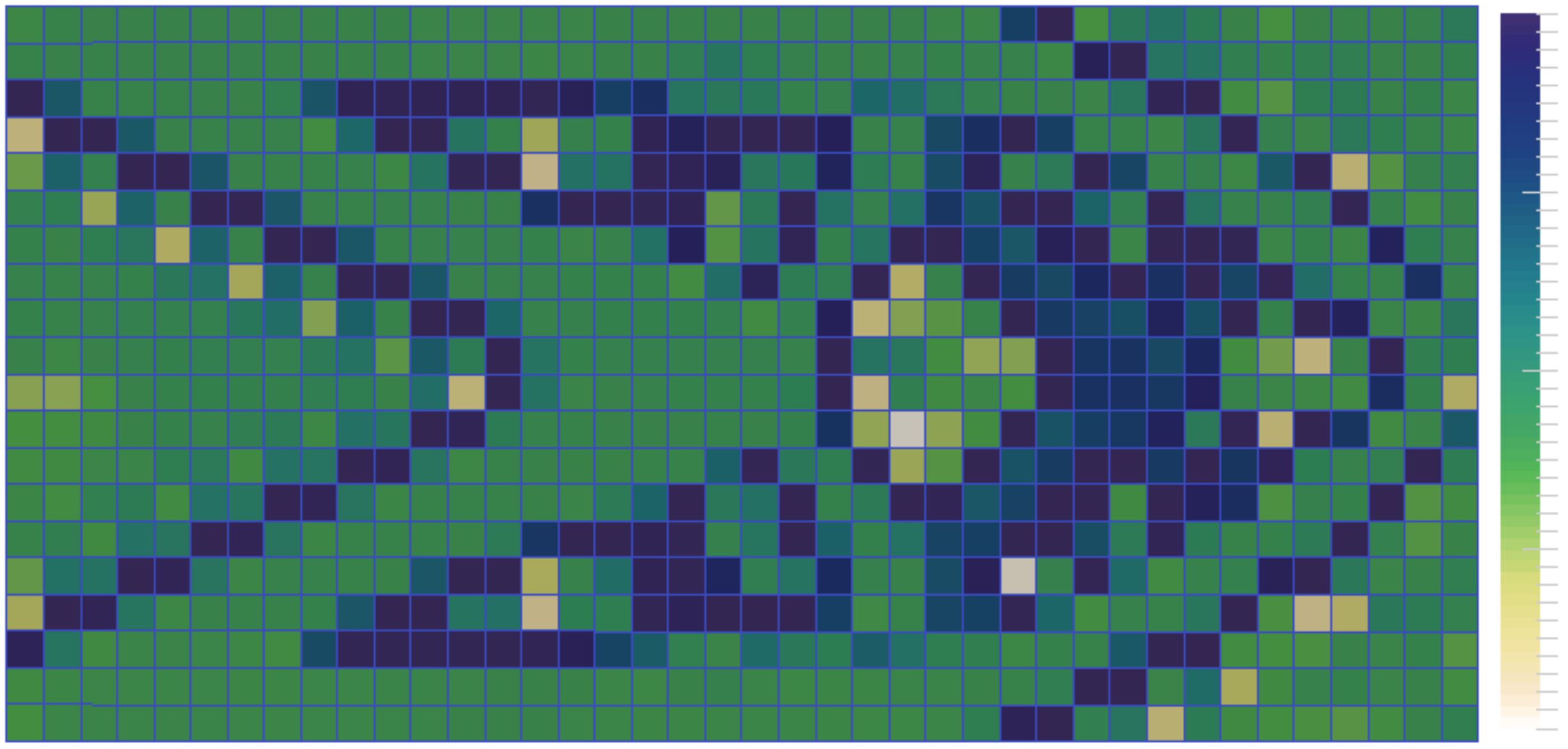}};
    \node[text width=0cm] at (4.2, 1.9) 
    {\tiny \color{darkgray}{0.99}};
    \node[text width=0cm] at (4.2, 1.32) 
    {\tiny \color{darkgray}{0.77}};
    \node[text width=0cm] at (4.2, 0.725) 
    {\tiny \color{darkgray}{0.55}};
\node[text width=0cm] at (4.2, 0.1) 
    {\tiny \color{darkgray}{0.33}};
\end{tikzpicture}
\caption{QR-error}
\label{fig_only_qr_b}
\end{subfigure}
\begin{subfigure}{0.32\textwidth}
\centering
\begin{tikzpicture}
    \node[anchor=south west,inner sep=0] at (0,0){
\includegraphics[scale=0.15]{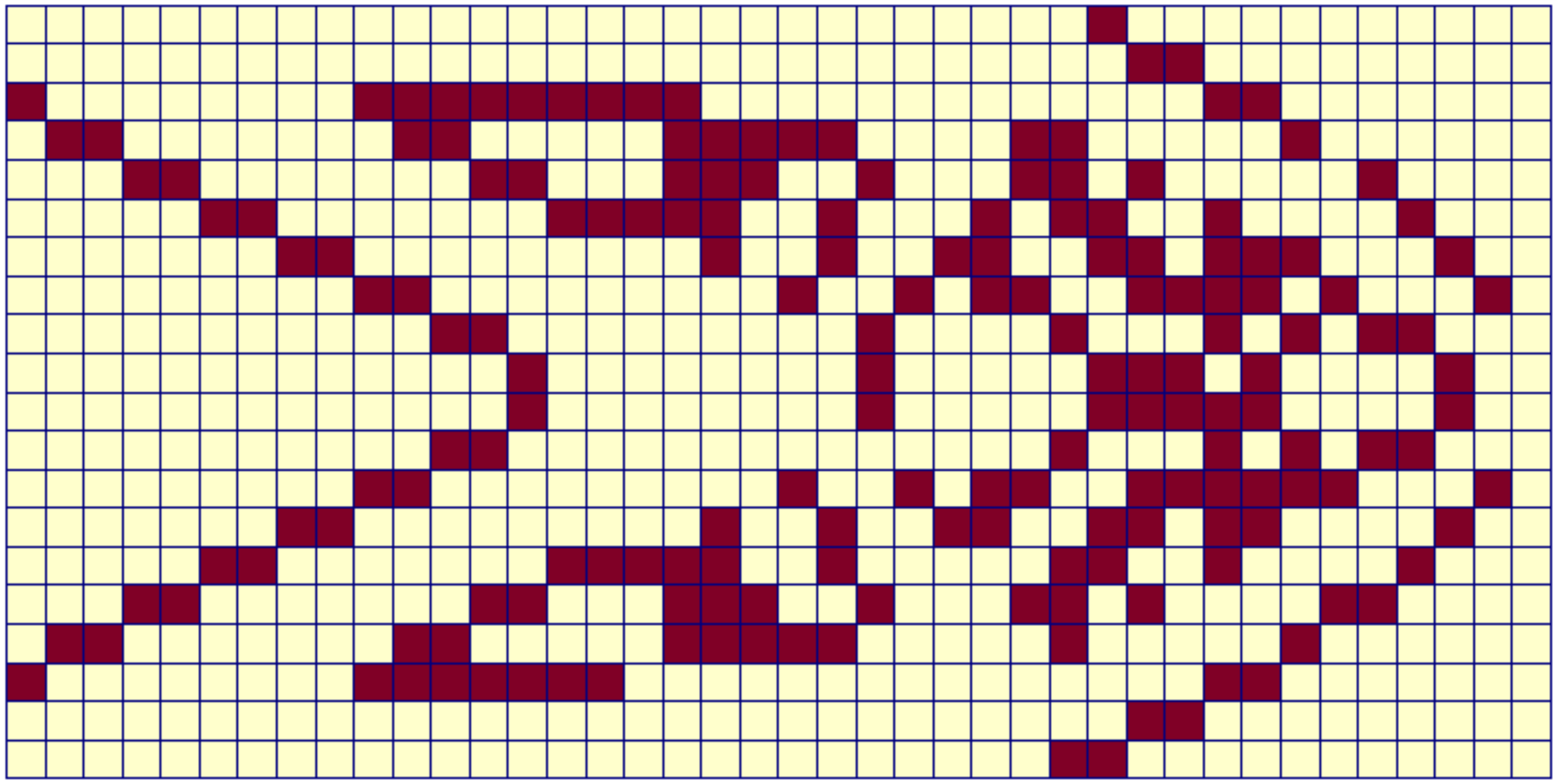}};
\end{tikzpicture}
\caption{Flagged cells}
\label{fig_only_qr_c}
\end{subfigure}
\caption{(a) Optimized design obtained after first cycle of $dp$-adaptive MTO run for a cantilever subjected to point load (Fig. \ref{cant_point_load_fig1}), (b) corresponding QR-error distribution plot obtained during the first adaptive cycle, and (c) domain showing the elements flagged for refinement using the QR-indicator. To avoid excessive refinement, only cells with error value larger than 0.9 have been flagged for refinement. The initial mesh comprises $40 \times 20$ Q2 finite elements with 16 design points and $4 \times 4$ density cells per element.}
\label{fig_only_qr}
\end{figure*}
Fig. \ref{fig_only_qr_a} and \ref{fig_only_qr_b} show an optimized design obtained after first cycle of MTO run, and the corresponding error distribution obtained using the QR-error indicator for the problem shown in Fig. \ref{cant_point_load_fig1}. Since the element-level test for QR-error is very conservative, it predicts higher error values compared to the actual full-scale TO problem \cite{Gupta2018}. Thus, to avoid undesired excessive increase in the values of $p$, we restrict the increment of $p$ by only 1 per adaptive cycle based on the QR-error test. Also, to avoid excessive spatial refinement per adaptive cycle, only the cells with error value larger than 0.9 are adaptively refined. The elements flagged for refinement are shown in Fig. \ref{fig_only_qr_c}. It is observed that the regions where the QR-patterns exist, have been flagged for refinement. Moreover, elements at the material boundaries, which are partially void or solid, also show high value of QR-error and are flagged.

An interesting observation in Fig. \ref{fig_only_qr_b} is that the elements which are completely void or solid also show QR-error values in the range 0.3-0.5. Although significant, the QR-error values in this range are relatively smaller than other parts of the domain and these elements do not get flagged for refinement. The reason for substantial QR-error values in these regions is the use of low order shape functions. For low values of $p$, the displacement solution for even a uniform density field may not be accurately modeled. When solving element-level FE problems with low shape function orders $p$ and $p+1$, it is observed that the modeling accuracy significantly improves when $p$ is increased. Due to this, nonzero large values of $\epsilon_k^Q$ are recorded in solid and void parts as well.

\subsection{$dp$-adaptivity algorithm}
\label{sec_dpalgo}
The different steps of $dp$-adaptivity have briefly been introduced in Section \ref{dp_general}. After treating the three indicators involved, here we discuss each of these steps in more details. Once a TO cycle has been completed, the optimized design is analyzed using the composite refinement criterion, and the following steps are carried out.
\begin{enumerate}
\item Once a cycle of TO run is completed, get the optimized design for $dp$-adaptivity.
\item Perform $p$-adaptivity based on analysis error criterion.
\begin{enumerate}
\item Update $\boldsymbol\Gamma^a = \{\Gamma_1^a, \Gamma_2^a, \hdots, \Gamma^a_{n_{el}}\}$ values for the whole analysis mesh (discussed in Section \ref{sec_ana_indic}), where $\Gamma_i^a$ is the analysis error indicator value for the $i^{\text{th}}$ MTO element.
\item Sort $\boldsymbol\Gamma^a$ in ascending order such that a corresponding ordered set $\tilde{\boldsymbol\Gamma}^a$ is obtained.
\item Set the refine/coarsen flag of the $k^{\text{th}}$ element $\Theta_k$ to -1 for the first $\alpha^d_c$ fraction of the MTO elements in $\tilde{\boldsymbol\Gamma}^a$, and $\Theta_k = 1$, for the last  $\alpha^a_r$ fraction of the elements. Here, $-1$ and 1 denote that the cell has been flagged for coarsening (decrease in $p$ value) and refinement (increase in $p$ value), respectively. For no refinement/coarsening, $\Theta_k$ is set to 0. 
\item Increase/decrease $p$-values based on flag $\Theta$.
\end{enumerate}

\item Refine/coarsen $p$ and $d$ values based on density-based refinement criterion.
\begin{enumerate}
\item Update $\boldsymbol\Gamma^d = \{\Gamma_1^d, \Gamma_2^d, \hdots, \Gamma^d_{n_{el}}\}$ values for the whole domain (discussed in \mbox{Section \ref{sec_des_indic}}), where $\Gamma_i^d$ is the density-based refinement indicator value for the $i^{\text{th}}$ MTO element.
\item Sort $\boldsymbol\Gamma^d$ in ascending order such that a corresponding ordered set $\tilde{\boldsymbol\Gamma}^d$ is obtained.
\item Update $p$-values by iterating over $k$ from 1 to $n_{el}$:
	\begin{enumerate}
	\item For the first $\alpha^d_c$ fraction of the elements in  $\tilde{\boldsymbol\Gamma}^d$, do:
		\begin{enumerate}
		\item if $p_k = 1$, set $\Theta_k = -2$. This helps to identify that the current element has been checked for coarsening. Since $p_k$ cannot be lower than 1, no coarsening is performed.
		\item if $p_k > 1$ and $\Theta_k = 0$, set $p_k = p_k - 1$.
		\end{enumerate}
	\item For the last $\alpha^d_r$ fraction of the elements in  $\tilde{\boldsymbol\Gamma}^d$, do:
		\begin{enumerate}
		\item if $\Theta_k = 0$ or $\Theta_k = -1$, set $p_k = p_k + 1$. This means that if the element has been coarsened or left untreated based on the analysis indicator above, then refine it. 
		\end{enumerate}
	\end{enumerate}
\item Reduce the difference of $p$-values between adjacent elements to a maximum of 2 at this point. This is achieved by iterating through the whole domain ($p_{\max} - p_{\min} - 2$) times, where $p_{\max}$ and $p_{\min}$ are the maximum and minimum values of $p$ in the domain. At every check, the correction is done by raising the lower value of $p$.
\item Update the design-field ($d$ values) by iterating over $k$ from 1 to $n_{el}$:
	\begin{enumerate}
	\item if $\Theta_k = -2$, set $d_k$ = 1. This situation occurs when $p_k = 1$, and the density-based indicator flags the cell for further coarsening.
	\item if $\Theta_k \neq -2$, set $d_k$ equal to the element-level upper bound for the $k^{\text{th}}$ element (based on \cite{Gupta2016a}).	Thus, $d_k = \text{DOFs} - r_b$, where $r_b$ denotes the number of rigid body modes for that element.	
	\end{enumerate}
\item Update the background mesh
	\begin{enumerate}
	\item Find maximum number of design variables per MTO element ($\max(d_{el})$).
	\item Find first perfect square (cube in 3D) number ($\bar{\bar{d}}$) greater than $\max(d_{el})$.
	\item Set the number of density cells per MTO element equal to $\bar{\bar{d}}$.
	\item Update projection radius $r$ for every MTO element (Eq. \ref{eq_projr}).
\end{enumerate}
\end{enumerate}
\item Update $p$ values to reduce the QR-error in every MTO element.
	\begin{enumerate}
	\item Iterate over $k$ from 1 to $n_{el}$, do:
	\begin{enumerate}
	\item Calculate the QR-error for the $k^{\text{th}}$ cell (discussed in Section \ref{sec_qr_error}).
	\item Update $p_k = p_k + 1$ for the $k^{\text{th}}$ element, if QR-error is greater than a certain error tolerance $\alpha_{\text{QR}}$.
	\end{enumerate}
	\end{enumerate}
\end{enumerate}
The $dp$-adaptive MTO cycle is complete once the domain has been adaptively refined based on the three indicators. With the new $dp$-refined mesh, the next cycle of TO is performed.

\section{Numerical tests}
\label{examples}

\subsection{Definition of test problems}
\label{define_test}

\begin{figure}
\centering
\begin{subfigure}{0.44\textwidth}
\centering
\begin{tikzpicture}
\node[inner sep=0pt](design1) at (0, 0)
{\includegraphics[scale=1.2]{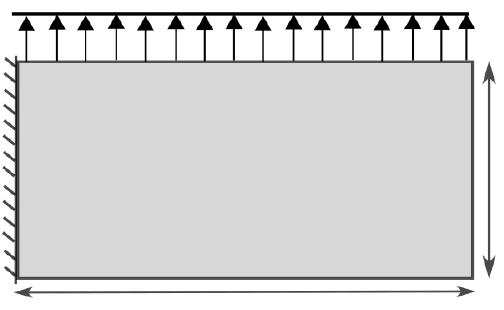}};
\node[] at (-0.1, 2.0) {$F$};
\node[] at (-0.2, -1.9) {$2L$};
\node[] at (3.1, -0.0) {$L$};
\end{tikzpicture}
\caption{A cantilever subjected to distributed load}
\label{fig_cant_dist}
\end{subfigure}
\begin{subfigure}{0.54\textwidth}
\centering
\begin{tikzpicture}
\node[inner sep=0pt](design1) at (0, 0)
{\includegraphics[scale=1.25, trim=0 -0.3cm 0 -0.2cm, clip=true]{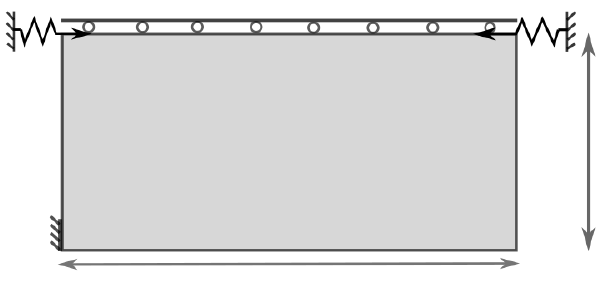}};
\node[] at (3.1, 0.85) {$k_{out}$};
\node[] at (-3.3, 0.85) {$k_{in}$};
\node[] at (-2.5, 1) {$f_{in}$};
\node[] at (2.3, 1) {$u_{out}$};
\node[] at (-0.2, -1.8) {$2L$};
\node[] at (3.4, -0.0) {$L$};
\node[] at (-3.65, -1.35) {$0.01L$};
\end{tikzpicture}
\caption{A compliant force inverter problem}
\label{fig_comp_inv}
\end{subfigure}
\caption{Problem domains and boundary conditions for a cantilever beam subjected to distributed load (left) and a force inverter (right). Here, $L = 1$ m, $F = \frac{0.5\text{N}}{L}$, $k_{in} = 1$ Nm$^{-1}$, $k_{out} = 0.001$ Nm$^{-1}$ and \mbox{$f_{in} = 1$ N}.}
\label{cant_dist_load_fig1}
\end{figure}
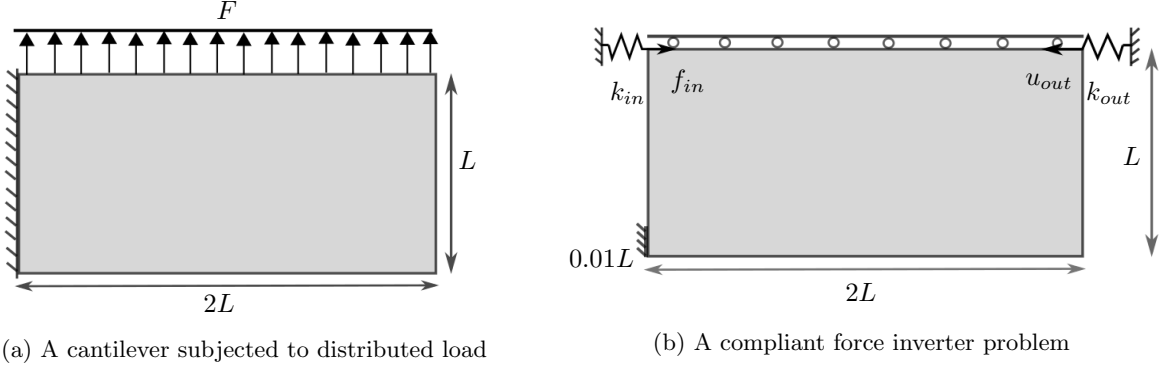

To demonstrate the applicability and effectiveness of $dp$-adaptivity, two test problems of minimum compliance and one compliant mechanism problem are considered \cite{Groen2016}. In this paper, only 2D problems are studied, whereas an extension to a 3D setting is a part of future work. Young's modulus $E_0$ is set to 1 Nm$^{-2}$, $\nu = 0.3$, and the SIMP penalization factor $q$ is set to 3.\ The domain in each case is discretized using an initial mesh of $40 \times 20$ MTO elements, comprising quadrilateral finite elements with shape functions of polynomial order 2 and $4 \times 4$ design points per element. The radius $R$ is set to 0.3$h$, where $h$ is the edge-length of any MTO element in the mesh. As a stopping criterion for all the test cases used in this paper, the optimization process for the $\tilde{k}^{\textrm{th}}$ cycle is terminated when the change in objective value between two consecutive iterations is less than $\Delta J_1 \times \gamma^{(\tilde{k} - 1)}$. Here, $\Delta J_1$ denotes the minimum required change in objective value between two consecutive iterations of the first MTO cycle, below which the optimization process terminates. For the subsequent cycles, the minimum required change in objective value is reduced by a factor of $\gamma$ at every MTO cycle. The adaptive stopping criterion used here allows to control the extent of design convergence per cycle. For the numerical examples used in this paper, $\Delta J_1$ and $\gamma$ are set to 0.04 and 0.6, respectively, and these values have been found to work well. Based on this, the first ($\tilde{k} = 1$) and second ($\tilde{k} = 2$) optimization cycles are terminated if the minimum changes in objective value are less than 0.04 and 0.024, respectively.

To validate the accuracy of the MTO modeling of the design, we use the method proposed in \cite{Groen2016}, where the obtained design is compared with a reference solution. For the reference solution, we discretize the domain using a high-resolution traditional TO mesh with elementwise constant densities. In this paper, the reference mesh comprises \mbox{320 $\times$ 160} finite elements and the polynomial order $p$ of the involved shape functions is set to 3. With this mesh configuration, the resolution of the reference domain is equal to the highest density resolution that has been used in the MTO problem.

For the first test problem, compliance needs to be minimized for a Michell beam cantilever subjected to a point load $F$ (Fig. \ref{cant_point_load_fig1}). For this case, $F = 1$ N and $L$ = 1 m. Three variants of this problem are used with maximum allowed material volume fractions set to 0.45, 0.2 and 0.1, to study the capability of the method in low volume fraction problems on coarse meshes. For the other problems used in this paper, only one volume constraint of 0.45 is considered. The second test problem is that of compliance minimization for a cantilever beam subjected to a distributed load (Fig. \ref{fig_cant_dist}), and it is ensured that the load is consistently distributed over the various cycles of adaptivity. Here,  \mbox{$F = \frac{0.5\text{N}}{L}$} and $L = 1$ m.  The distributed load tends to generate a lot of fine structures locally, and the resultant design was earlier found to be prone to QR artefacts \cite{Groen2016}, which makes it an interesting problem. For both these problems, the objective functional of \mbox{Eq. \ref{eq_comp_1}} with $\mathbf{z = f}$. The third case is a compliant mechanism problem where a force inverter needs to be designed, such that for a point load $f_{in}$ at one end, the displacement $u_{out}$ at the other end is maximized \mbox{(Fig. \ref{fig_comp_inv})}. Here, spring stiffnesses $k_{in}$ and $k_{out}$ are set to 1 Nm$^{-1}$ and 0.001 Nm$^{-1}$, respectively. For the force inverter, $\mathbf{z}$ in \mbox{Eq. \ref{eq_comp_1}} is a vector of zeros with 1 contained at the DOF where $u_{out}$ needs to be maximized. Thus, $\mathbf{z} = [0 \hdots 0 \enskip 1 \enskip 0 \hdots 0]^{\intercal}$. The flexure hinges that are formed in this compliant mechanism problem will have sub-element resolution, and this aspect makes also this problem an interesting test for our method. \\

\subsection{Results}

Here, we discuss the results obtained for the three test problems using a $dp$-adaptive MTO scheme. To provide an understanding of the computational advantage of the proposed method, a comparison of CPU times is performed for the designs obtained using the proposed method as well as those obtained using the conventional MTO scheme discussed in \cite{Groen2016}. \mbox{Groen \emph{et al.} \cite{Groen2016}} have shown that by using the MTO approach, the computational time can already be reduced by factors of up to 2.9 and 32 for 2D and 3D problems, respectively, compared to the traditional TO approach. In this paper, we demonstrate the potential of $dp$-adaptive MTO schemes for 2D problems, and for this purpose, we will compare its performance with the non-adaptive MTO scheme, implemented in the same framework and evaluated on the same computing hardware.
\subsubsection{Compliance minimization for point load}
\begin{figure}
\centering
\begin{subfigure}{0.48\linewidth}
\centering
\includegraphics[scale=0.22]{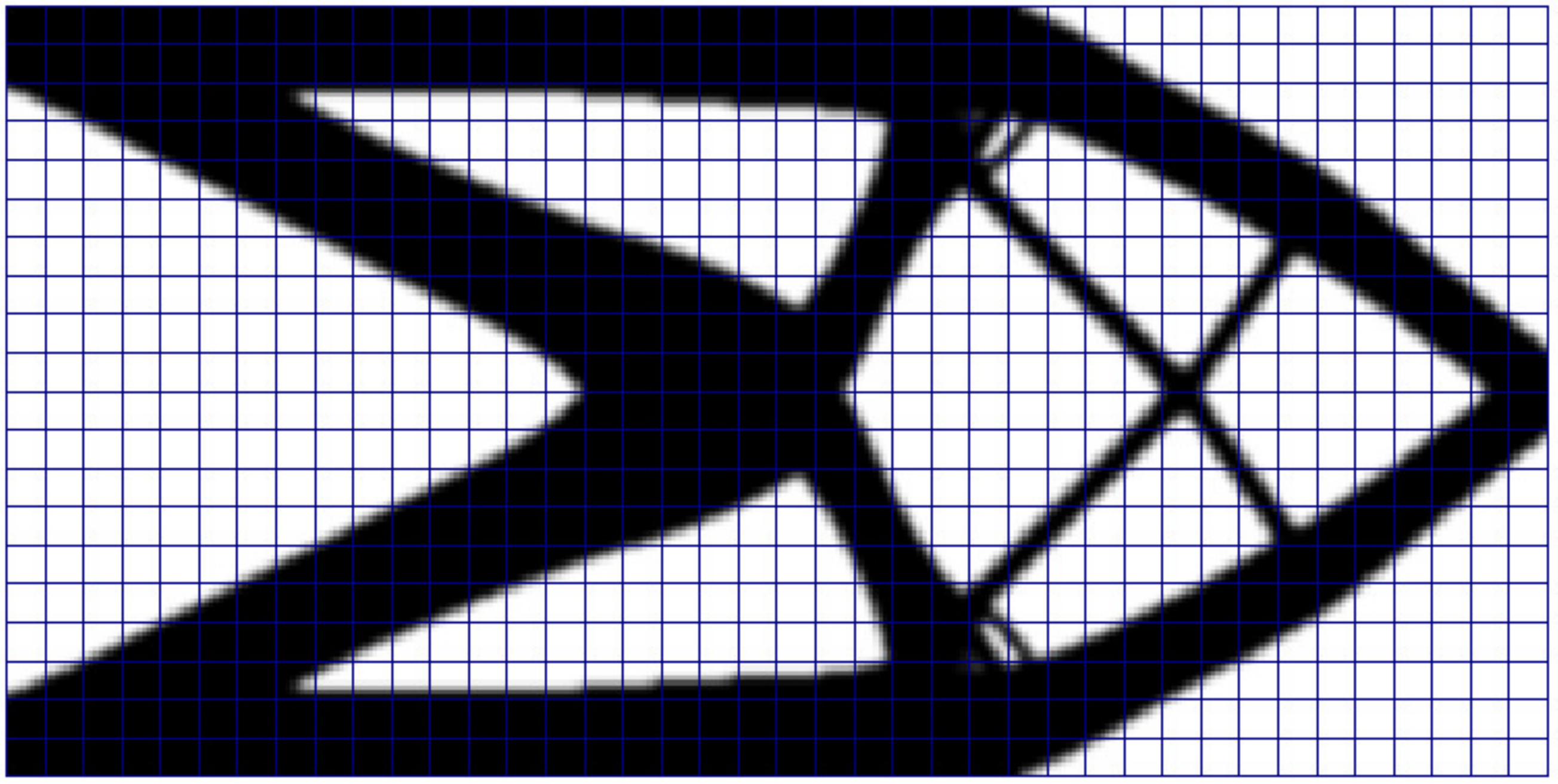}
\caption{MTO ($\mathcal{J}_0 = 72.39$J)}
\label{point_cant45_1a}
\end{subfigure}
\begin{subfigure}{0.48\linewidth}
\centering
\includegraphics[scale=0.22]{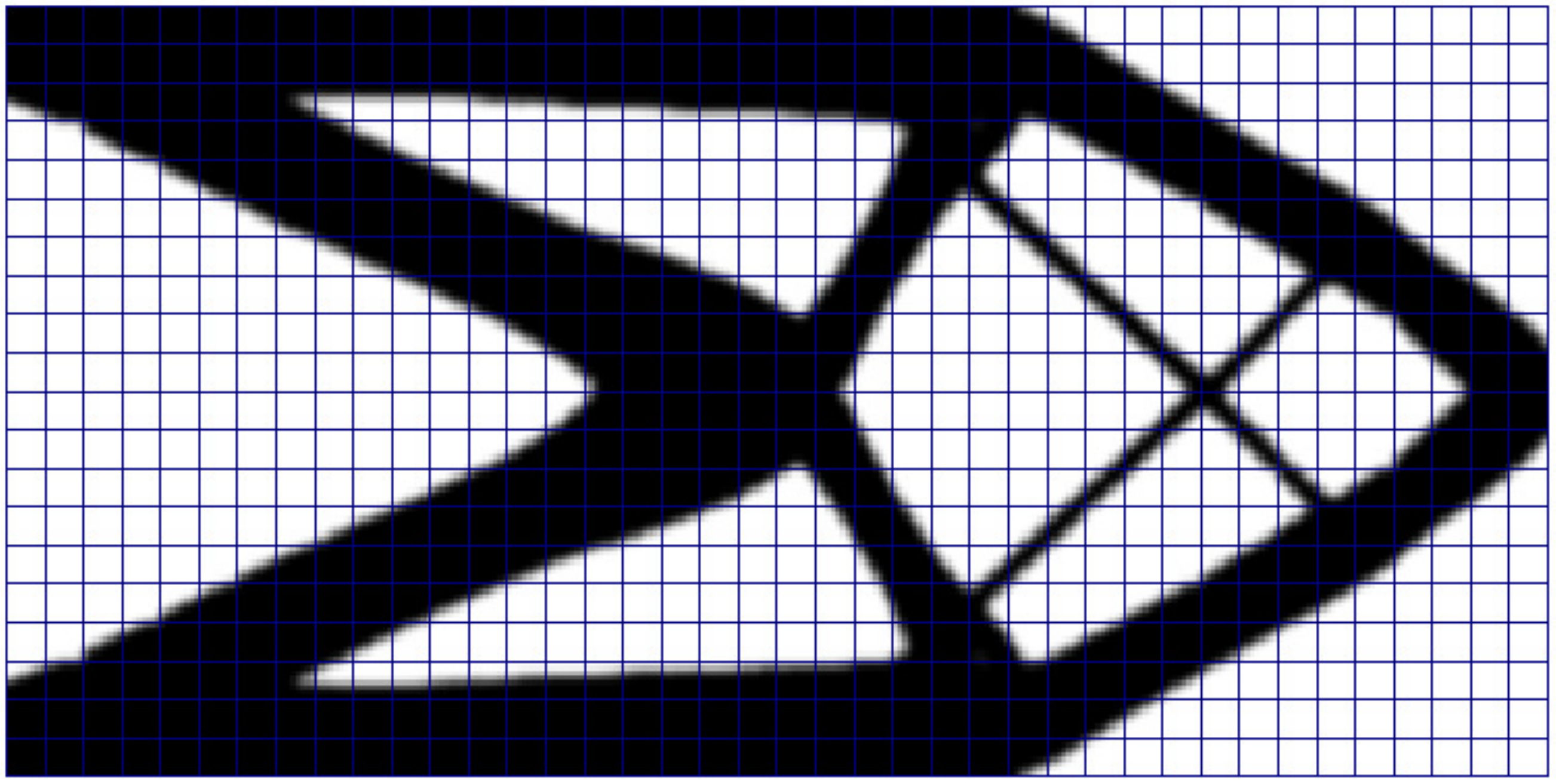}
\caption{$dp$-adaptive MTO ($\mathcal{J} = 70.92$J)}
\label{point_cant45_1b}
\end{subfigure}
\caption{Optimized cantilever designs for the point load case shown in Fig. \ref{cant_point_load_fig1}, obtained using (a) a uniform MTO mesh and (b) $dp$-adaptive MTO approach. The maximum permissible material volume fraction is set to 0.45. A 4.5-fold speed-up as well as a superior objective value are obtained using $dp$-adaptivity. Additional information related to this test case is listed in Table \ref{table_mto_runs}.}
\label{point_cant45_1}
\end{figure}
\begin{table}
\caption{Numerical findings of several $dp$-adaptive MTO cases. For all the cases, the domain has been discretized using $40 \times 20$ MTO elements, and the initial polynomial order of the shape funtions is set to 2 for every element. Each MTO element initially consists of 16 design points and the projection radius $R$ is set to $0.3h$, where $h$ denotes element size. The maximum permissible values for shape function order $p_{\textrm{max}}$ and number of designs points $d_{\textrm{max}}$  are set to 5 and 64, respectively. For the reference solution, a globally uniform mesh comprising $320 \times 160$ finite elements with $p = 3$ is used. Below, $V_0$ denotes maximum allowed volume fraction of material, $\mathcal{J}$ and $\mathcal{J}_0$ are the objective values for $dp$-adaptive MTO run and the non-adaptive MTO run, and $\mathcal{J}^*$ denotes the reference solution. The $N_d$ and DOFs denote number of design points and free degrees of freedom employed in the last cycle of $dp$-adaptive MTO run.}
\begin{tabular}{ l | c | c | c | c | c | c | c}
  Problem &  Definition & $V_0$ & Speed-up & $\mathcal{J}/\mathcal{J}_0$ & $\mathcal{J}/\mathcal{J}^*$ & $N_d$ & DOFs\\
  \hline
  \hline
      \multirow{4}{*}{Minimum compliance} & \multirow{3}{*}{point load} & 0.45 & 4.5 & 0.98 & 0.98 & 22935 & 17262\\      
         &  & 0.20 & 8.3 & 0.93 & 0.98 & 20056 & 15096\\
         & & 0.10 & 10.0 & 1.03 & 0.96 & 19590 & 15186\\ \cline{2-8}
         & distributed load & 0.45 & 4.6 & 0.98 & 1.0 & 22636 & 16932\\
         \hline 
         Compliant mechanism & - & 0.45 & 6.2 & 1.01$^{\dagger}$ & 1.0 & 23375 & 17516\\
	\hline
	\multicolumn{8}{l}{\footnotesize $^{\dagger}$This case refers to a maximization problem, where a value higher than 1 denotes that the $dp$-adaptive MTO approach} \\
	\multicolumn{8}{l}{\footnotesize performed better over the non-adaptive MTO scheme.}\\
\end{tabular}
\label{table_mto_runs}
\end{table}
\indent \vspace{0.5em} \\ Fig \ref{point_cant45_1} shows two optimized cantilever designs obtained for the problem shown in Fig. \ref{cant_point_load_fig1}. The first design (Fig. \ref{point_cant45_1a}) has been obtained using the traditional non-adaptive MTO scheme, and the other (Fig. \ref{point_cant45_1b}) by our $dp$-adaptive approach. For the two cases, the maximum allowed material volume fraction $V_0$ is set to 0.45. Visually, the designs  differ only slightly. \mbox{Table \ref{table_mto_runs}} provides the details on various parameters related to MTO cases for the two optimized designs. The first remarkable observation regarding the $dp$-adaptive MTO result is the reduced computational cost. Adding the $dp$-adaptive framework to the existing MTO allows a reduction in computational cost by a factor of 4.5. This reduction in cost is mainly due to the reduced number of design variables $N_d$ and free DOFs used in the $dp$-adaptive MTO case. While the uniformly refined mesh used in MTO comprises 51200 design points and 40400 free DOFs, only 22935 design points and 17262 free DOFs are used in the final (4$^{\text{th}}$) cycle  of the $dp$-adaptive MTO run, \emph{i.e.} a reduction by over 50\%. The free DOFs and number of design variables used in the earlier cycles are even lower (Table \ref{table_mto_stats2}). 

Another reason that accounts for the speed-up is the reduced number of iterations required in the final cycle of the $dp$-adaptive method under the same stopping criterion as used for the non-adaptive MTO method. The convergence of the TO process is significantly affected by the choice of the initial design \cite{vanSchoubroeck2018}. In our approach, each preceding cycle, after refinement/coarsening, provides a high quality initial design for the next one. Since the design converges significantly in the first 3 cycles itself using less refined meshes, only 18 iterations are needed in the final cycle, while the non-adaptive MTO scheme uses a total of 56 iterations. Table \ref{table_mto_stats2} provides the details related to the $dp$-adaptive MTO run for this case. It is observed that Cycles 1 and 2 use a higher number of iterations. However, since the number of design variables and free DOFs are lower during these cycles, the associated computational cost is not very high.

\begin{table}
\caption{Parameters related to $dp$-adaptive MTO run for the point load cantilever design problem shown in Fig. \ref{cant_point_load_fig1}. The material volume fraction $V_0$ has been set to 0.45 for this case.}
\centering
\begin{tabular}{ r | r | r | r | r}
  Cycle &  DOFs & $N_d$ & Iterations & $\mathcal{J}/\mathcal{J}^*$ \\
  \hline
  \hline
  1 & 6560 & 12800 & 67 & 0.86\\
 2 & 7204 & 10646 & 34 & 0.97\\      
3 & 12818 & 16256 & 17 & 0.98 \\
4 & 17262 & 22935 & 18 & 0.98\\ 
	\hline
\end{tabular}
\label{table_mto_stats2}
\end{table}

In terms of performance, the cantilever design obtained from the $dp$-adaptive approach slightly outperforms the design obtained using non-adaptive MTO. The obtained performance ratio $\mathcal{J}/\mathcal{J}_0$ is equal to 0.98, where $\mathcal{J}$ and $\mathcal{J}_0$ denote the compliance objective values obtained using the proposed method and non-adaptive MTO, respectively. From Table \ref{table_mto_stats2}, it is observed that the global solution accuracy $\mathcal{J}/\mathcal{J}^* = 0.98$, where $\mathcal{J}$ and $\mathcal{J}^*$ refer to the objective values reported using adaptive MTO and that evaluated using the reference mesh, respectively. Since solution accuracy is close to 1, it is implied that the final optimized design is correct and free from artefacts. Moreover, we see that with every cycle of refinement, the global solution accuracy has improved. Thus, the $dp$-adaptive MTO method allows to obtain designs with a desired analysis accuracy. 

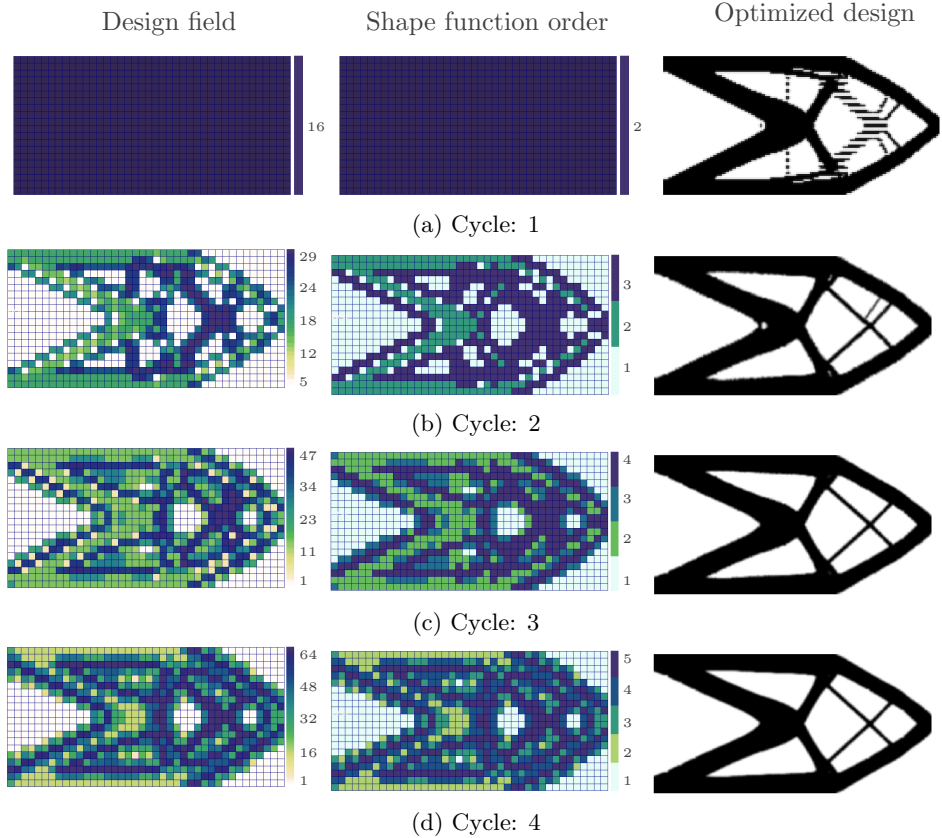
\begin{figure*}[h]
\centering
\begin{subfigure}{1\textwidth}
\centering
\hspace{0.5em}
\begin{tikzpicture}
    \node[anchor=south west,inner sep=0] at (0,0){
\includegraphics[scale=0.14]{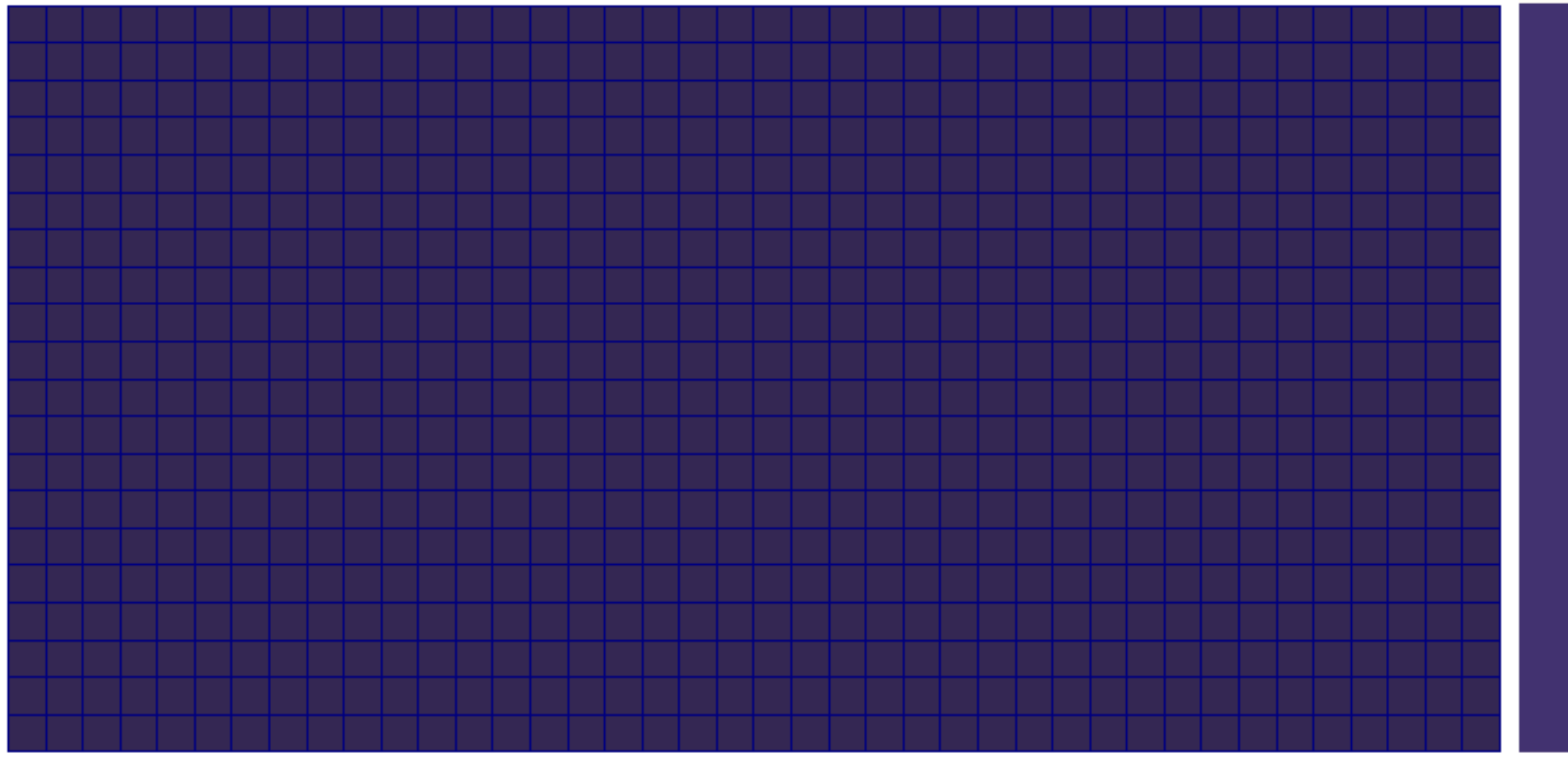}};
    \node[text width=0cm] at (3.93, 0.92) 
    {\tiny \color{darkgray}{16}};
    \node[text width=2cm] at (2.2, 2.3) 
    {\normalsize \color{darkgray}{Design field}};
\end{tikzpicture}\hspace{0.5em}
\begin{tikzpicture}
    \node[anchor=south west,inner sep=0] at (0,0){
\includegraphics[scale=0.14]{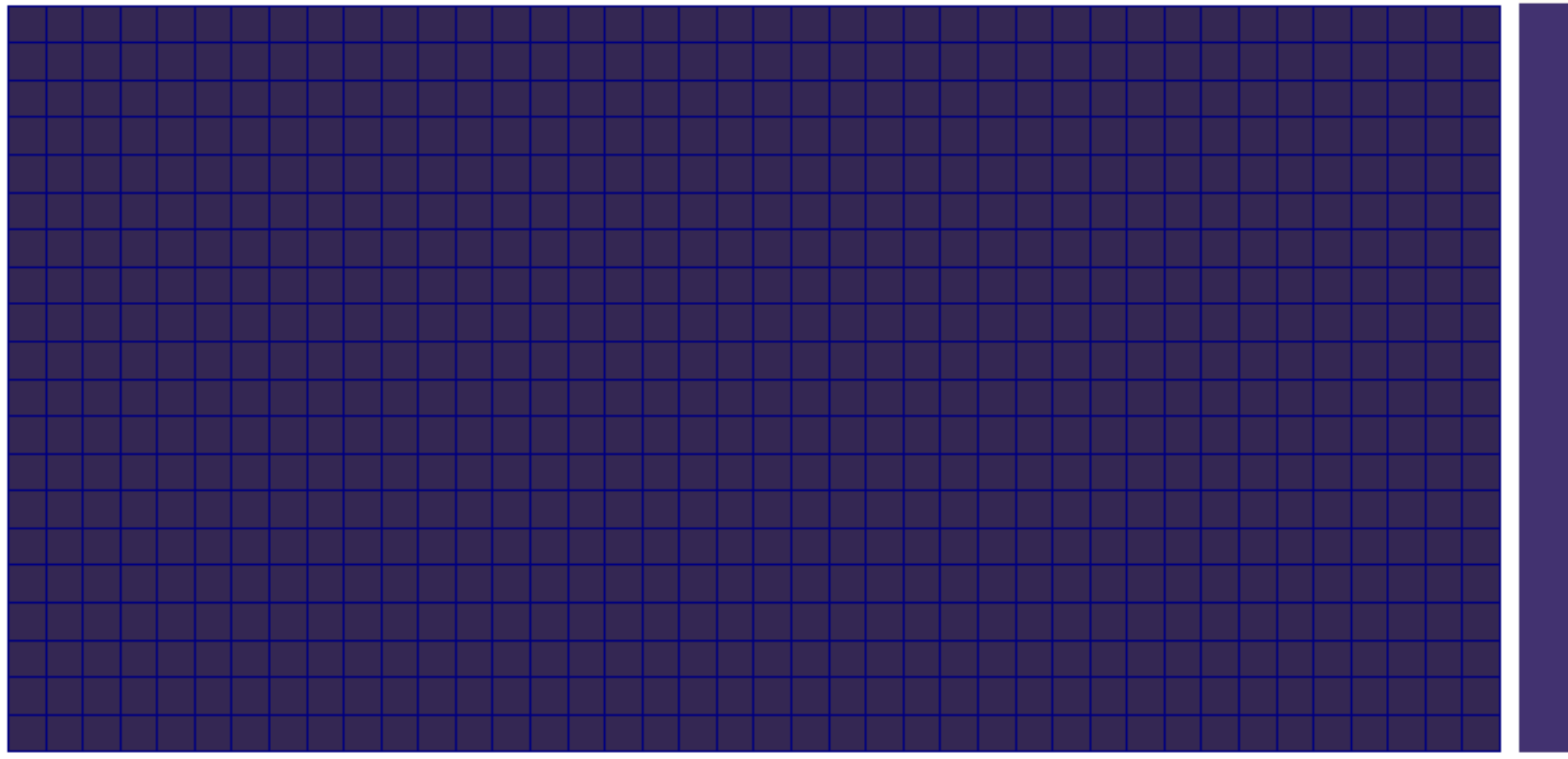}};
    \node[text width=0cm] at (3.93, 0.92) 
    {\tiny \color{darkgray}{2}};
    \node[text width=3.7cm] at (2.23, 2.3) 
    {\normalsize \color{darkgray}{Shape function order}};
\end{tikzpicture}
\begin{tikzpicture}
    \node[anchor=south west,inner sep=0] at (0,0){
\includegraphics[scale=0.14]{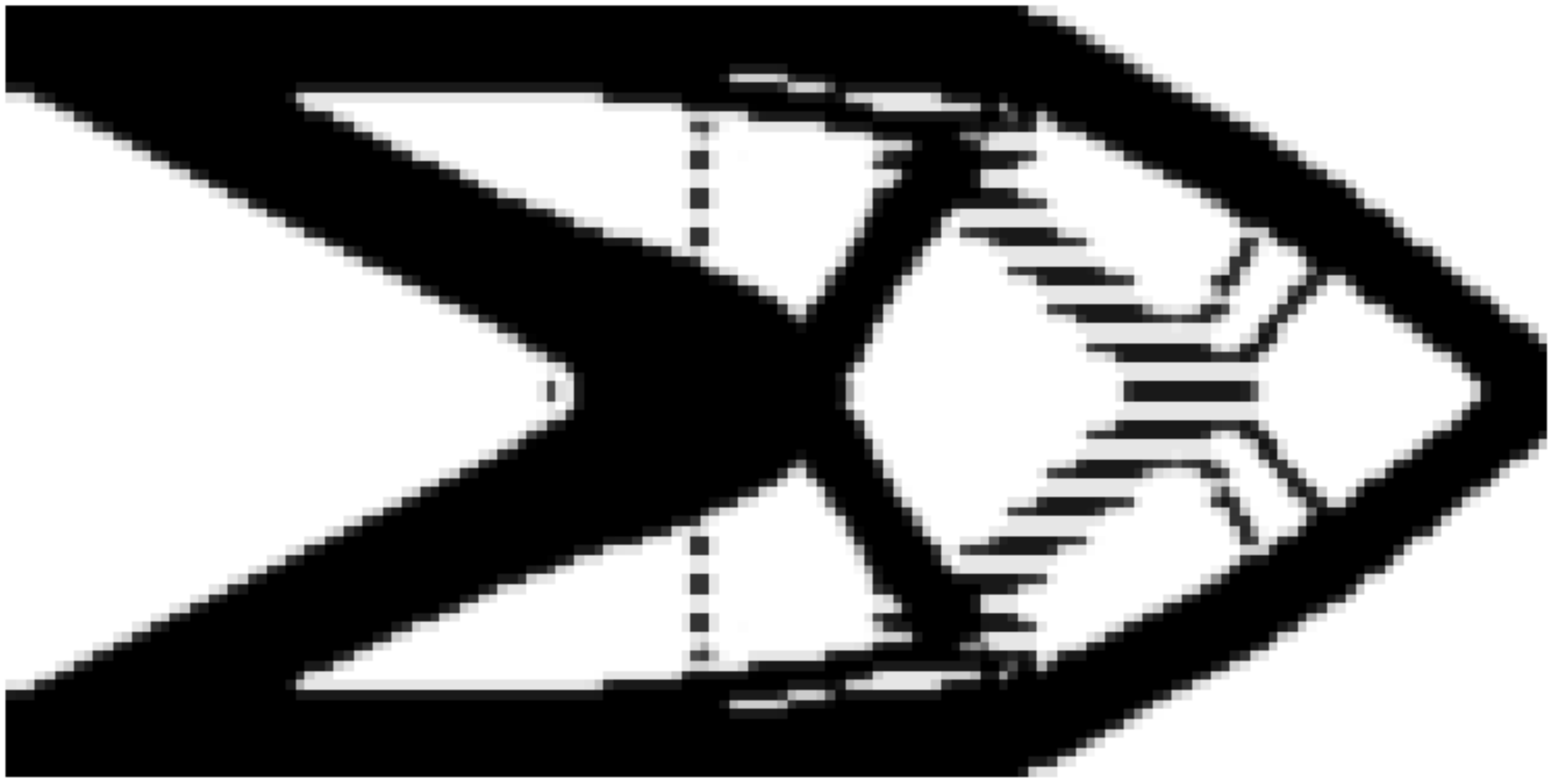}};
    \node[text width=3cm] at (2.2, 2.42) 
    {\normalsize \color{darkgray}{Optimized design}};
\end{tikzpicture}\hspace{0.5em}
\caption{Cycle: 1}
\end{subfigure}
\begin{subfigure}{1\textwidth}
\centering
\begin{tikzpicture}
    \node[anchor=south west,inner sep=0] at (0,0){
\includegraphics[scale=0.14]{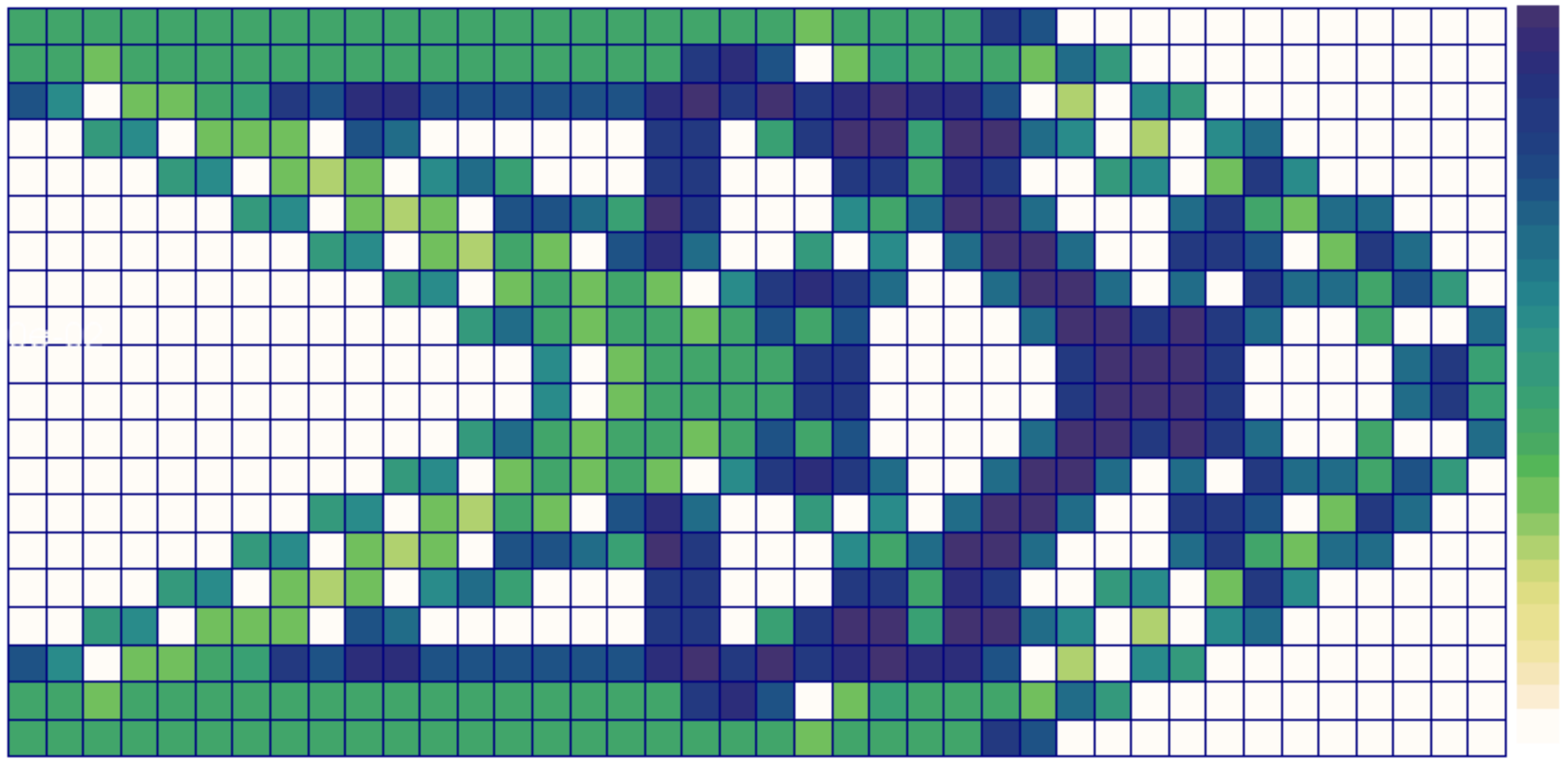}};
\node[text width=0cm] at (3.9, 1.77) 
    {\tiny \color{darkgray}{29}};
    \node[text width=0cm] at (3.9, 1.35) 
    {\tiny \color{darkgray}{24}};
    \node[text width=0cm] at (3.9, 0.92) 
    {\tiny \color{darkgray}{18}};
    \node[text width=0cm] at (3.9, 0.47) 
    {\tiny \color{darkgray}{12}};
\node[text width=0cm] at (3.9, 0.1) 
    {\tiny \color{darkgray}{5}};
\end{tikzpicture}\hspace{0.5em}
\begin{tikzpicture}
    \node[anchor=south west,inner sep=0] at (0,0){
\includegraphics[scale=0.14]{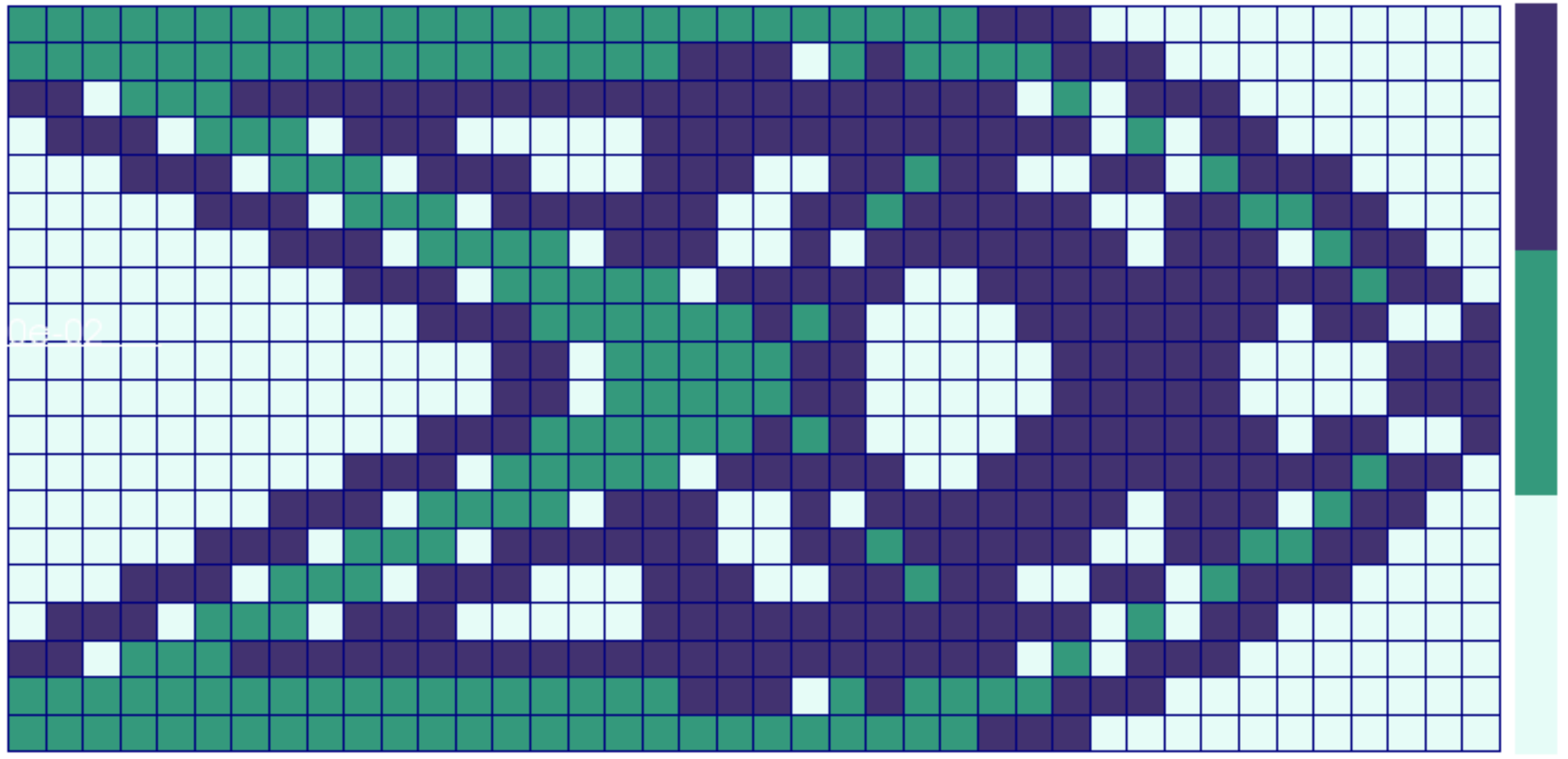}};
\node[text width=0cm] at (3.9, 1.47) 
    {\tiny \color{darkgray}{3}};
    \node[text width=0cm] at (3.9, 1.35) 
    {\tiny \color{darkgray}{}};
    \node[text width=0cm] at (3.9, 0.92) 
    {\tiny \color{darkgray}{2}};
    \node[text width=0cm] at (3.9, 0.47) 
    {\tiny \color{darkgray}{}};
\node[text width=0cm] at (3.9, 0.37) 
    {\tiny \color{darkgray}{1}};
\end{tikzpicture}\hspace{0.5em}
\begin{tikzpicture}
    \node[anchor=south west,inner sep=0] at (0,0){
\includegraphics[scale=0.14]{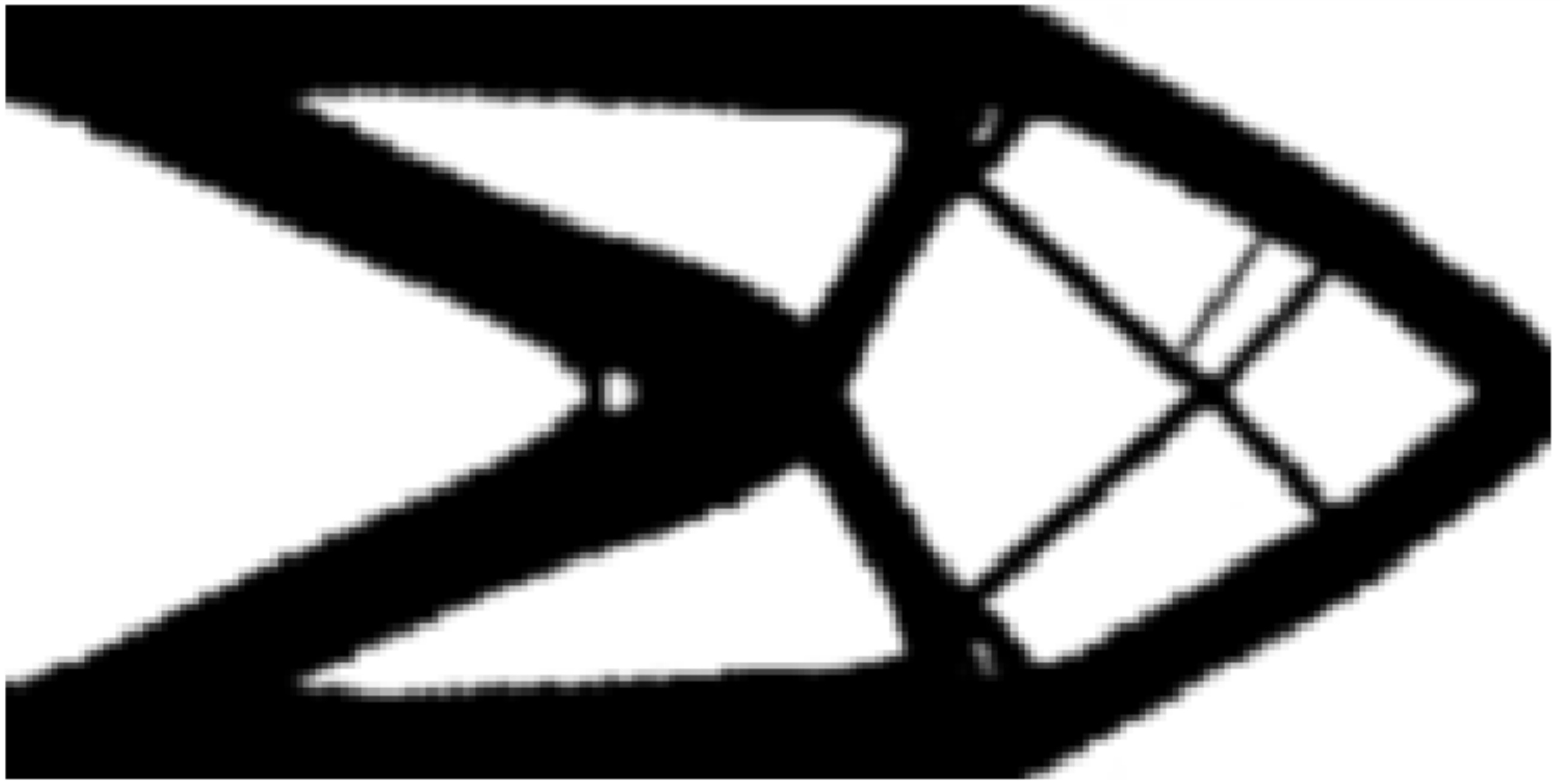}};
\end{tikzpicture}\hspace{0.5em}
\caption{Cycle: 2}
\end{subfigure}
\begin{subfigure}{1\textwidth}
\centering
\begin{tikzpicture}
    \node[anchor=south west,inner sep=0] at (0,0){
\includegraphics[scale=0.14]{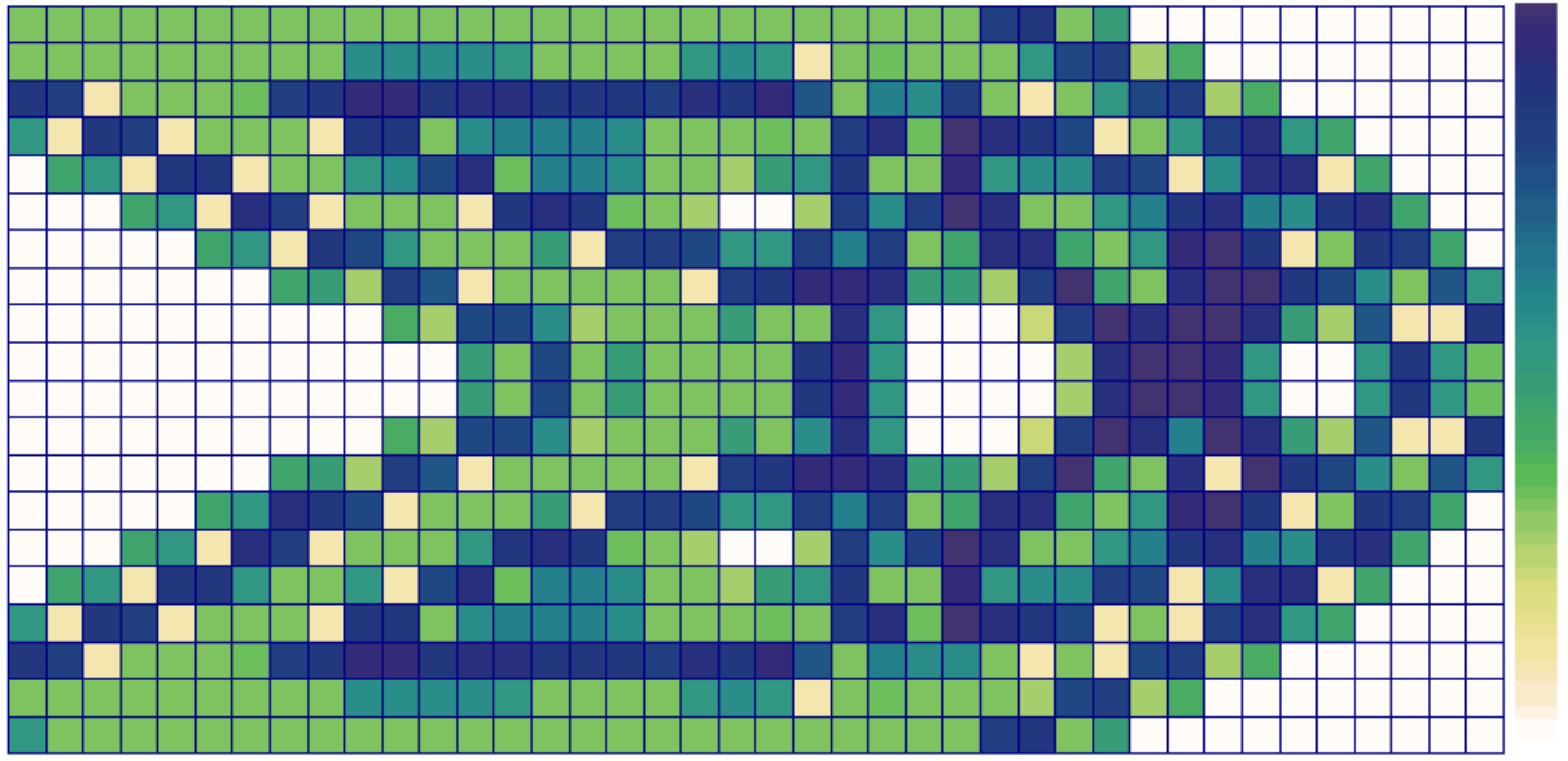}};
\node[text width=0cm] at (3.9, 1.77) 
    {\tiny \color{darkgray}{47}};
    \node[text width=0cm] at (3.9, 1.35) 
    {\tiny \color{darkgray}{34}};
    \node[text width=0cm] at (3.9, 0.92) 
    {\tiny \color{darkgray}{23}};
    \node[text width=0cm] at (3.9, 0.49) 
    {\tiny \color{darkgray}{11}};
\node[text width=0cm] at (3.9, 0.1) 
    {\tiny \color{darkgray}{1}};
\end{tikzpicture}\hspace{0.5em}
\begin{tikzpicture}
    \node[anchor=south west,inner sep=0] at (0,0){
\includegraphics[scale=0.14]{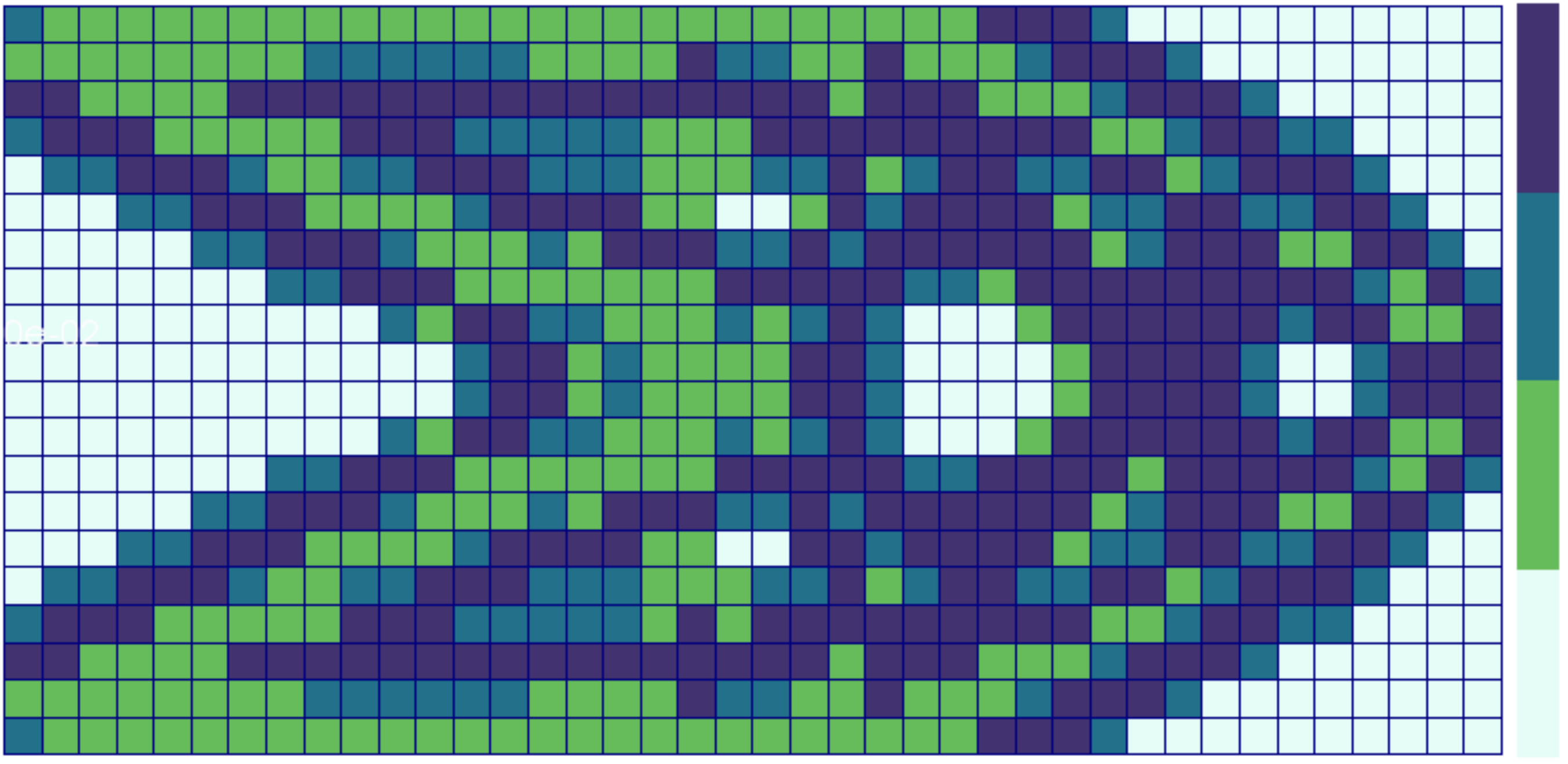}};
\node[text width=0cm] at (3.9, 1.77) 
    {\tiny \color{darkgray}{4}};
    \node[text width=0cm] at (3.9, 1.25) 
    {\tiny \color{darkgray}{3}};
    \node[text width=0cm] at (3.9, 0.72) 
    {\tiny \color{darkgray}{2}};
\node[text width=0cm] at (3.9, 0.15) 
    {\tiny \color{darkgray}{1}};
\end{tikzpicture}\hspace{0.5em}
\begin{tikzpicture}
    \node[anchor=south west,inner sep=0] at (0,0){
\includegraphics[scale=0.14]{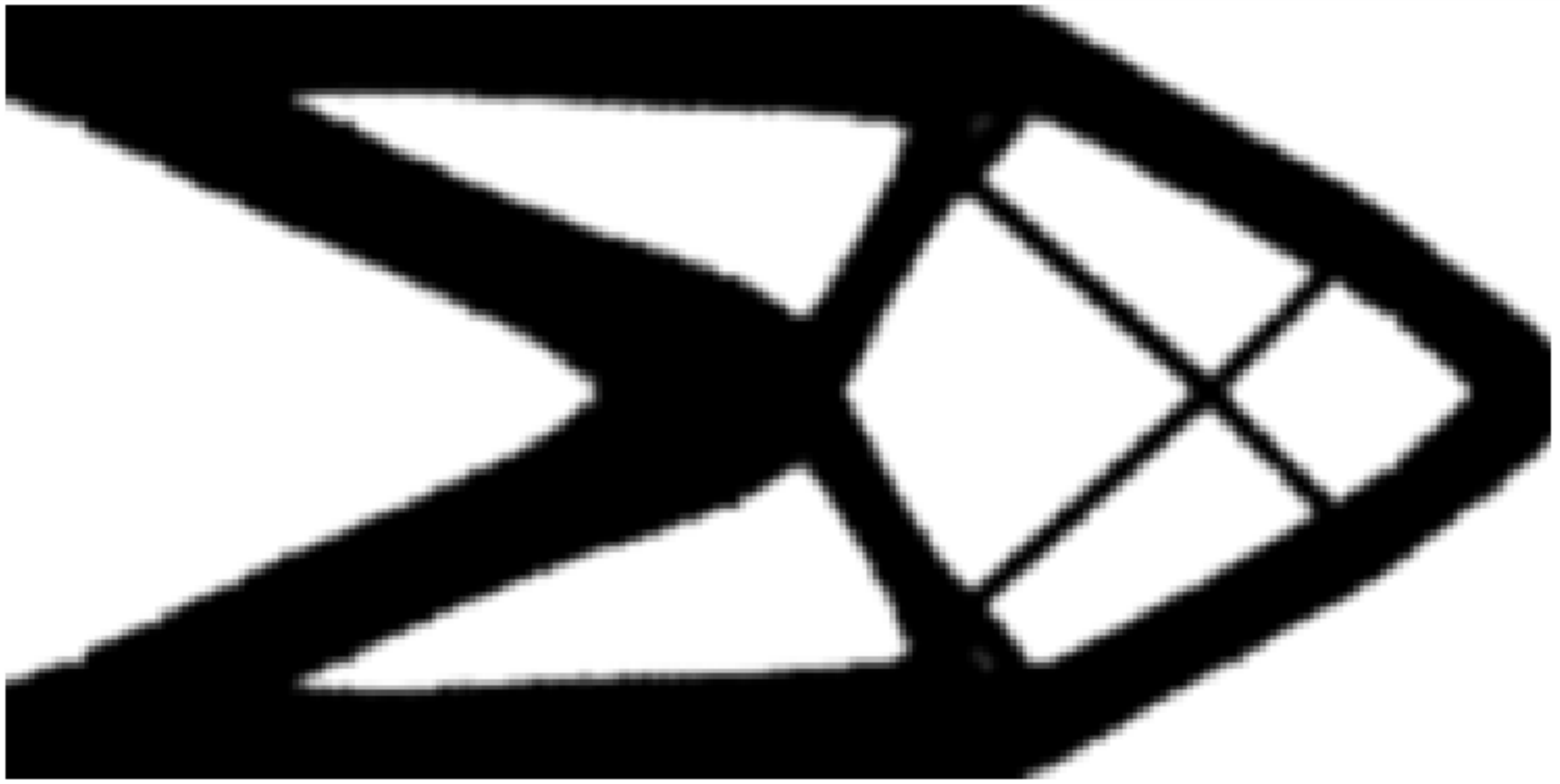}};
\end{tikzpicture}\hspace{0.5em}
\caption{Cycle: 3}
\end{subfigure}
\begin{subfigure}{1\textwidth}
\begin{tikzpicture}
    \node[anchor=south west,inner sep=0] at (0,0){
\includegraphics[scale=0.14]{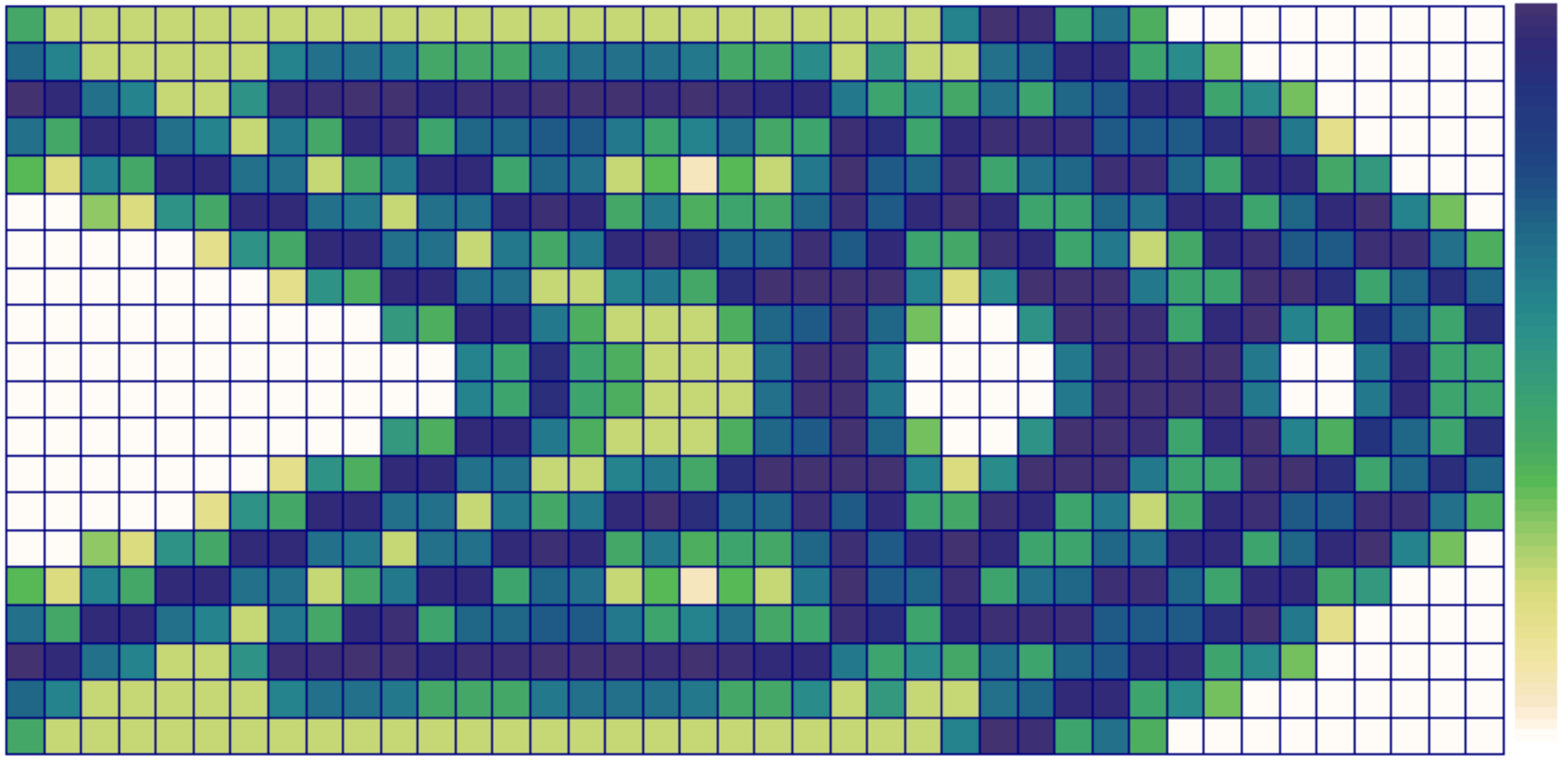}};
\node[text width=0cm] at (3.9, 1.77) 
    {\tiny \color{darkgray}{64}};
    \node[text width=0cm] at (3.9, 1.35) 
    {\tiny \color{darkgray}{48}};
    \node[text width=0cm] at (3.9, 0.92) 
    {\tiny \color{darkgray}{32}};
    \node[text width=0cm] at (3.9, 0.47) 
    {\tiny \color{darkgray}{16}};
\node[text width=0cm] at (3.9, 0.1) 
    {\tiny \color{darkgray}{1}};
\end{tikzpicture}\hspace{0.5em}
\centering
\begin{tikzpicture}
    \node[anchor=south west,inner sep=0] at (0,0){
\includegraphics[scale=0.14]{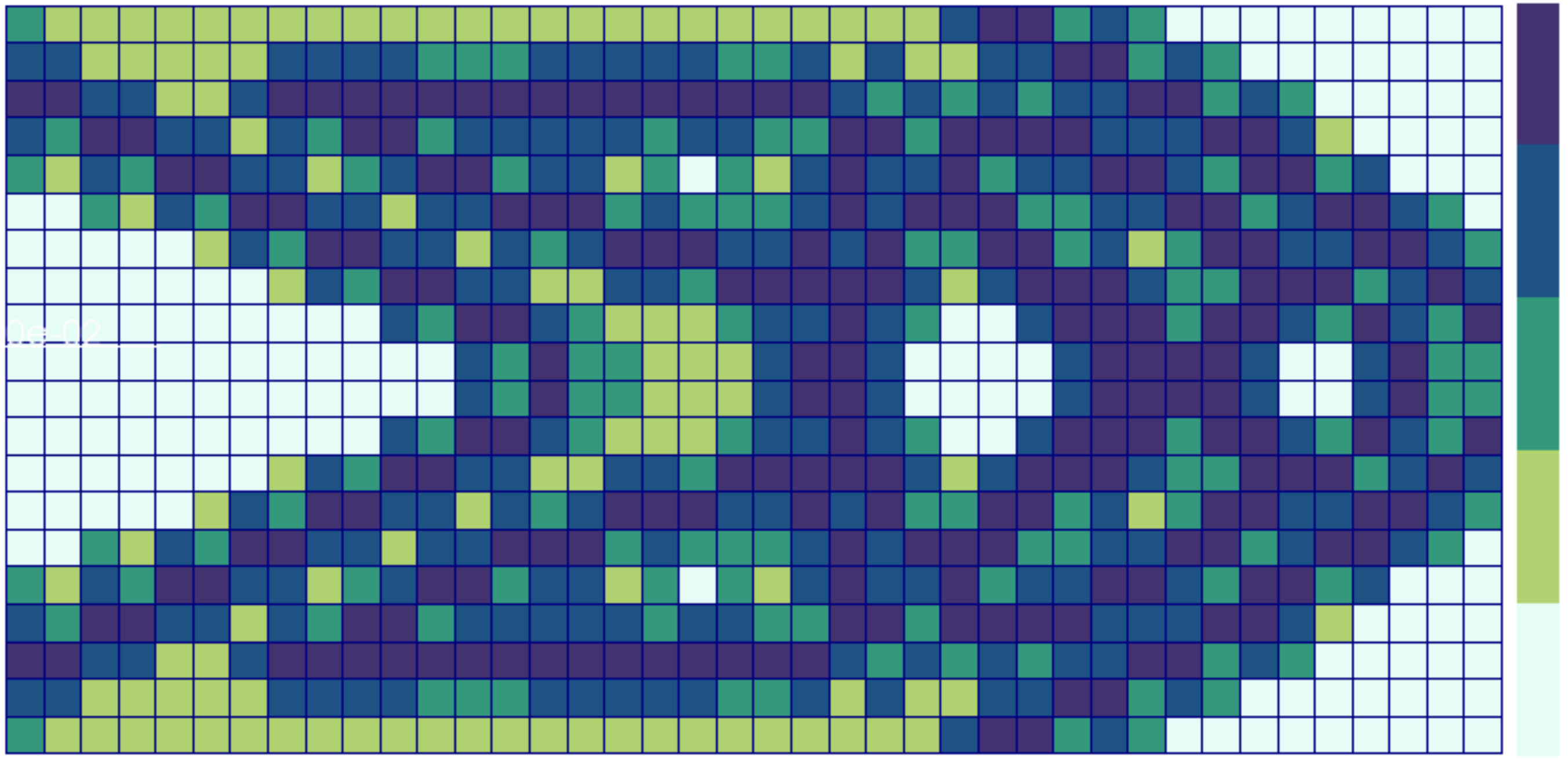}};
\node[text width=0cm] at (3.9, 1.77) 
    {\tiny \color{darkgray}{5}};
    \node[text width=0cm] at (3.9, 1.35) 
    {\tiny \color{darkgray}{4}};
    \node[text width=0cm] at (3.9, 0.92) 
    {\tiny \color{darkgray}{3}};
    \node[text width=0cm] at (3.9, 0.52) 
    {\tiny \color{darkgray}{2}};
\node[text width=0cm] at (3.9, 0.15) 
    {\tiny \color{darkgray}{1}};
\end{tikzpicture}\hspace{0.5em}
\begin{tikzpicture}
    \node[anchor=south west,inner sep=0] at (0,0){
\includegraphics[scale=0.14]{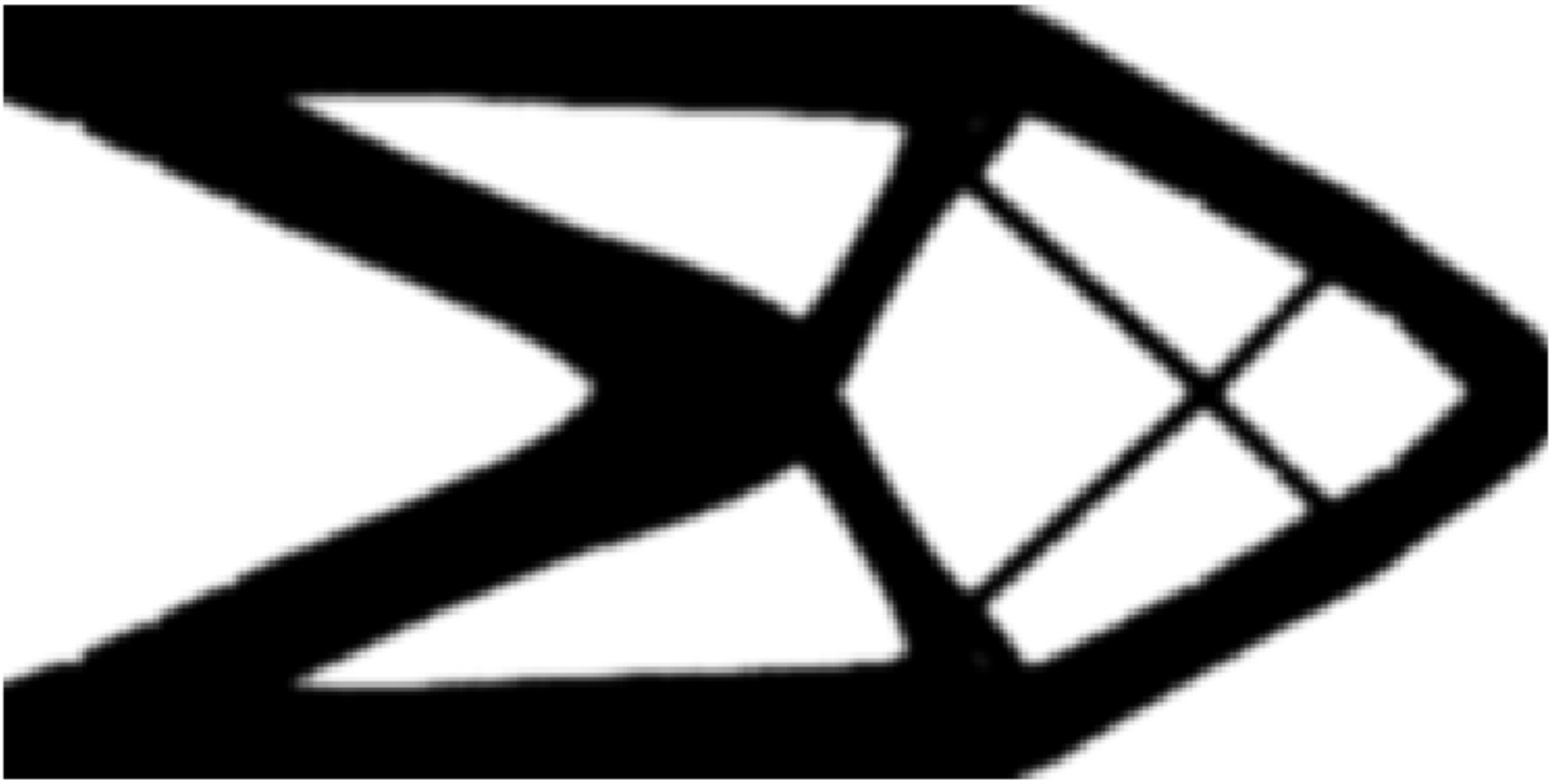}};
\end{tikzpicture}\hspace{0.5em}
\caption{Cycle: 4}
\end{subfigure}
\caption{Optimized designs (right), and the respective shape function orders (middle) and design field (left) obtained for 4 cycles of $dp$-adaptive MTO run for a cantilever beam subjected to point load (Fig. \ref{cant_point_load_fig1}). The initial mesh  is uniform and each element has shape functions of polynomial order 2 and 16 design points per element. The maximum allowed shape function order and number of design points are restricted to 5 and 64 per element, respectively. }
\label{fig_pointcant1_prog}
\end{figure*}

Fig. \ref{fig_pointcant1_prog} shows the distributions of shape function order and design points as well as the optimized designs for 4 cycles of the $dp$-adaptive MTO run of this case. It can be seen that refinement mainly occurs near the edges of the structure, and coarsening occurs as desired in solid and void parts. The optimized design in Cycle 1 consists of disconnected features, which are primarily the QR-patterns arising from the limitations of low order polynomial shape functions in those parts of the design \cite{Gupta2018}. Over the next cycles, $p$-refinement occurs in those regions and the QR-patterns are eliminated.  Since the design points are distributed in the domain using $k$-means clustering without symmetry constraints, the distribution of design points itself can be asymmetrical, which in Cycle 2 leads to an asymmetrical design. An example of such asymmetry can be observed in the optimized design of Cycle 2, which gradually disappears over the next cycle.

\begin{figure}
\centering
\begin{subfigure}{0.48\linewidth}
\centering
\includegraphics[scale=0.22]{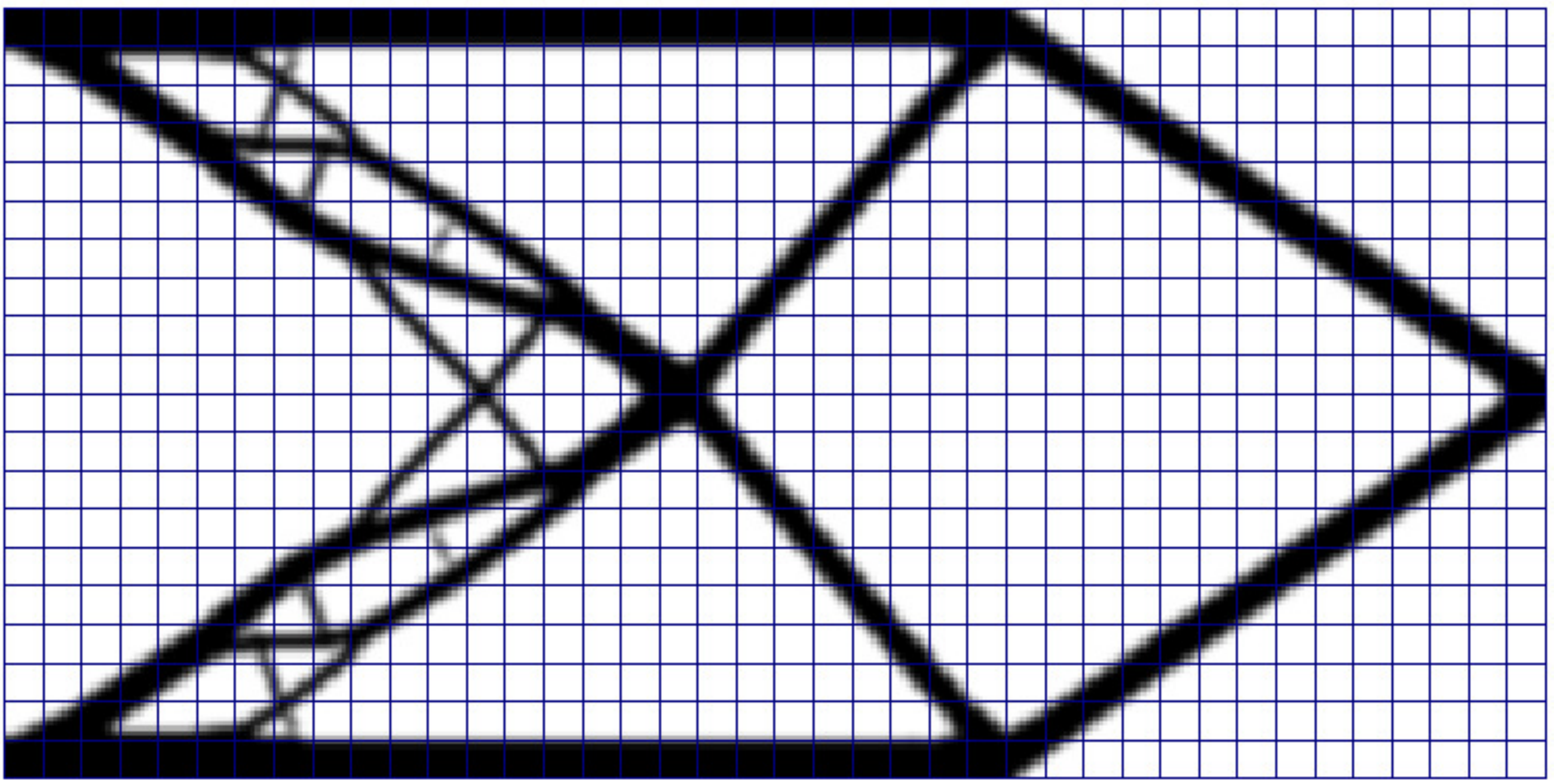}
\caption{MTO ($\mathcal{J}_0 = 175.72$J)}
\label{point_cant20_1a}
\end{subfigure}
\begin{subfigure}{0.48\linewidth}
\centering
\includegraphics[scale=0.22]{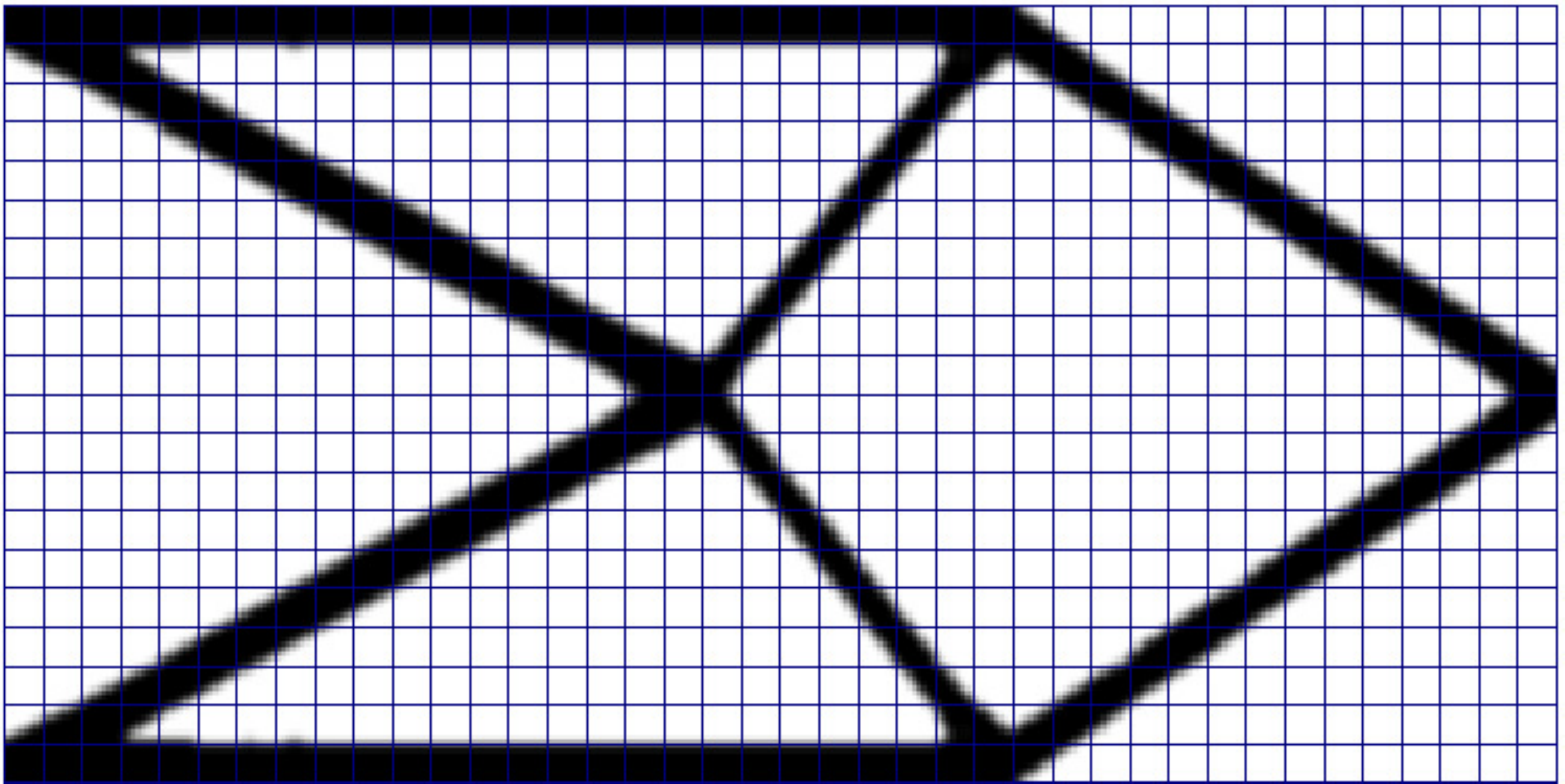}
\caption{$dp$-adaptive MTO ($\mathcal{J} = 163.39$J)}
\label{point_cant20_1b}
\end{subfigure}
\caption{Optimized cantilever designs for the point load case shown in Fig. \ref{cant_point_load_fig1}, obtained using a uniform MTO mesh (left) and $dp$-adaptive MTO approach (right). The maximum permissible material volume fraction is set to 0.20. A speed-up of 8.3 times is obtained using $dp$-adaptivity. Additional information related to this test case is listed in Table \ref{table_mto_runs}.}
\label{point_cant20_1}
\end{figure}
\begin{figure}
\centering
\begin{subfigure}{0.48\linewidth}
\centering
\includegraphics[scale=0.22]{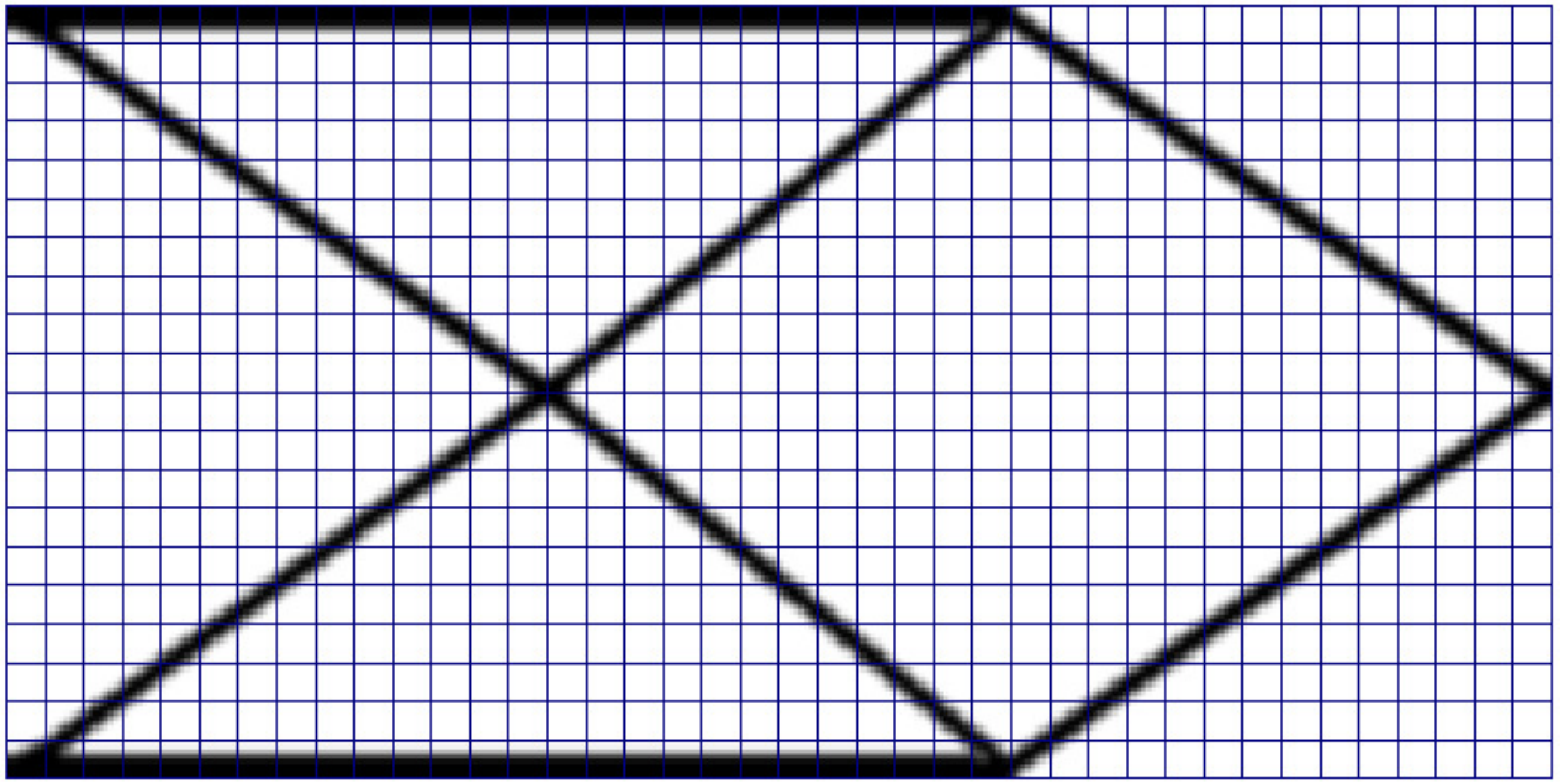}
\caption{MTO ($\mathcal{J}_0 = 410.39$J)}
\end{subfigure}
\begin{subfigure}{0.48\linewidth}
\centering
\includegraphics[scale=0.22]{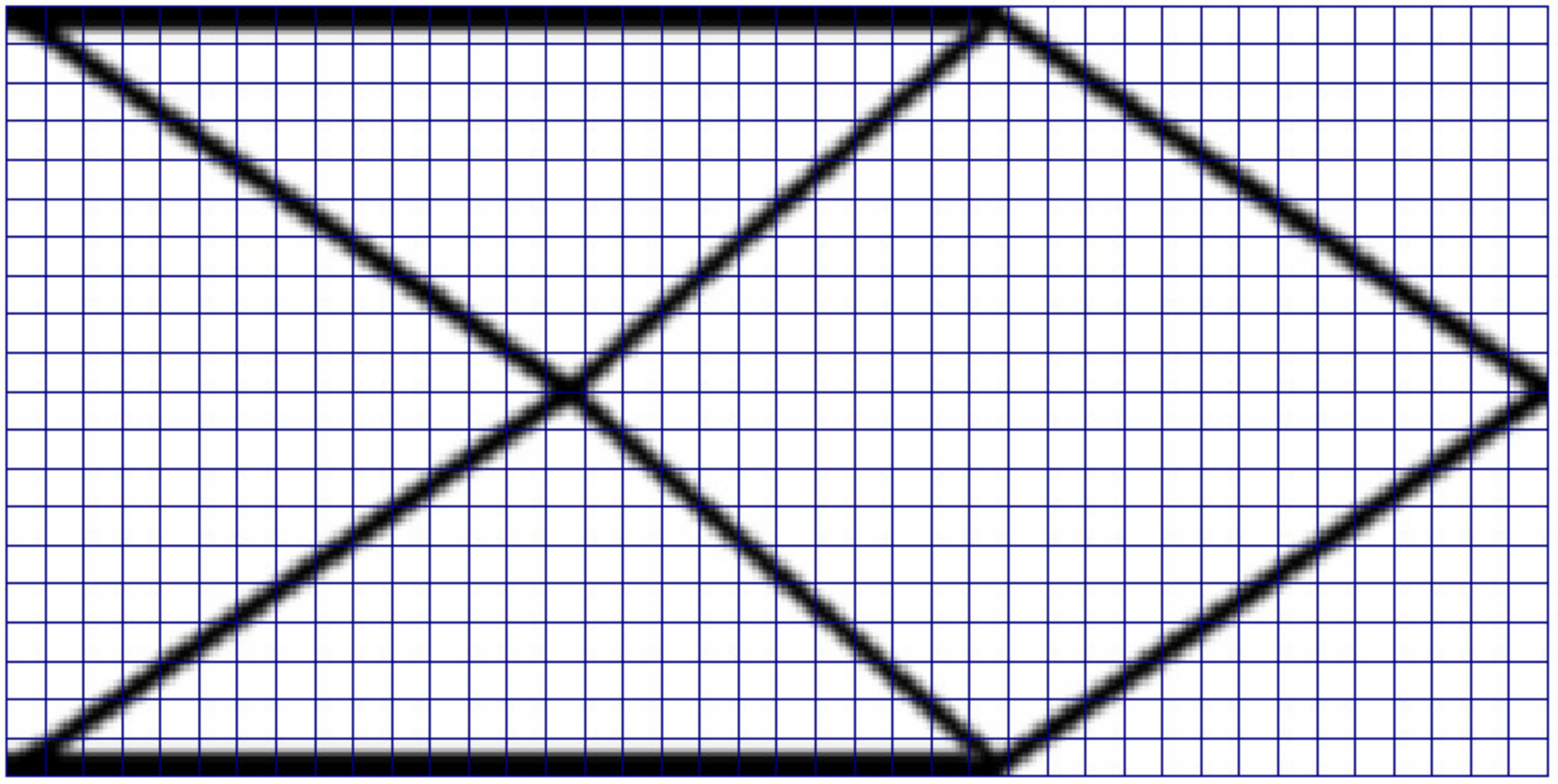}
\caption{$dp$-adaptive MTO ($\mathcal{J} = 423.23$J)}
\end{subfigure}
\caption{Optimized cantilever designs for the point load case shown in Fig. \ref{cant_point_load_fig1}, obtained using a uniform MTO mesh (left) and $dp$-adaptive MTO approach (right). The maximum permissible material volume fraction is set to 0.10. A 10-fold speed-up is obtained using $dp$-adaptivity. Additional information related to this test case is listed in Table \ref{table_mto_runs}.}
\label{point_cant10_1}
\end{figure}
\begin{figure*}[!htb]
\centering
\begin{subfigure}{1\textwidth}
\centering
\hspace{0.5em}
\begin{tikzpicture}
    \node[anchor=south west,inner sep=0] at (0,0){
\includegraphics[scale=0.14]{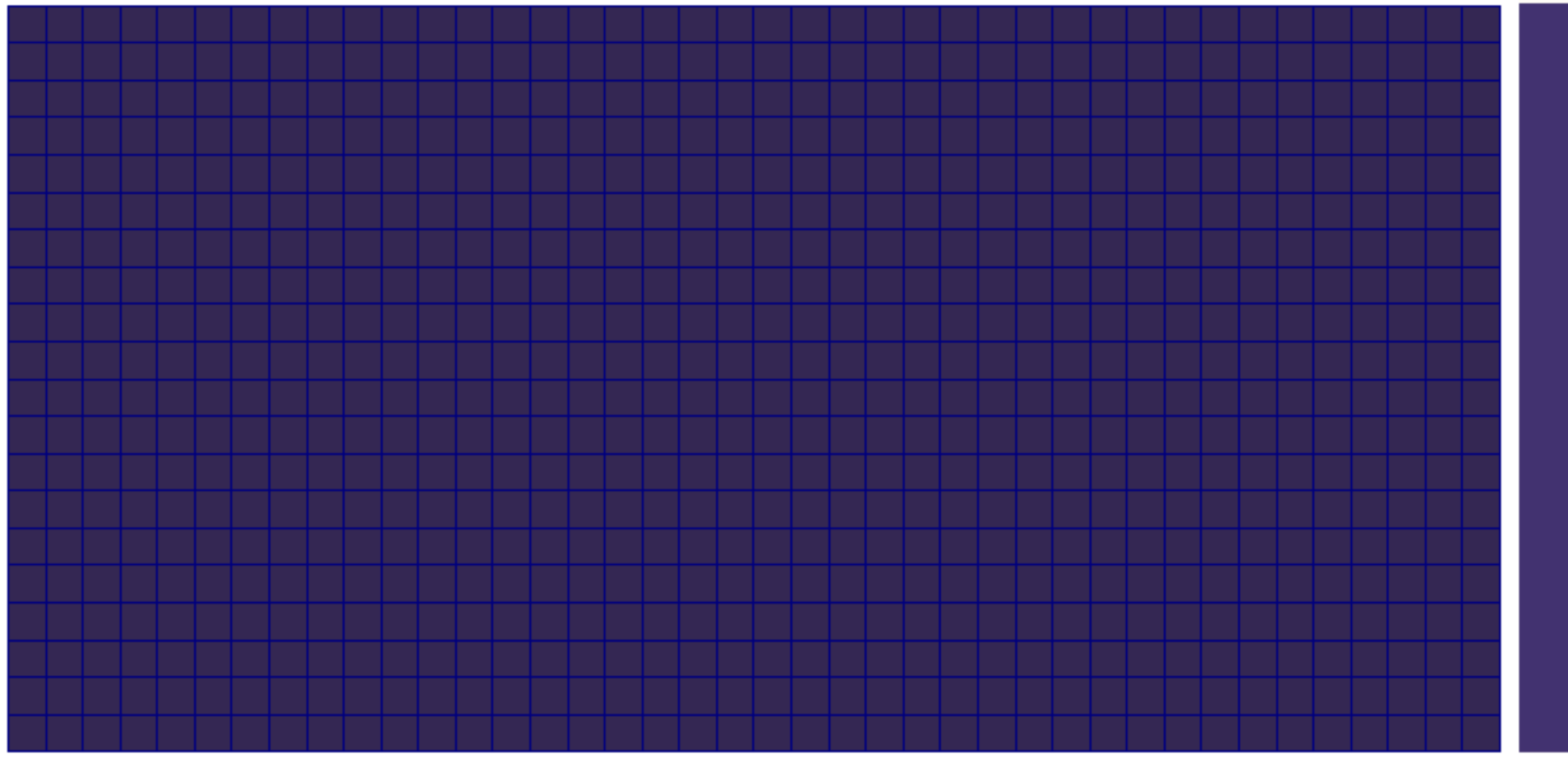}};
    \node[text width=0cm] at (3.93, 0.92) 
    {\tiny \color{darkgray}{16}};
    \node[text width=2cm] at (2.2, 2.3) 
    {\normalsize \color{darkgray}{Design field}};
\end{tikzpicture}\hspace{0.5em}
\begin{tikzpicture}
    \node[anchor=south west,inner sep=0] at (0,0){
\includegraphics[scale=0.14]{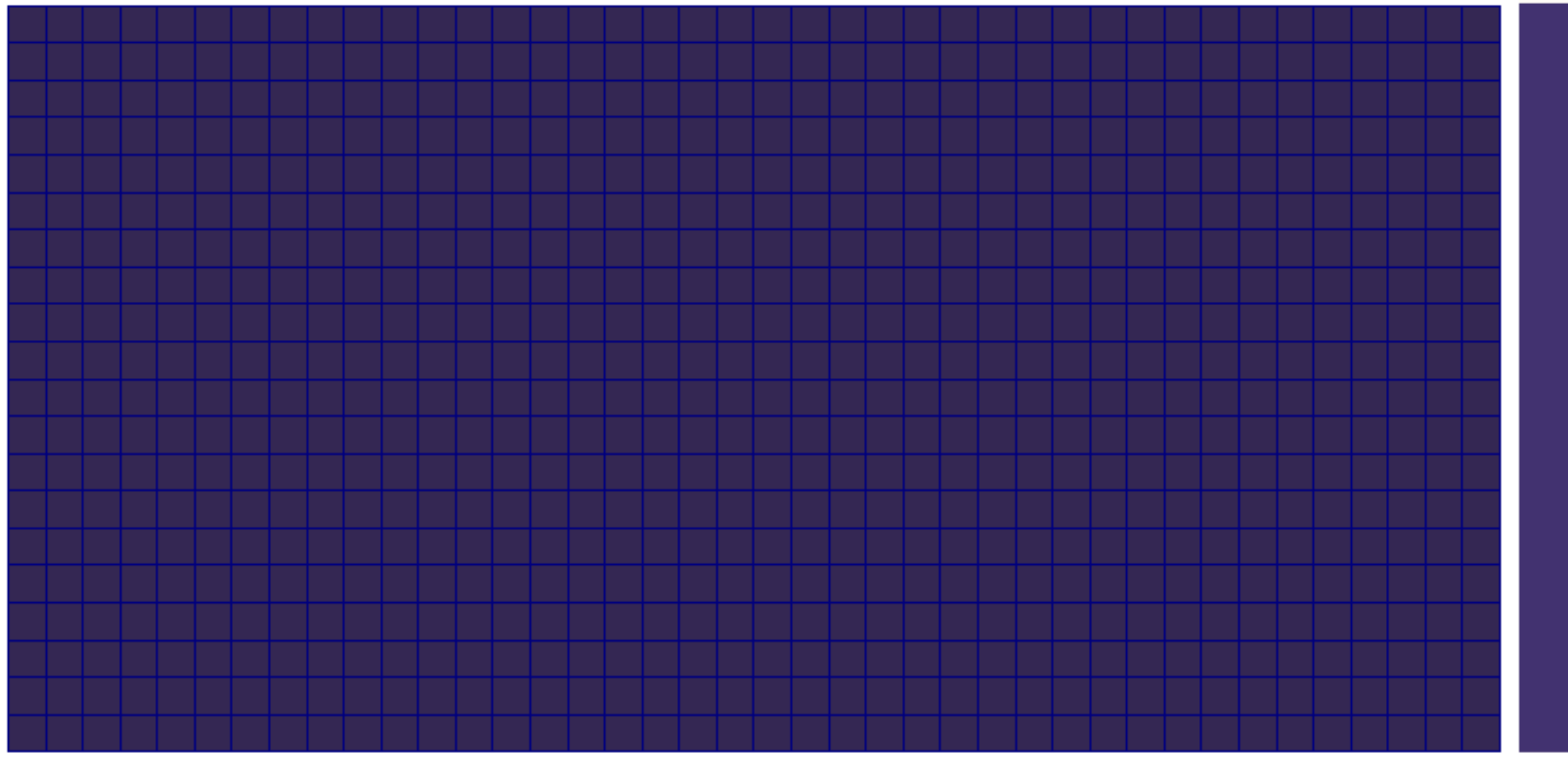}};
    \node[text width=0cm] at (3.93, 0.92) 
    {\tiny \color{darkgray}{2}};
    \node[text width=3.7cm] at (2.23, 2.3) 
    {\normalsize \color{darkgray}{Shape function order}};
\end{tikzpicture}
\begin{tikzpicture}
    \node[anchor=south west,inner sep=0] at (0,0){
\includegraphics[scale=0.14]{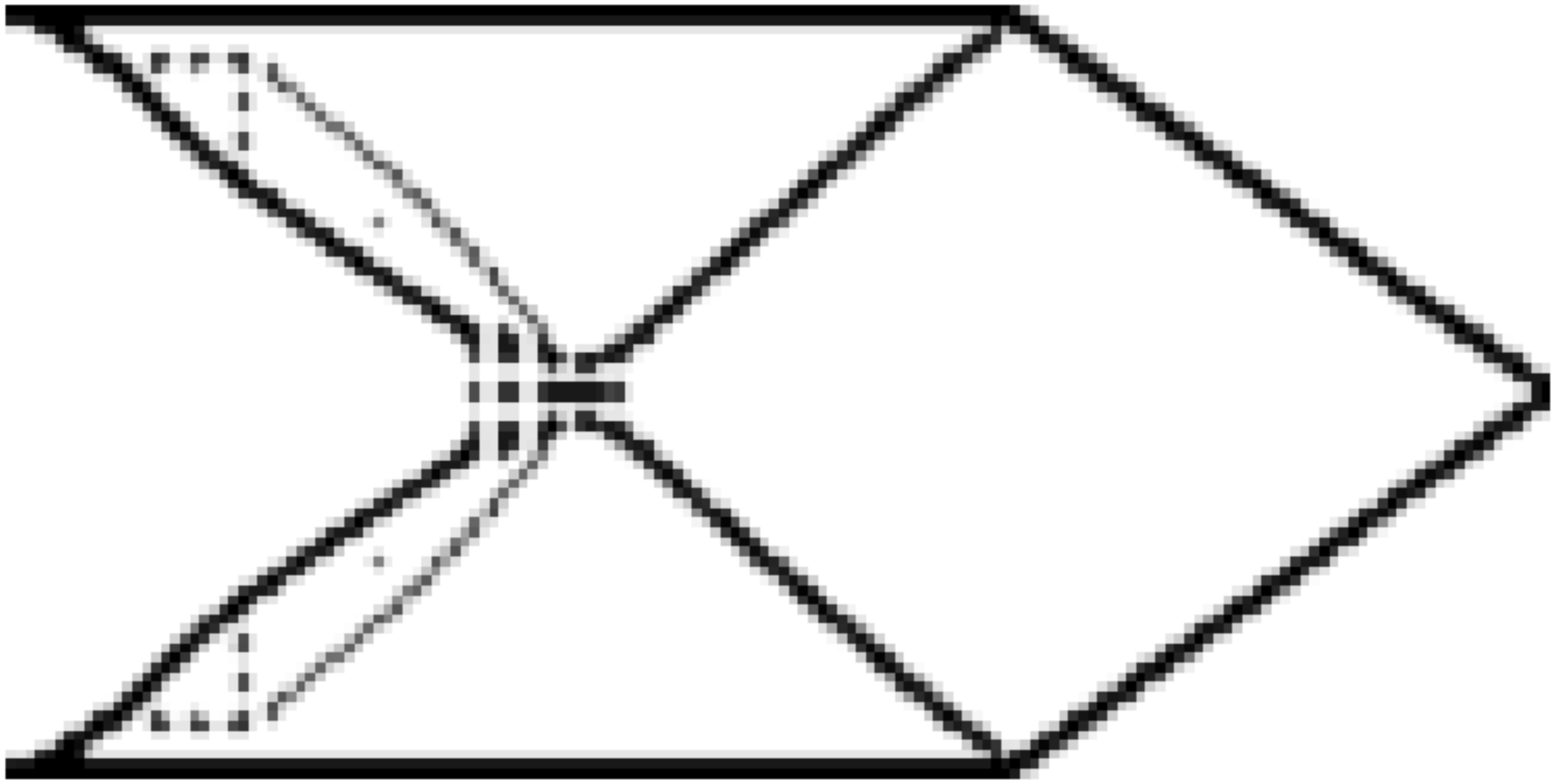}};
    \node[text width=3cm] at (2.2, 2.42) 
    {\normalsize \color{darkgray}{Optimized design}};
\end{tikzpicture}\hspace{0.5em}
\caption{Cycle: 1}
\end{subfigure}
\begin{subfigure}{1\textwidth}
\centering
\begin{tikzpicture}
    \node[anchor=south west,inner sep=0] at (0,0){
\includegraphics[scale=0.14]{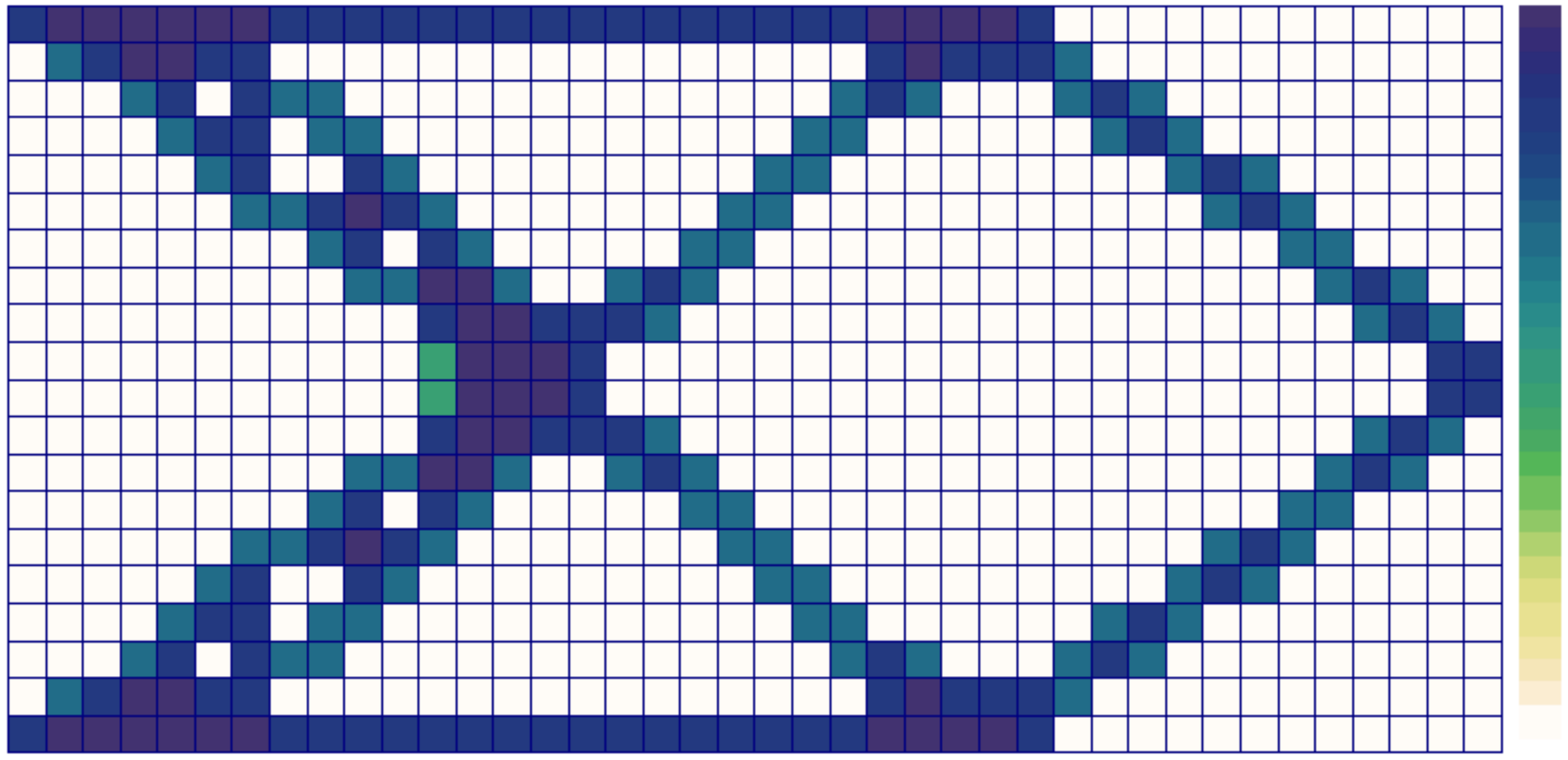}};
\node[text width=0cm] at (3.9, 1.77) 
    {\tiny \color{darkgray}{29}};
    \node[text width=0cm] at (3.9, 1.35) 
    {\tiny \color{darkgray}{24}};
    \node[text width=0cm] at (3.9, 0.92) 
    {\tiny \color{darkgray}{18}};
    \node[text width=0cm] at (3.9, 0.47) 
    {\tiny \color{darkgray}{12}};
\node[text width=0cm] at (3.9, 0.1) 
    {\tiny \color{darkgray}{5}};
\end{tikzpicture}\hspace{0.5em}
\begin{tikzpicture}
    \node[anchor=south west,inner sep=0] at (0,0){
\includegraphics[scale=0.14]{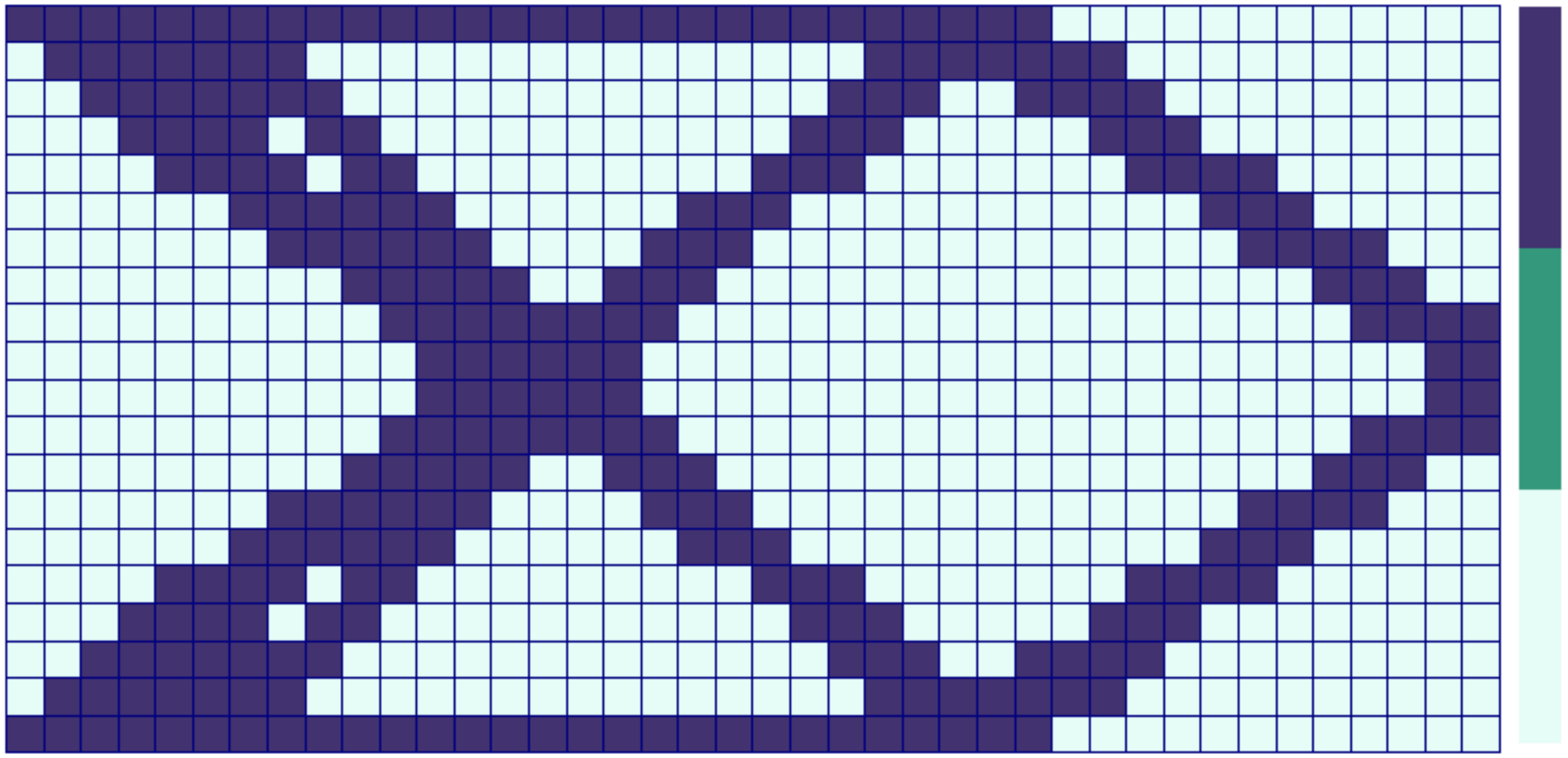}};
\node[text width=0cm] at (3.9, 1.47) 
    {\tiny \color{darkgray}{3}};
    \node[text width=0cm] at (3.9, 1.35) 
    {\tiny \color{darkgray}{}};
    \node[text width=0cm] at (3.9, 0.92) 
    {\tiny \color{darkgray}{2}};
    \node[text width=0cm] at (3.9, 0.47) 
    {\tiny \color{darkgray}{}};
\node[text width=0cm] at (3.9, 0.37) 
    {\tiny \color{darkgray}{1}};
\end{tikzpicture}\hspace{0.5em}
\begin{tikzpicture}
    \node[anchor=south west,inner sep=0] at (0,0){
\includegraphics[scale=0.14]{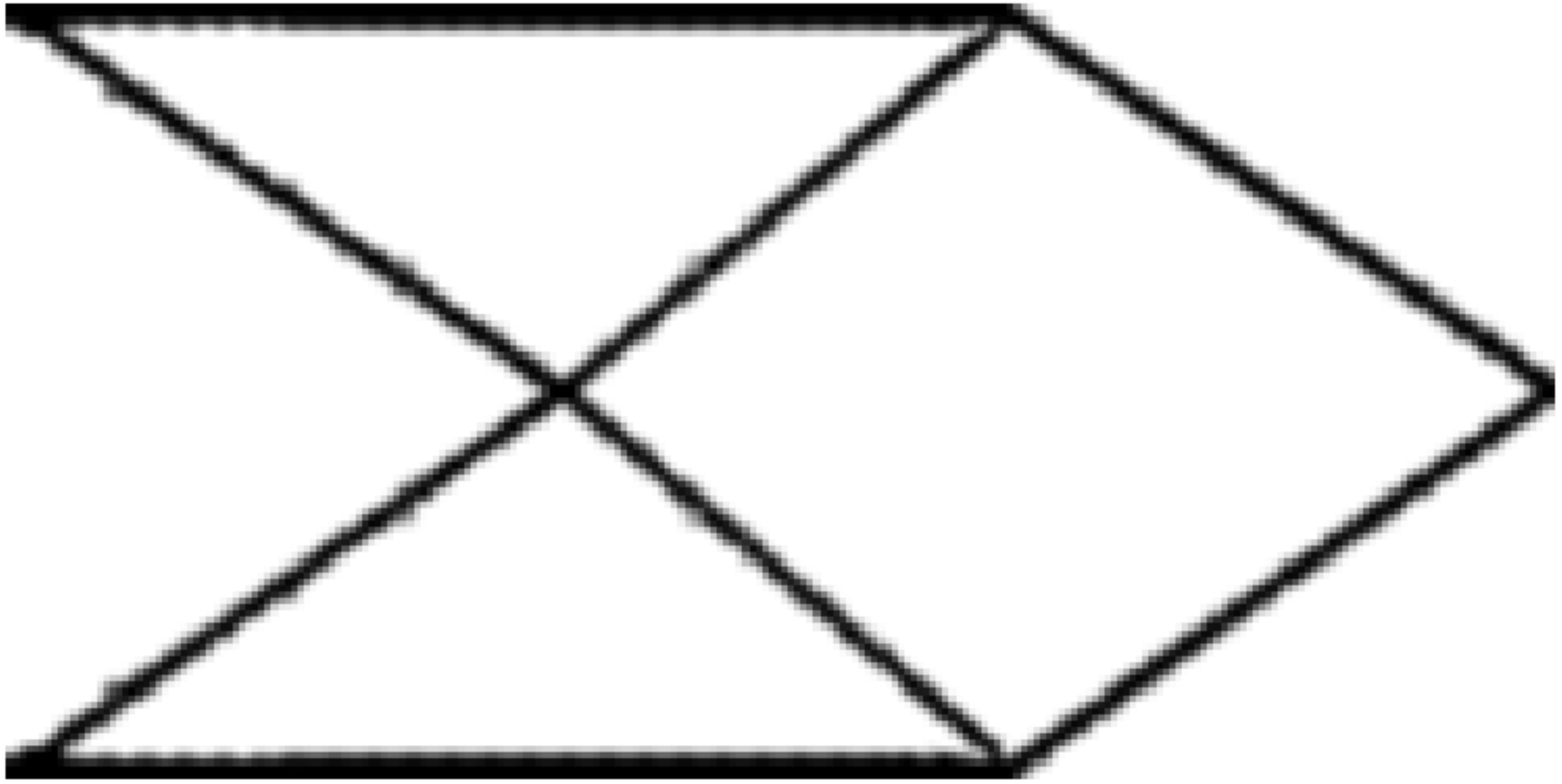}};
\end{tikzpicture}\hspace{0.5em}
\caption{Cycle: 2}
\end{subfigure}
\begin{subfigure}{1\textwidth}
\centering
\begin{tikzpicture}
    \node[anchor=south west,inner sep=0] at (0,0){
\includegraphics[scale=0.14]{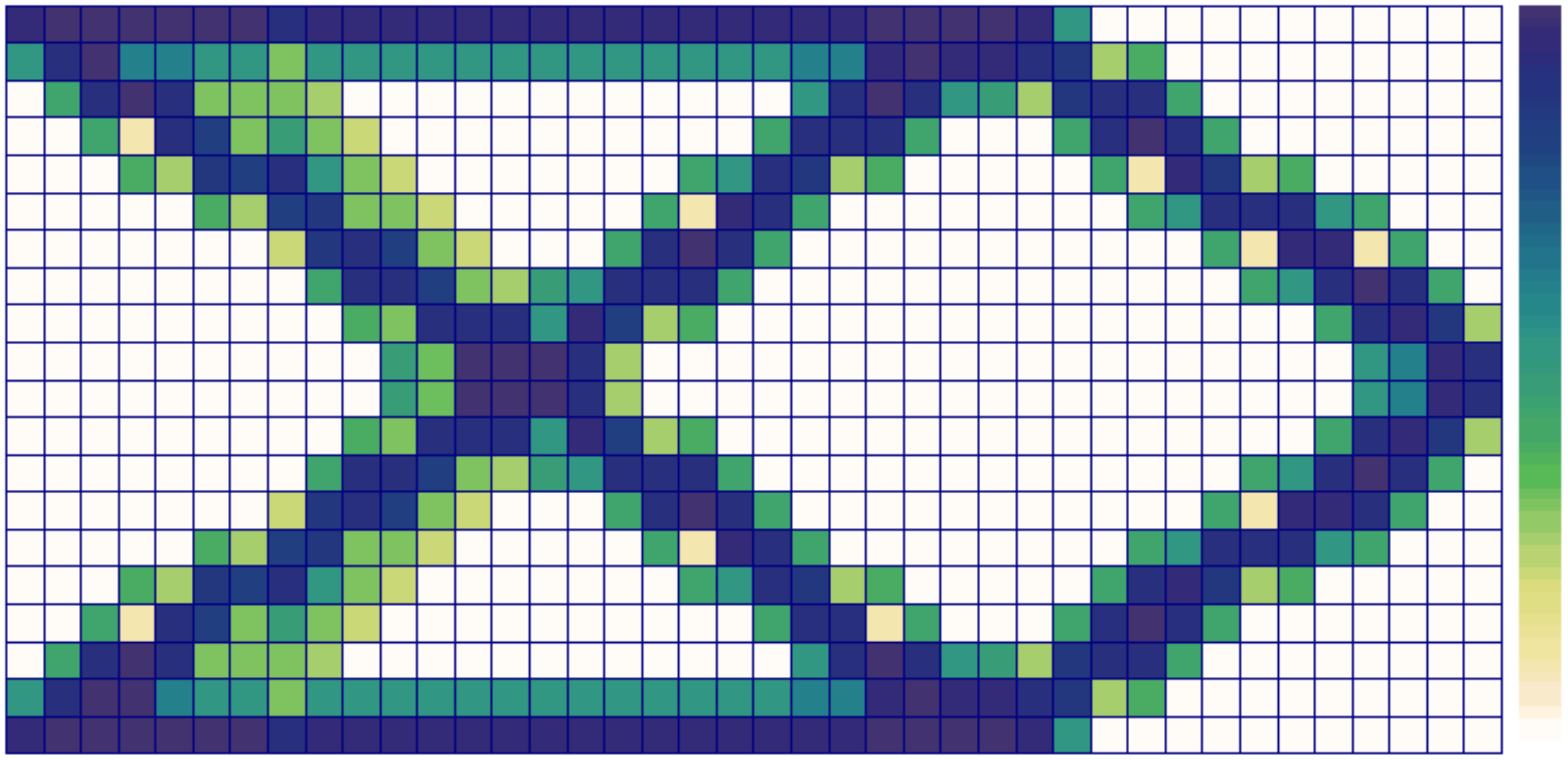}};
\node[text width=0cm] at (3.9, 1.77) 
    {\tiny \color{darkgray}{47}};
    \node[text width=0cm] at (3.9, 1.35) 
    {\tiny \color{darkgray}{34}};
    \node[text width=0cm] at (3.9, 0.92) 
    {\tiny \color{darkgray}{23}};
    \node[text width=0cm] at (3.9, 0.49) 
    {\tiny \color{darkgray}{11}};
\node[text width=0cm] at (3.9, 0.1) 
    {\tiny \color{darkgray}{1}};
\end{tikzpicture}\hspace{0.5em}
\begin{tikzpicture}
    \node[anchor=south west,inner sep=0] at (0,0){
\includegraphics[scale=0.14]{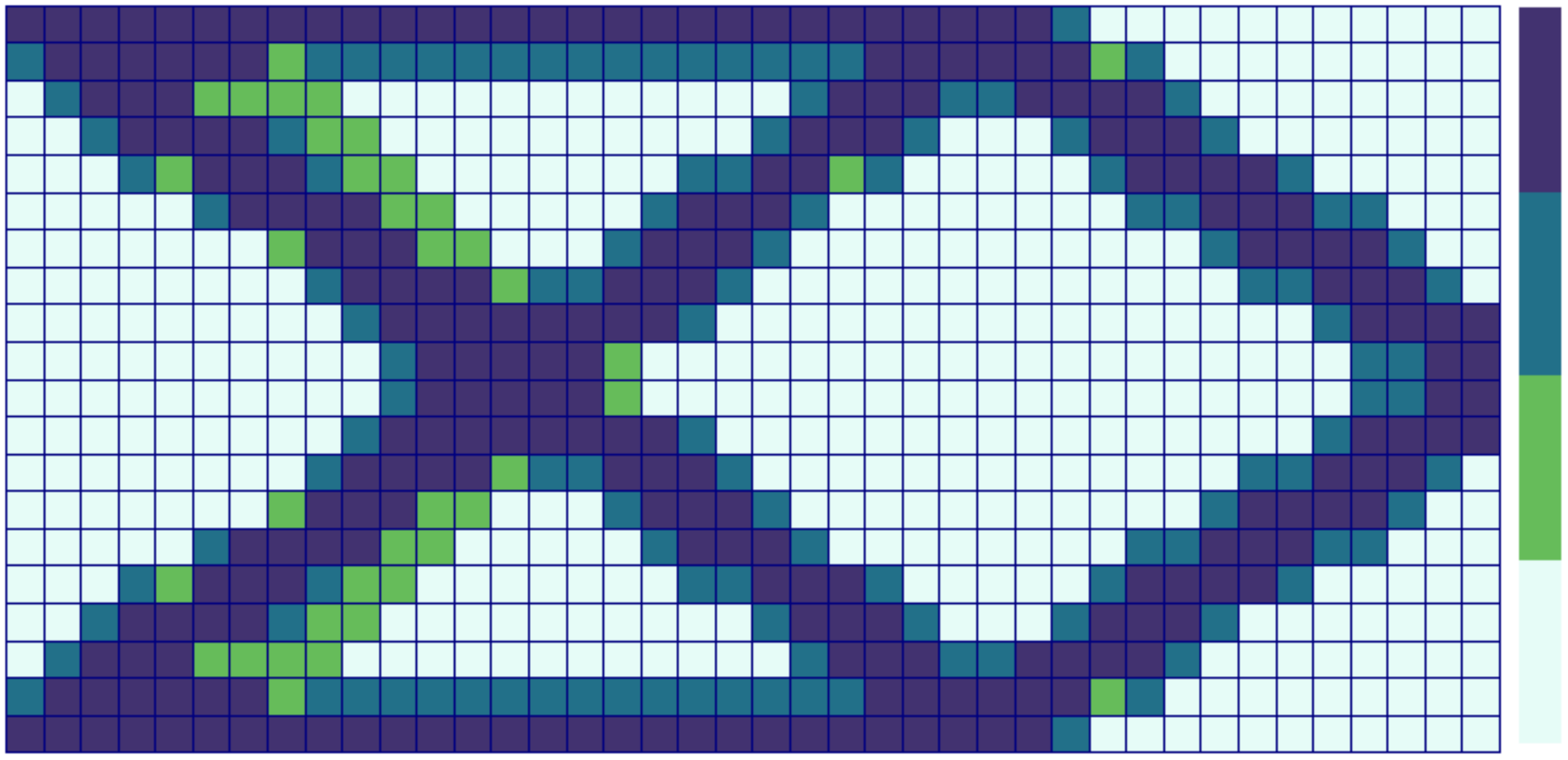}};
\node[text width=0cm] at (3.9, 1.77) 
    {\tiny \color{darkgray}{4}};
    \node[text width=0cm] at (3.9, 1.25) 
    {\tiny \color{darkgray}{3}};
    \node[text width=0cm] at (3.9, 0.72) 
    {\tiny \color{darkgray}{2}};
\node[text width=0cm] at (3.9, 0.15) 
    {\tiny \color{darkgray}{1}};
\end{tikzpicture}\hspace{0.5em}
\begin{tikzpicture}
    \node[anchor=south west,inner sep=0] at (0,0){
\includegraphics[scale=0.14]{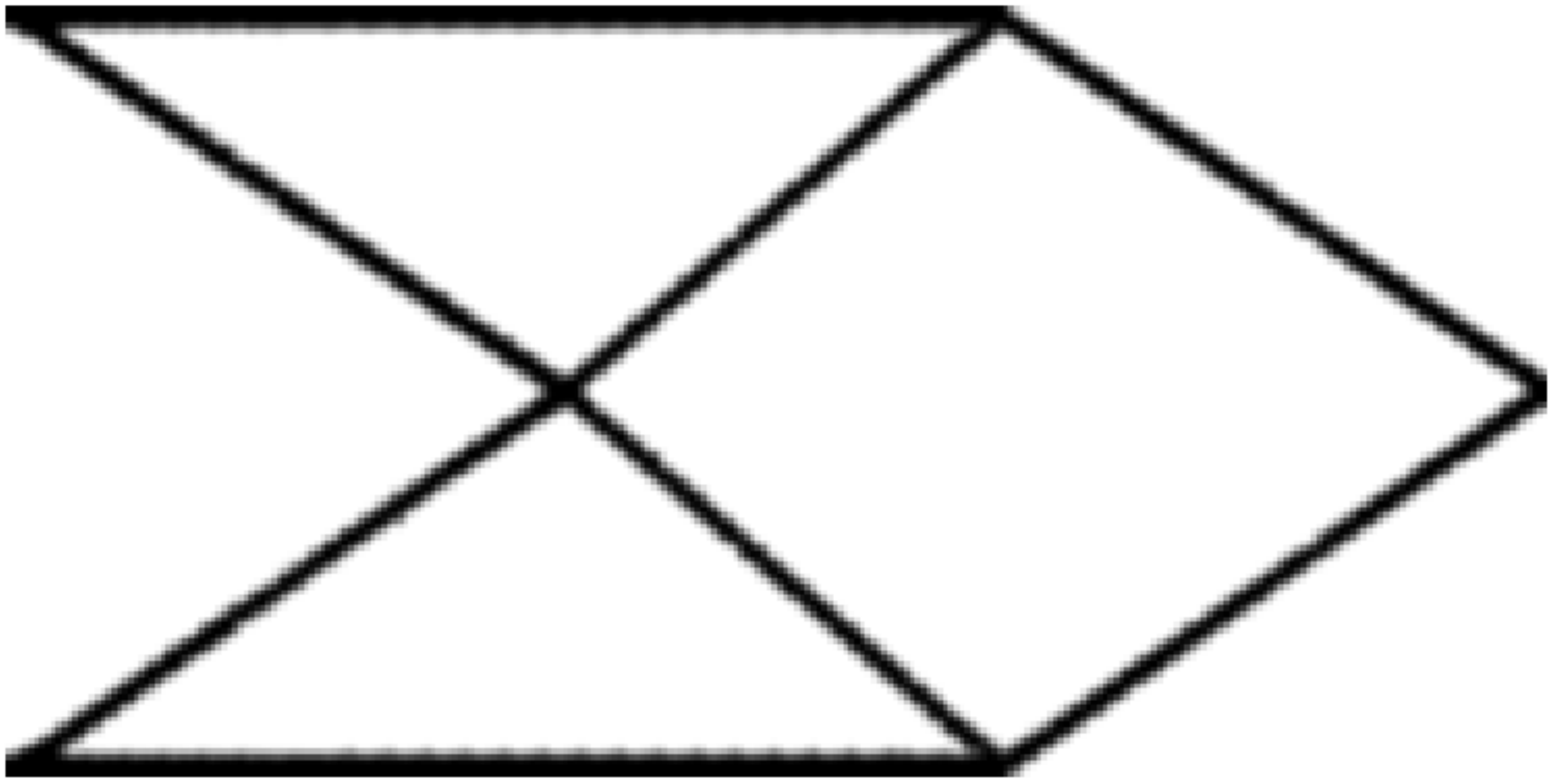}};
\end{tikzpicture}\hspace{0.5em}
\caption{Cycle: 3}
\end{subfigure}
\begin{subfigure}{1\textwidth}
\begin{tikzpicture}
    \node[anchor=south west,inner sep=0] at (0,0){
\includegraphics[scale=0.14]{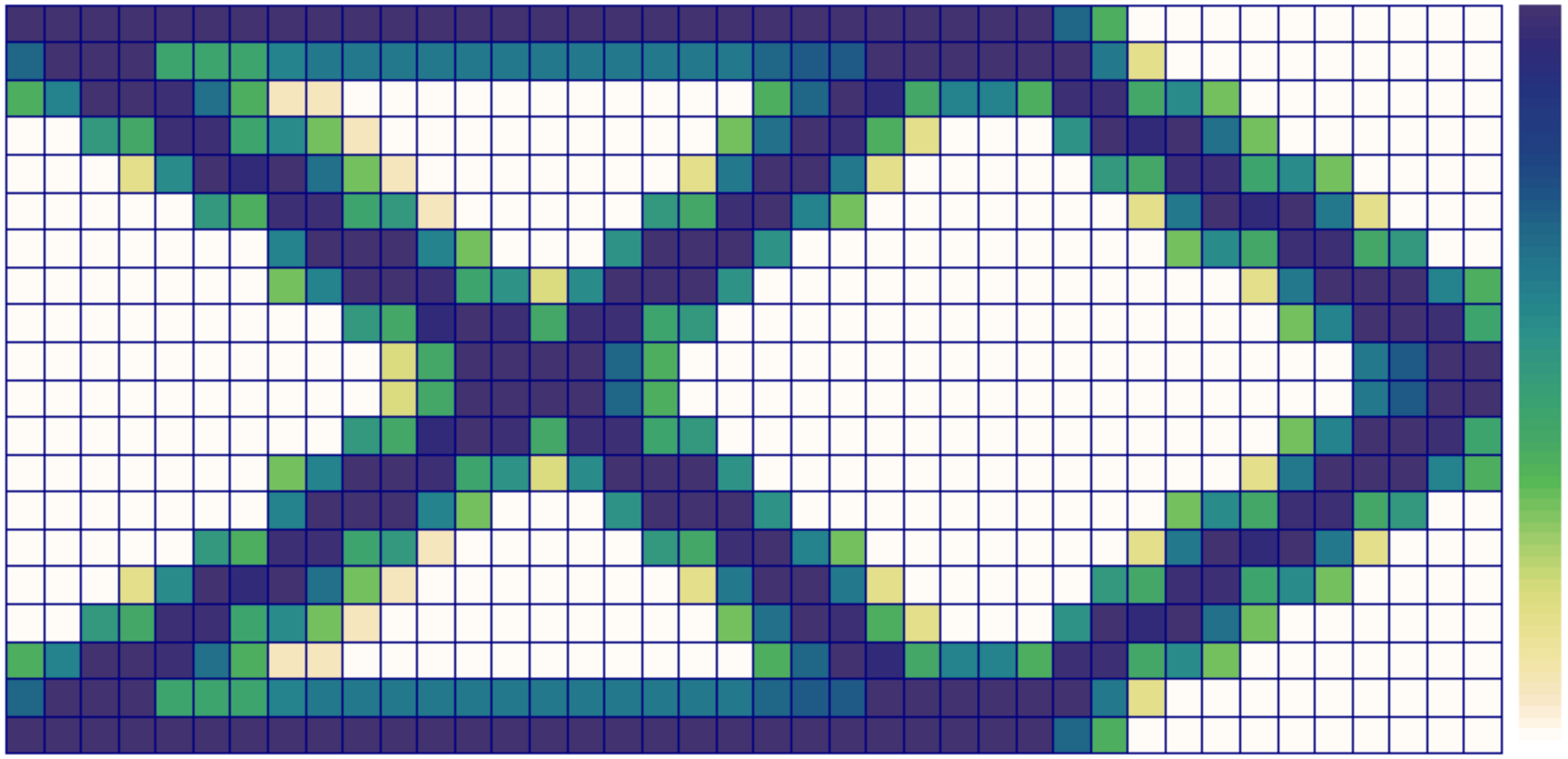}};
\node[text width=0cm] at (3.9, 1.77) 
    {\tiny \color{darkgray}{64}};
    \node[text width=0cm] at (3.9, 1.35) 
    {\tiny \color{darkgray}{48}};
    \node[text width=0cm] at (3.9, 0.92) 
    {\tiny \color{darkgray}{32}};
    \node[text width=0cm] at (3.9, 0.47) 
    {\tiny \color{darkgray}{16}};
\node[text width=0cm] at (3.9, 0.1) 
    {\tiny \color{darkgray}{1}};
\end{tikzpicture}\hspace{0.5em}
\centering
\begin{tikzpicture}
    \node[anchor=south west,inner sep=0] at (0,0){
\includegraphics[scale=0.14]{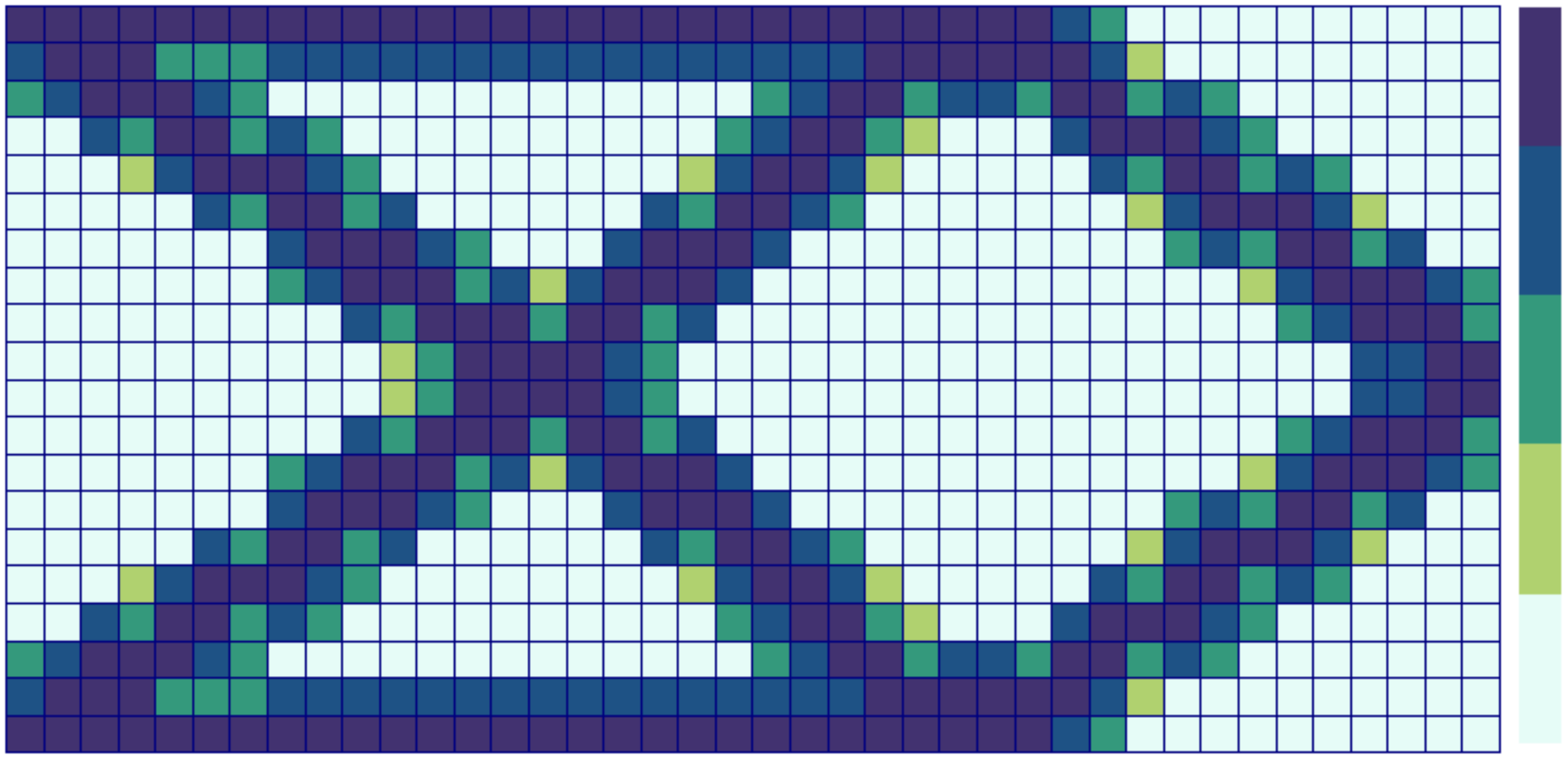}};
\node[text width=0cm] at (3.9, 1.77) 
    {\tiny \color{darkgray}{5}};
    \node[text width=0cm] at (3.9, 1.35) 
    {\tiny \color{darkgray}{4}};
    \node[text width=0cm] at (3.9, 0.92) 
    {\tiny \color{darkgray}{3}};
    \node[text width=0cm] at (3.9, 0.52) 
    {\tiny \color{darkgray}{2}};
\node[text width=0cm] at (3.9, 0.15) 
    {\tiny \color{darkgray}{1}};
\end{tikzpicture}\hspace{0.5em}
\begin{tikzpicture}
    \node[anchor=south west,inner sep=0] at (0,0){
\includegraphics[scale=0.14]{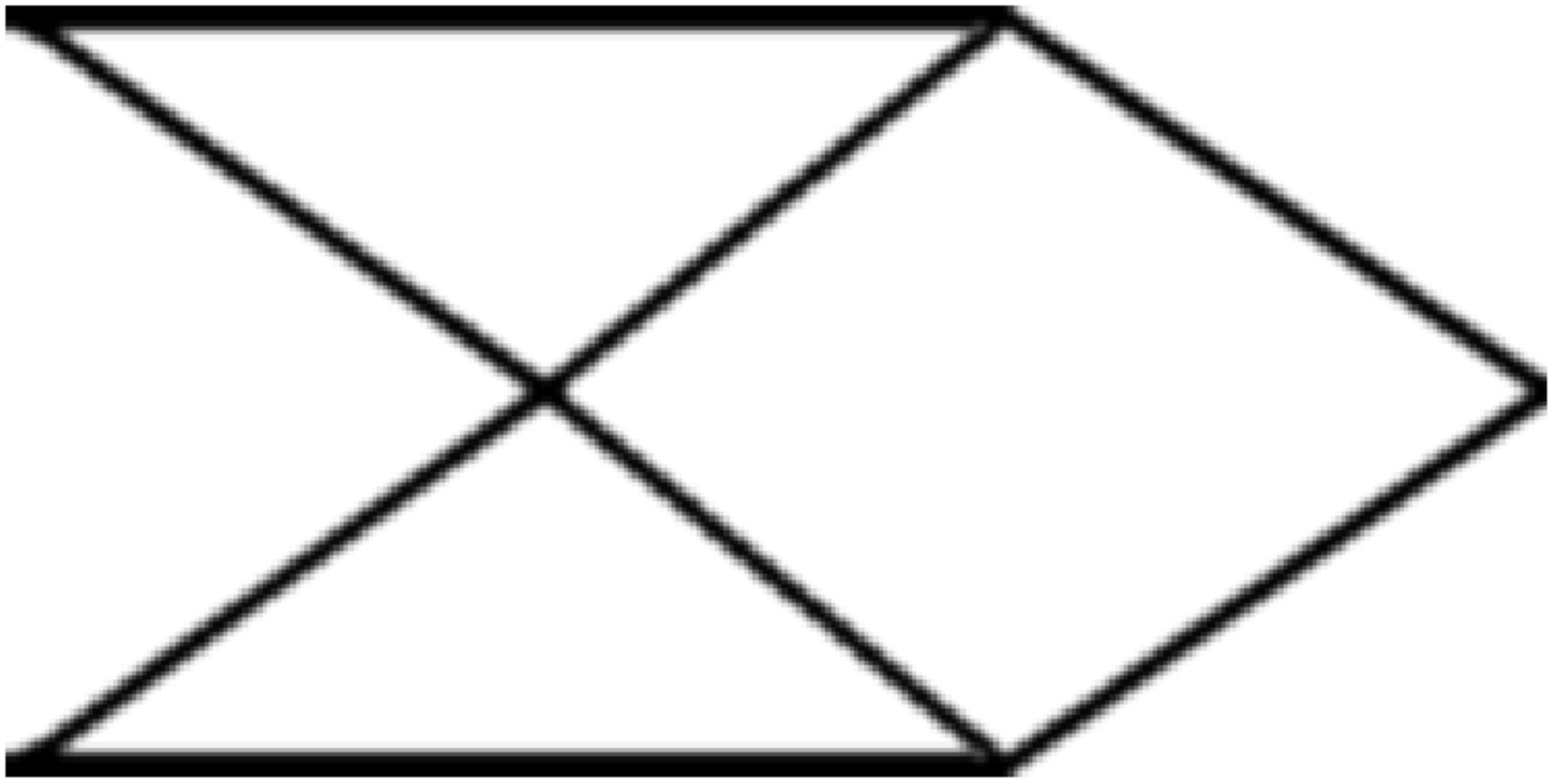}};
\end{tikzpicture}\hspace{0.5em}
\caption{Cycle: 4}
\end{subfigure}
\caption{Optimized designs (right) and the respective shape function orders (middle) and design field (left) obtained for 4 cycles of $dp$-adaptive MTO run for a cantilever beam subjected to point load (Fig. \ref{cant_point_load_fig1}). The initial mesh  is uniform and each element has shape functions of polynomial order 2 and 16 design points per element. The maximum allowed shape function order and number of design points are restricted to 5 and 64 per element, respectively.}
\label{fig_pointcant3_prog}
\end{figure*}

In general, TO problems involving lower volume fractions of material are more difficult in terms of convergence. Moreover, for problems involving low volume fractions of material, a significant part of the domain comprises voids, and in turn does not require a fine mesh resolution. Clearly, for such scenarios, $dp$-adaptivity could be potentially beneficial. To investigate this,  we study two additional cases of the point load cantilever beam involving lower values of $V_0$. 

Fig. \ref{point_cant20_1} shows the optimized designs for $V_0 = 0.20$ using conventional MTO (Fig. \ref{point_cant20_1a}) and \mbox{$dp$-adaptive} method (Fig. \ref{point_cant20_1b}), respectively. For $V_0 = 0.20$, the computational time advantage has increased to a factor of 8.3. Also, it is seen that the design obtained using the non-adaptive MTO method differs significantly from the result of $dp$-adaptivity. Moreover, in terms of performance, the design obtained using $dp$-adaptivity is relatively less compliant. The ratio $\mathcal{J}/\mathcal{J}_0$ is equal to 0.93. The compliance accuracy of the design obtained using the proposed method is found to be 0.98.

\indent As another test case for lower volume fractions, the point load cantilever problem is examined with $V_0 = 0.10$. Fig. \ref{point_cant10_1} shows the optimized designs for this volume fraction obtained using the conventional MTO method and $dp$-adaptive MTO, respectively. It is observed that for this volume fraction, the relative reduction in computational cost is even higher. Compared to the conventional MTO, a speed-up of 10 times is observed. The increase in speed-up is mainly due to the reduced number of free DOFs and design points, and the lower number of iterations required for convergence compared to the non-adaptive MTO. For this case, it is observed that $\mathcal{J}/\mathcal{J}_0$ is 1.03, which implies that the design obtained using $dp$-adaptivity is slightly inferior to that obtained using the non-adaptive version. The analysis accuracy is also slightly lower than in the previous cases, with $\mathcal{J}/\mathcal{J}^* = 0.96$.\\
\indent An understanding on the convergence of the $dp$-adaptive MTO process for $V_0 = 0.10$ can be obtained from Fig. \ref{fig_pointcant3_prog}. In the first cycle, the design distribution and shape function orders are uniform for the whole mesh. Similar to the case of $V_0 = 0.45$, it is observed that QR-patterns are formed here as well, which are removed by refinement in later cycles. Compared to Fig. \ref{fig_pointcant1_prog}, it is observed that only a small part of the domain gets refined. Because of the low volume fraction of material used, a significant part of the domain comprises mainly of void regions, which do not require refinement. For the non-adaptive as well as the $dp$-adaptive versions of MTO, it is observed that the convergence of the optimization problem slows down significantly when very low material volume fractions are used. For example, for the same error tolerance, the number of iterations required in the final cycle of $dp$-adaptive method for $V_0 = $ 0.45 and 0.10 are 18 and 82, respectively. Our observations on the effect of material volume fraction on the convergence of TO process align with the results reported in \cite{Rojas2015}, where similar results have been obtained over a set of numerical experiments. 

\subsubsection{Compliance minimization for distributed load}
%
\begin{figure}
\centering
\begin{subfigure}{0.48\linewidth}
\centering
\includegraphics[scale=0.22]{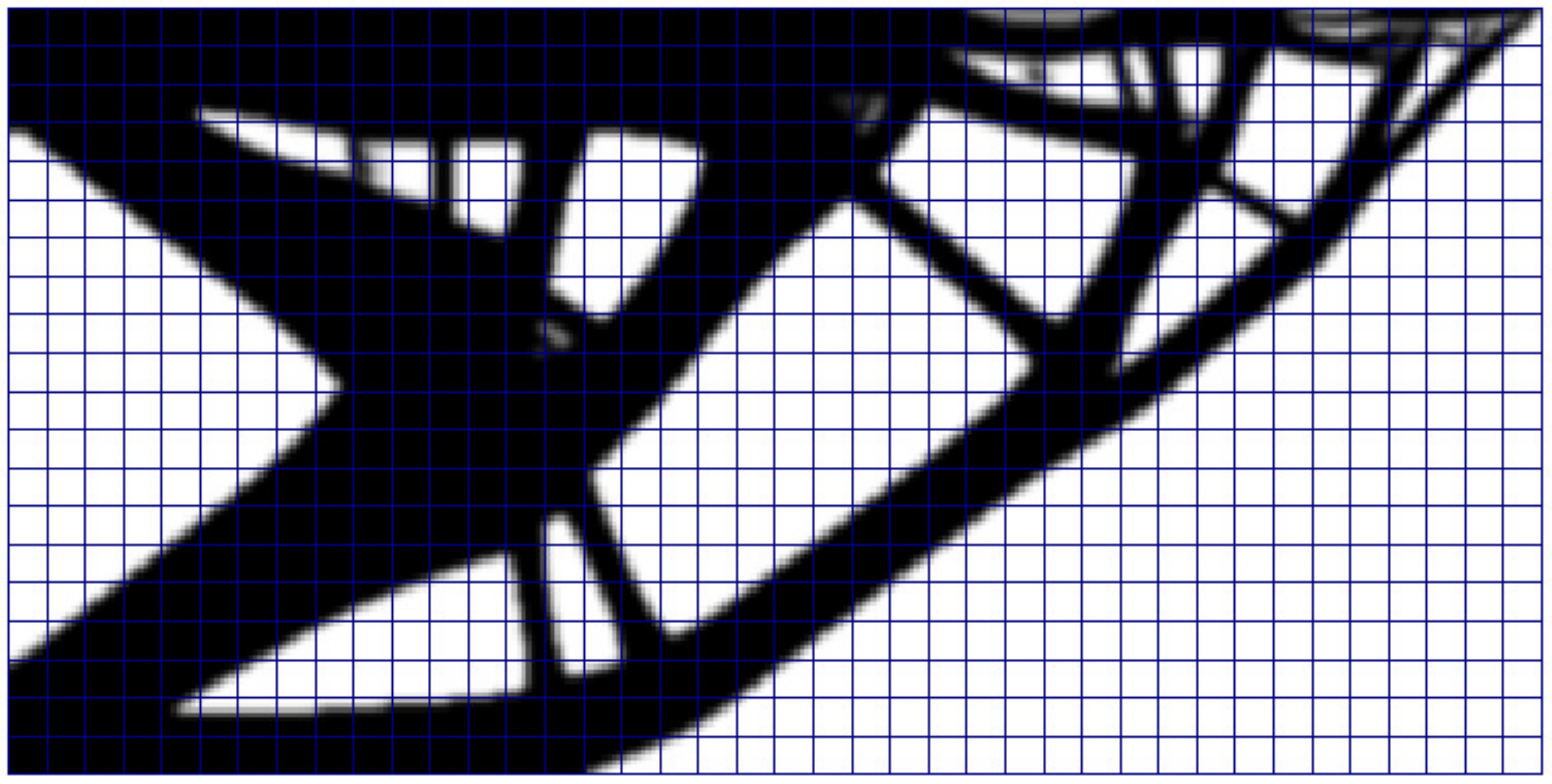}
\caption{MTO ($\mathcal{J}_0 = 12.97J$)}
\label{fig_cant_dist2a}
\end{subfigure}
\begin{subfigure}{0.48\linewidth}
\centering
\includegraphics[scale=0.22]{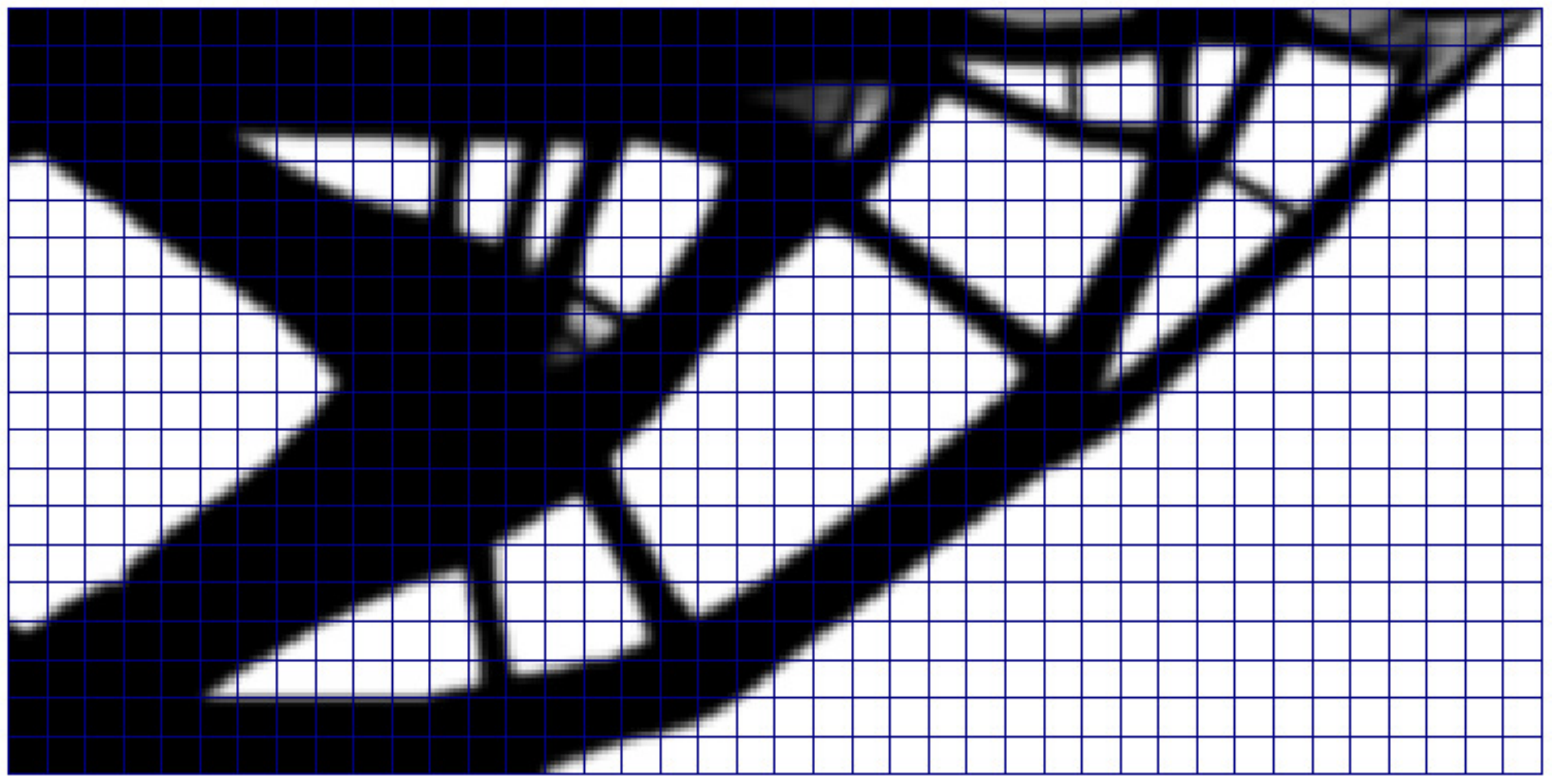}
\caption{$dp$-adaptive MTO ($\mathcal{J} = 12.66J$)}
\label{fig_cant_dist2b}
\end{subfigure}
\caption{Optimized cantilever designs for the distributed-load case shown in Fig.\ref{fig_cant_dist}, obtained using a uniform MTO mesh (left) and $dp$-adaptive MTO approach (right). A 4.6-fold speed-up is obtained using $dp$-adaptivity.}
\label{fig_cant_dist2}
\end{figure}
\indent \vspace{0.5em} \\ For the cantilever beam subjected to a distributed load (Fig. \ref{fig_cant_dist}), $V_0$ is set to 0.45. Fig. \ref{fig_cant_dist2} shows the optimized designs obtained using a uniform MTO mesh (Fig. \ref{fig_cant_dist2a}) and the $dp$-adaptive approach (Fig. \ref{fig_cant_dist2b}). The information on the two runs is listed in Table \ref{table_mto_runs}. As in the case of the point load cantilever, the designs obtained using the non-adaptive and adaptive variants of MTO are very similar. In terms of performance, a speed-up of 4.6 times is observed, and the accuracy of the obtained solution is close to 1. The obtained $\mathcal{J}/\mathcal{J}_0$ value is 0.98, which implies that the $dp$-adaptive MTO found a slightly stiffer design.

For both the designs, there exists a small region near the top right boundary which comprises intermediate densities and is not improved even with refinement. With $dp$-adaptive MTO, this region is more prominent. Among the possible reasons, one explanation could be that the distributed load applied on the upper boundary of the domain requires support material in those parts. In the absence of material near the upper boundary, the load point can get disconnected, which leads to a high overall compliance value for the structure. We observe that the optimizer is not inclined towards adding much solid material in these parts of the domain. Due to this, gray regions are formed, representing fine structural features beyond the design resolution. These intermediate densities can be suppressed by the use of methods such as modified Heaviside projection as has been demonstrated in \cite{Groen2016}, or simply by adding a solid non-design region at the top surface. 

\begin{figure}
\centering
\begin{subfigure}{0.48\linewidth}
\centering
\includegraphics[scale=0.22]{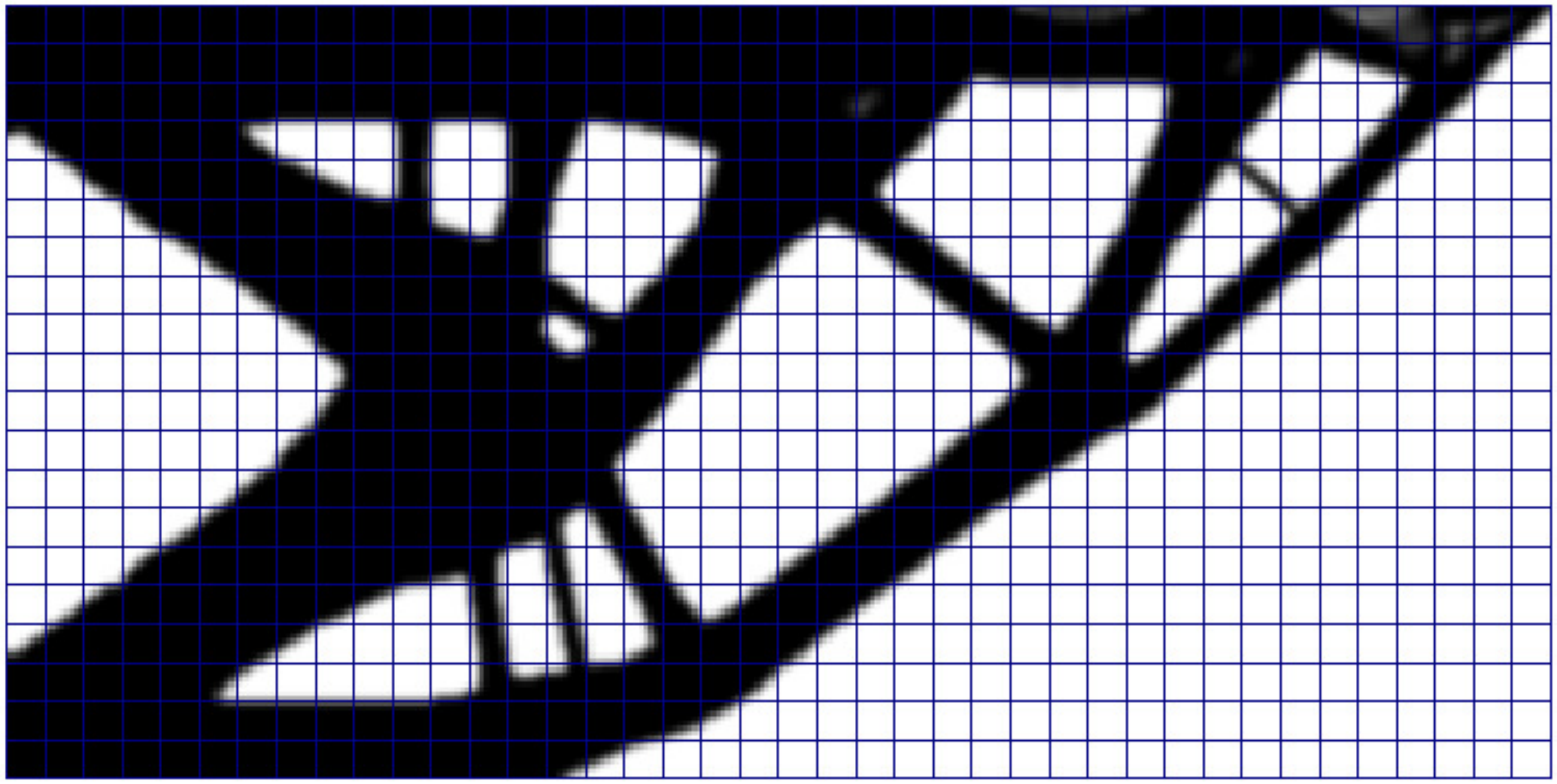}
\caption{$dp$-adaptive MTO ($q = $3, 4, 5 and 6)}
\label{fig_cant_dist2a_adap_p}
\end{subfigure}
\begin{subfigure}{0.48\linewidth}
\centering
\includegraphics[scale=0.22]{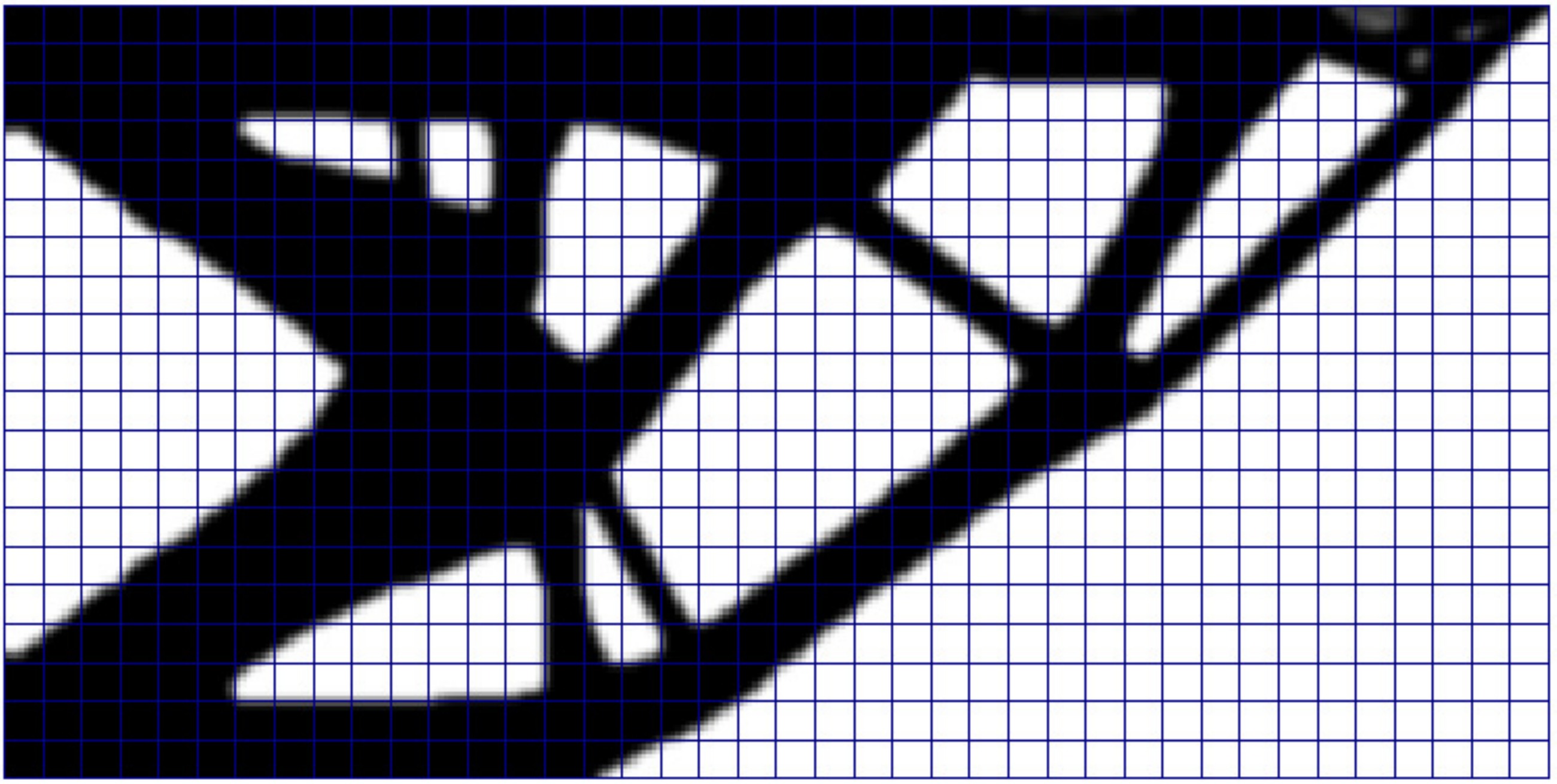}
\caption{$dp$-adaptive MTO ($q = $3, 5, 7 and 9)}
\label{fig_cant_dist2b_adap_p}
\end{subfigure}
\caption{Optimized cantilever designs for the distributed-load case shown in Fig.\ref{fig_cant_dist}, obtained using $dp$-adaptive MTO approach. For both the cases, adaptive penalization has been used. For the 4 cycles of the $dp$-adaptive MTO run, the values of $q$ used have been reported in the sub-captions. }
\label{fig_cant_dist2_adap_p}
\end{figure}

Using a stronger penalization on the intermediate densities at the later cycles of MTO has also been found to help in reducing the gray areas. Fig. \ref{fig_cant_dist2_adap_p} shows two optimized designs for this cantilever problem obtained using adaptive penalization schemes. For the first case (Fig. \ref{fig_cant_dist2a_adap_p}), the initial value of $q$ is 3 and it is increased by 1 at every cycle. For the second case (Fig. \ref{fig_cant_dist2b_adap_p}), the increment is by 2 at every cycle. It is observed that with stronger penalization on the intermediate densities, the gray regions are significantly reduced.

\begin{figure*}
\centering
\begin{subfigure}{1\textwidth}
\centering
\begin{tikzpicture}
    \node[anchor=south west,inner sep=0] at (0,0){
\includegraphics[scale=0.14]{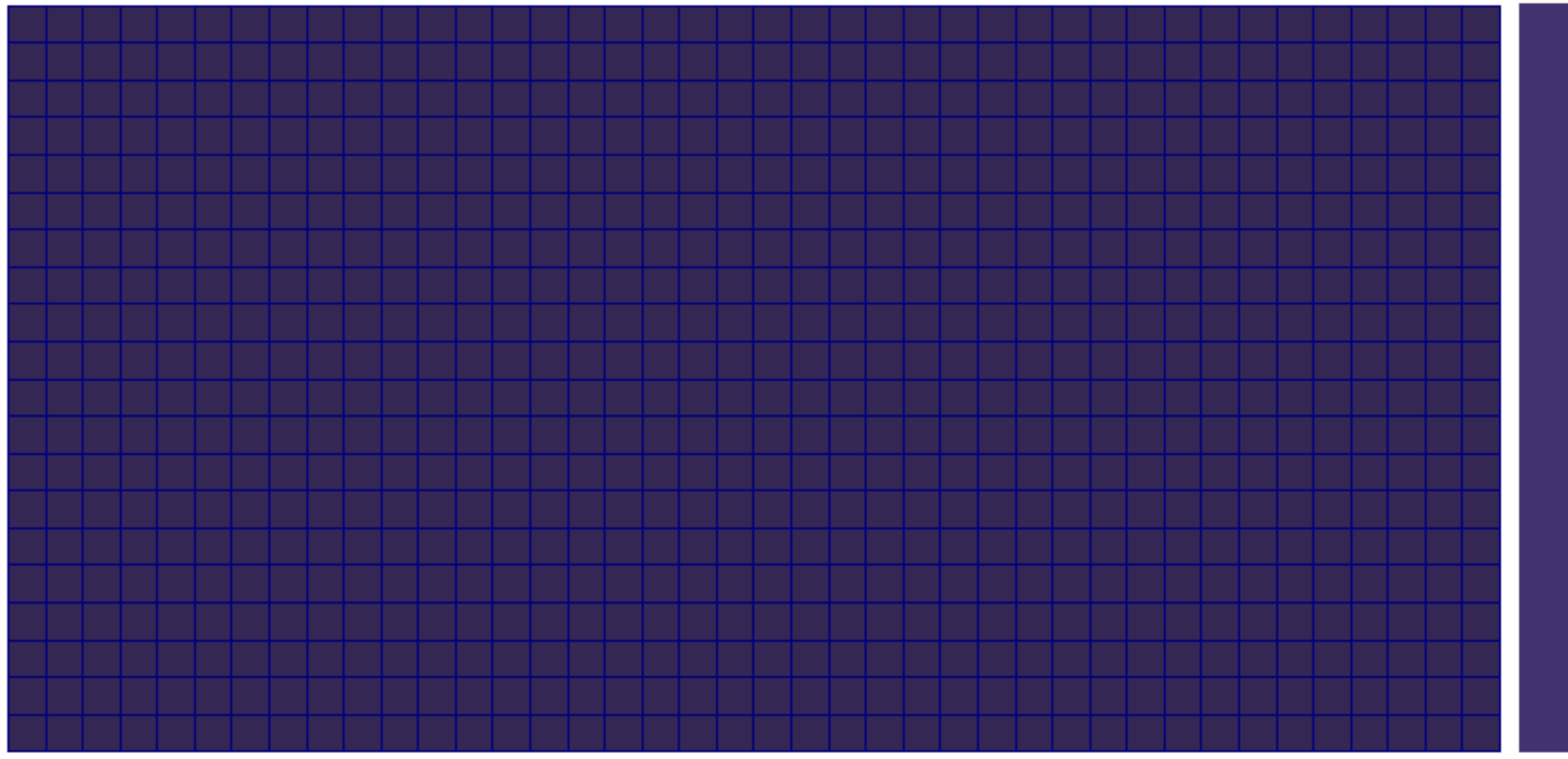}};
\node[text width=0cm] at (3.9, 1.77) 
    {\tiny \color{darkgray}{16}};
    \node[text width=0cm] at (3.9, 1.35) 
    {\tiny \color{darkgray}{}};
    \node[text width=0cm] at (3.9, 0.92) 
    {\tiny \color{darkgray}{16}};
    \node[text width=0cm] at (3.9, 0.49) 
    {\tiny \color{darkgray}{}};
\node[text width=0cm] at (3.9, 0.1) 
    {\tiny \color{darkgray}{16}};
    \node[text width=2cm] at (2.2, 2.3) 
    {\normalsize \color{darkgray}{Design field}};
\end{tikzpicture}\hspace{0.5em}
\begin{tikzpicture}
    \node[anchor=south west,inner sep=0] at (0,0){
\includegraphics[scale=0.14]{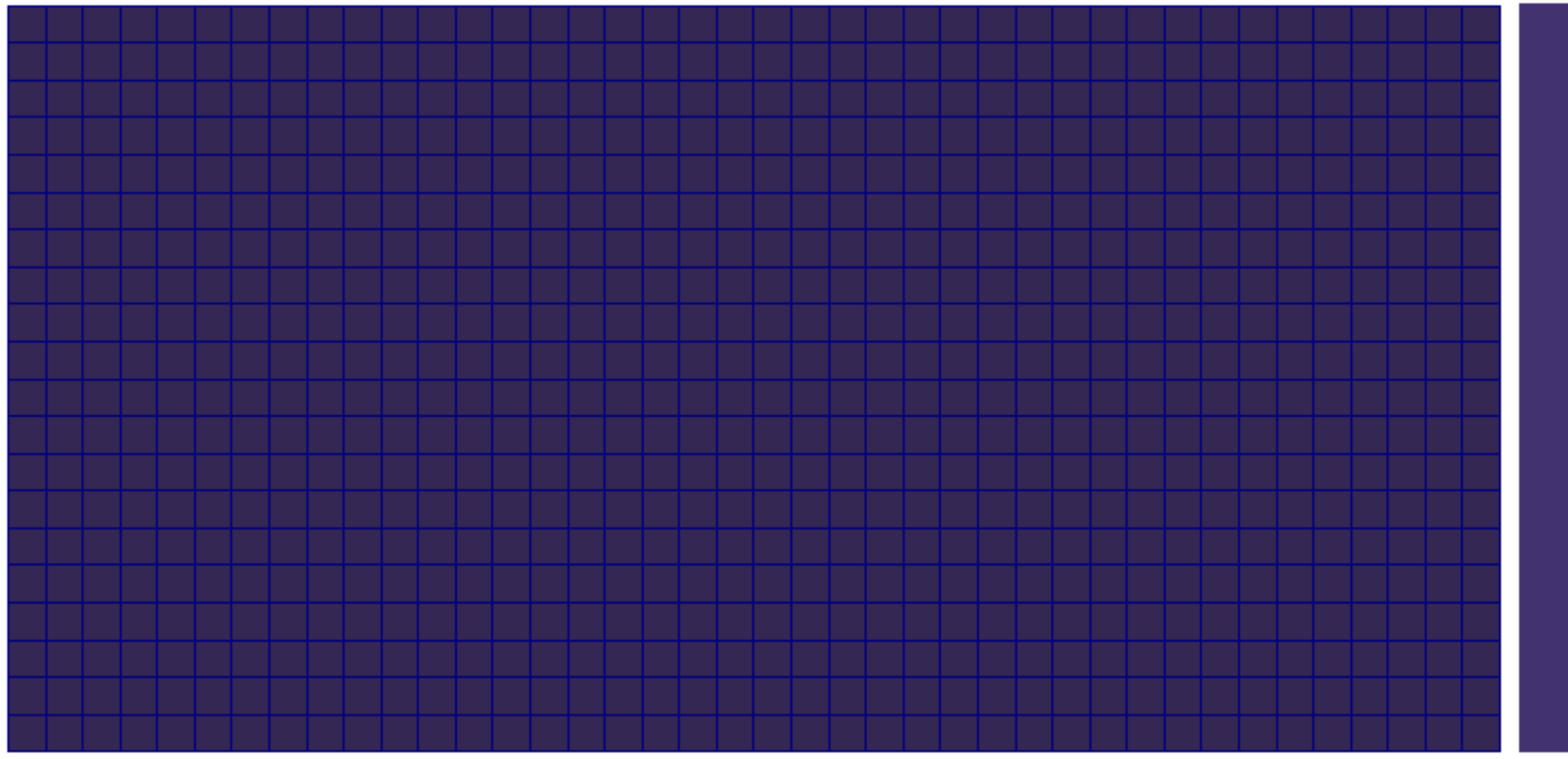}};
\node[text width=0cm] at (3.9, 1.77) 
    {\tiny \color{darkgray}{2}};
    \node[text width=0cm] at (3.9, 1.35) 
    {\tiny \color{darkgray}{}};
    \node[text width=0cm] at (3.9, 0.92) 
    {\tiny \color{darkgray}{2}};
    \node[text width=0cm] at (3.9, 0.49) 
    {\tiny \color{darkgray}{}};
\node[text width=0cm] at (3.9, 0.1) 
    {\tiny \color{darkgray}{2}};
    \node[text width=3.7cm] at (2.23, 2.3) 
    {\normalsize \color{darkgray}{Shape function order}};
\end{tikzpicture}\hspace{0.5em}
\begin{tikzpicture}
    \node[anchor=south west,inner sep=0] at (0,0){
\includegraphics[scale=0.15]{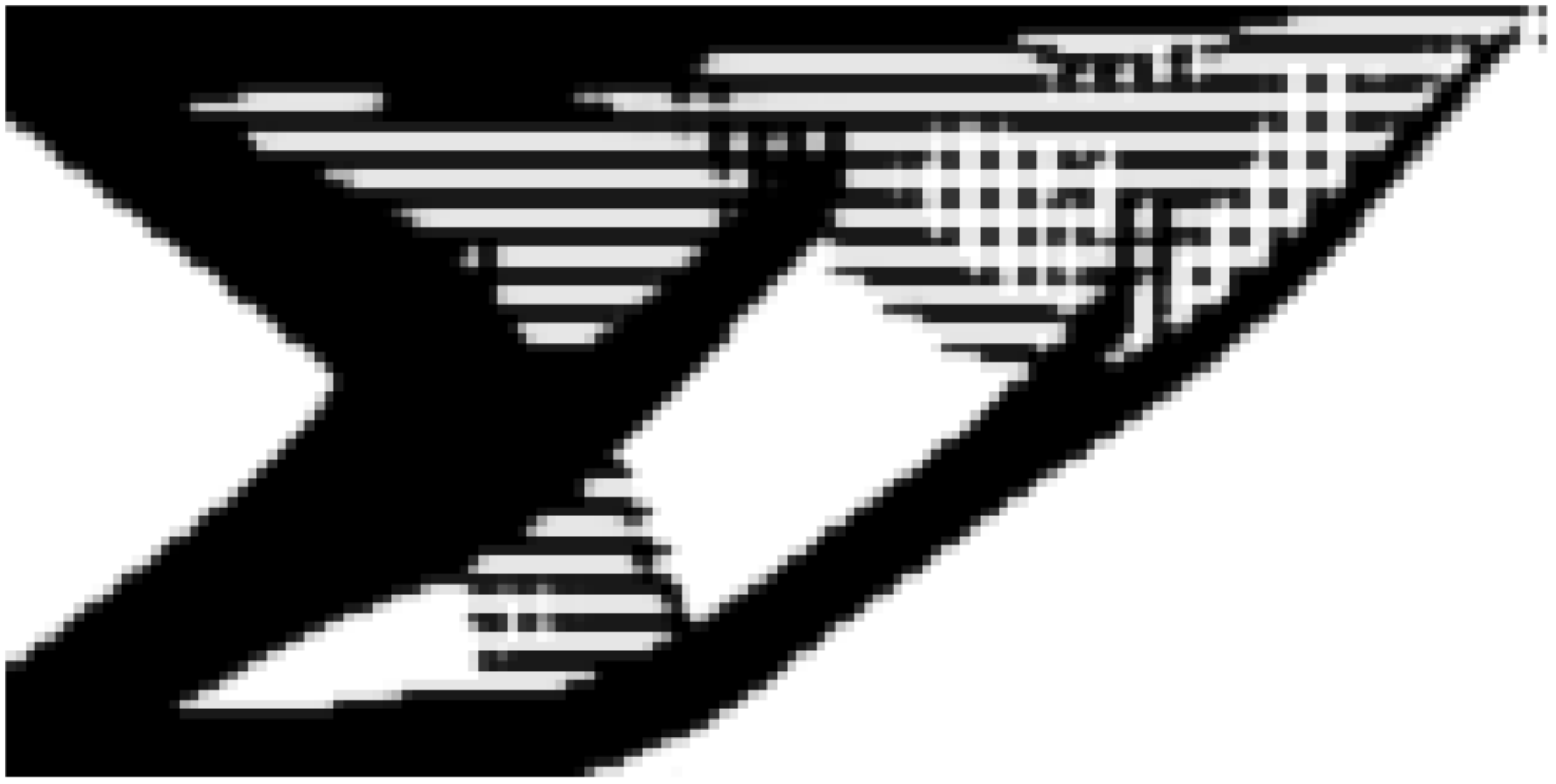}};
	\node[text width=3cm] at (2.23, 2.42) 
    {\normalsize \color{darkgray}{Optimized design}};
\end{tikzpicture}\hspace{0.5em}
\caption{Cycle: 1}
\end{subfigure}
\begin{subfigure}{1\textwidth}
\centering
\begin{tikzpicture}
    \node[anchor=south west,inner sep=0] at (0,0){
\includegraphics[scale=0.14]{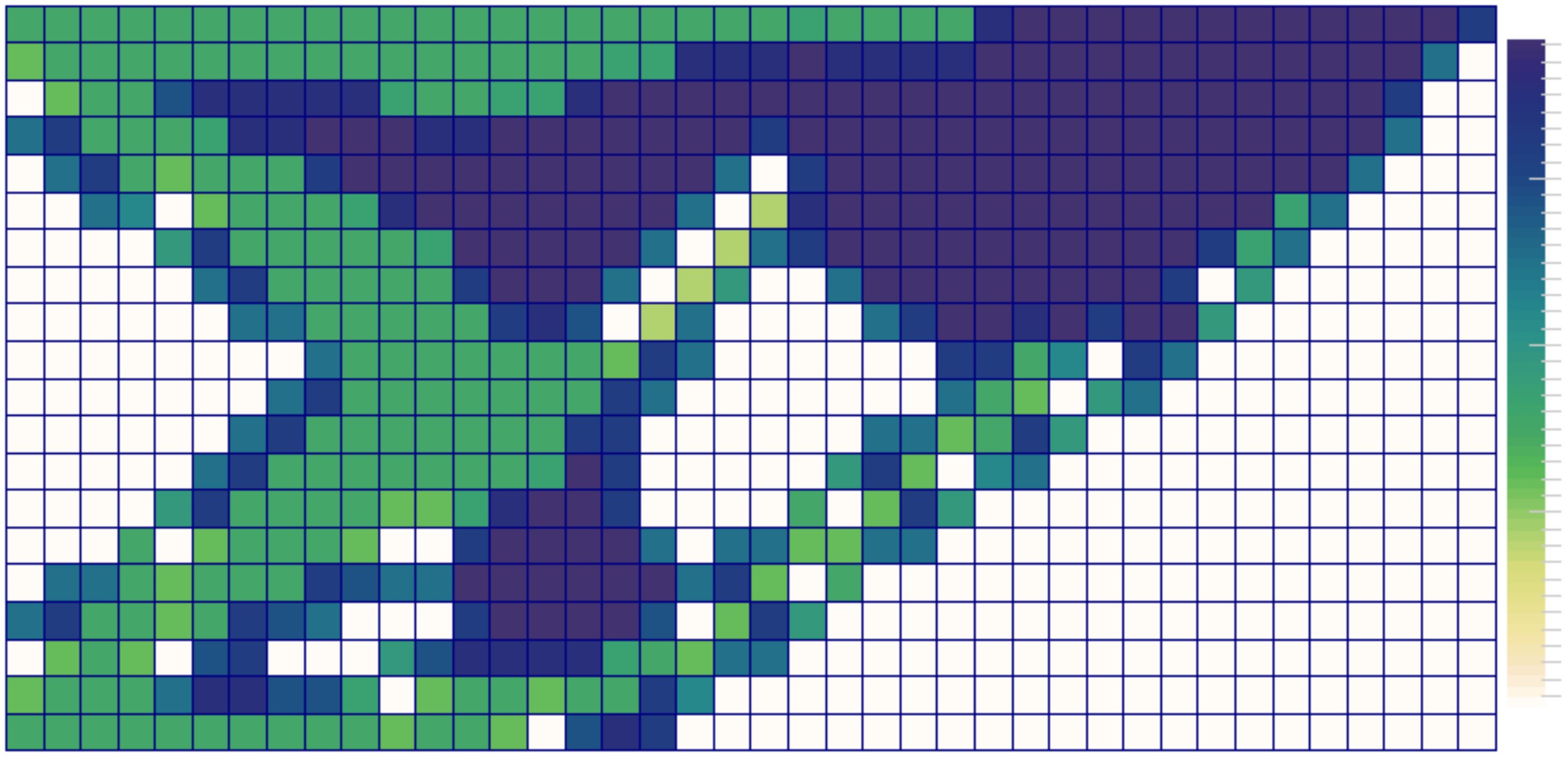}};
\node[text width=0cm] at (3.9, 1.77) 
    {\tiny \color{darkgray}{29}};
    \node[text width=0cm] at (3.9, 1.35) 
    {\tiny \color{darkgray}{24}};
    \node[text width=0cm] at (3.9, 0.92) 
    {\tiny \color{darkgray}{18}};
    \node[text width=0cm] at (3.9, 0.47) 
    {\tiny \color{darkgray}{12}};
\node[text width=0cm] at (3.9, 0.1) 
    {\tiny \color{darkgray}{5}};
\end{tikzpicture}\hspace{0.5em}
\begin{tikzpicture}
    \node[anchor=south west,inner sep=0] at (0,0){
\includegraphics[scale=0.14]{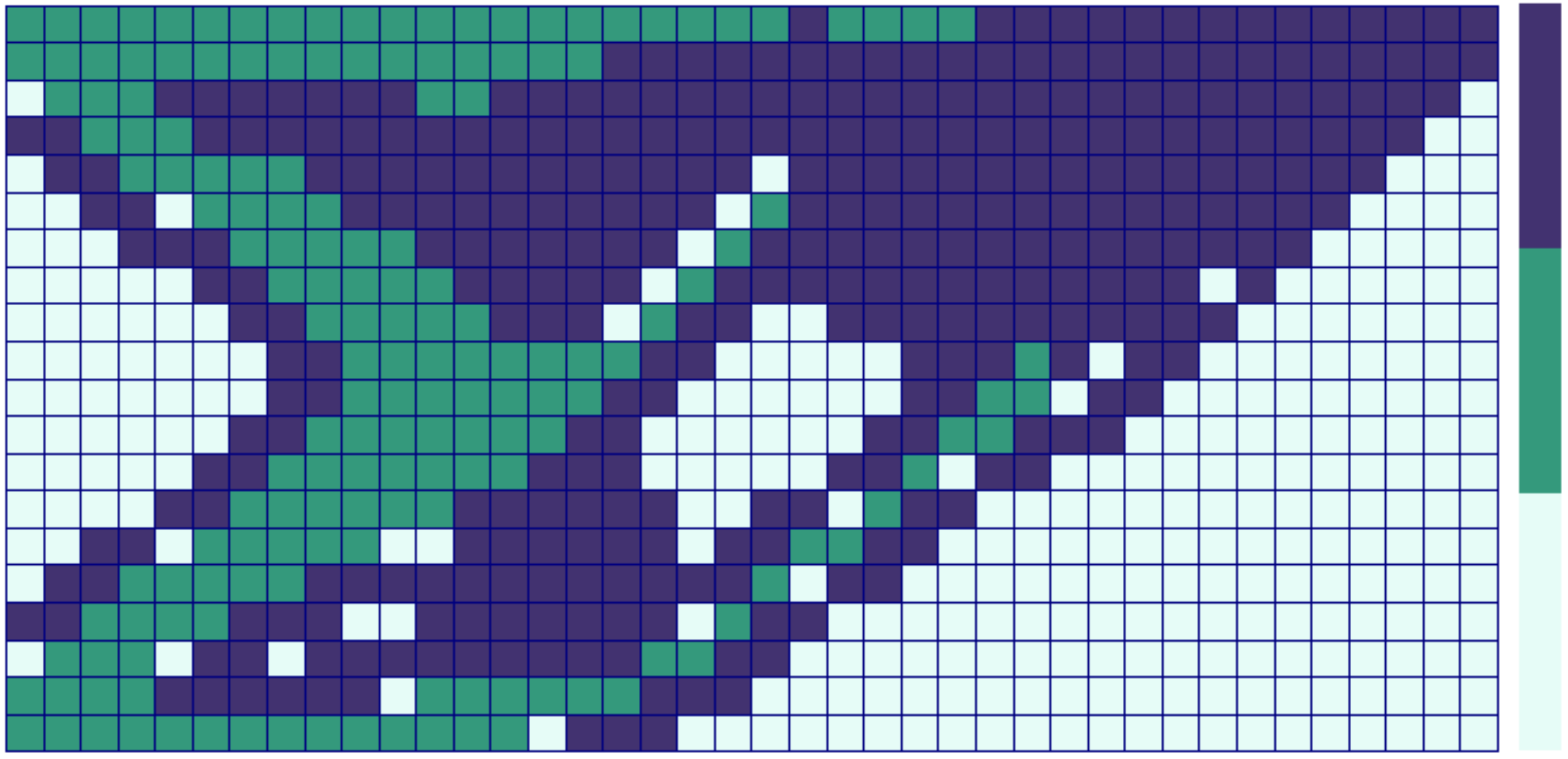}};
\node[text width=0cm] at (3.9, 1.77) 
    {\tiny \color{darkgray}{3}};
    \node[text width=0cm] at (3.9, 1.35) 
    {\tiny \color{darkgray}{}};
    \node[text width=0cm] at (3.9, 0.92) 
    {\tiny \color{darkgray}{2}};
    \node[text width=0cm] at (3.9, 0.47) 
    {\tiny \color{darkgray}{}};
\node[text width=0cm] at (3.9, 0.1) 
    {\tiny \color{darkgray}{1}};
\end{tikzpicture}\hspace{0.5em}
\begin{tikzpicture}
    \node[anchor=south west,inner sep=0] at (0,0){
\includegraphics[scale=0.15]{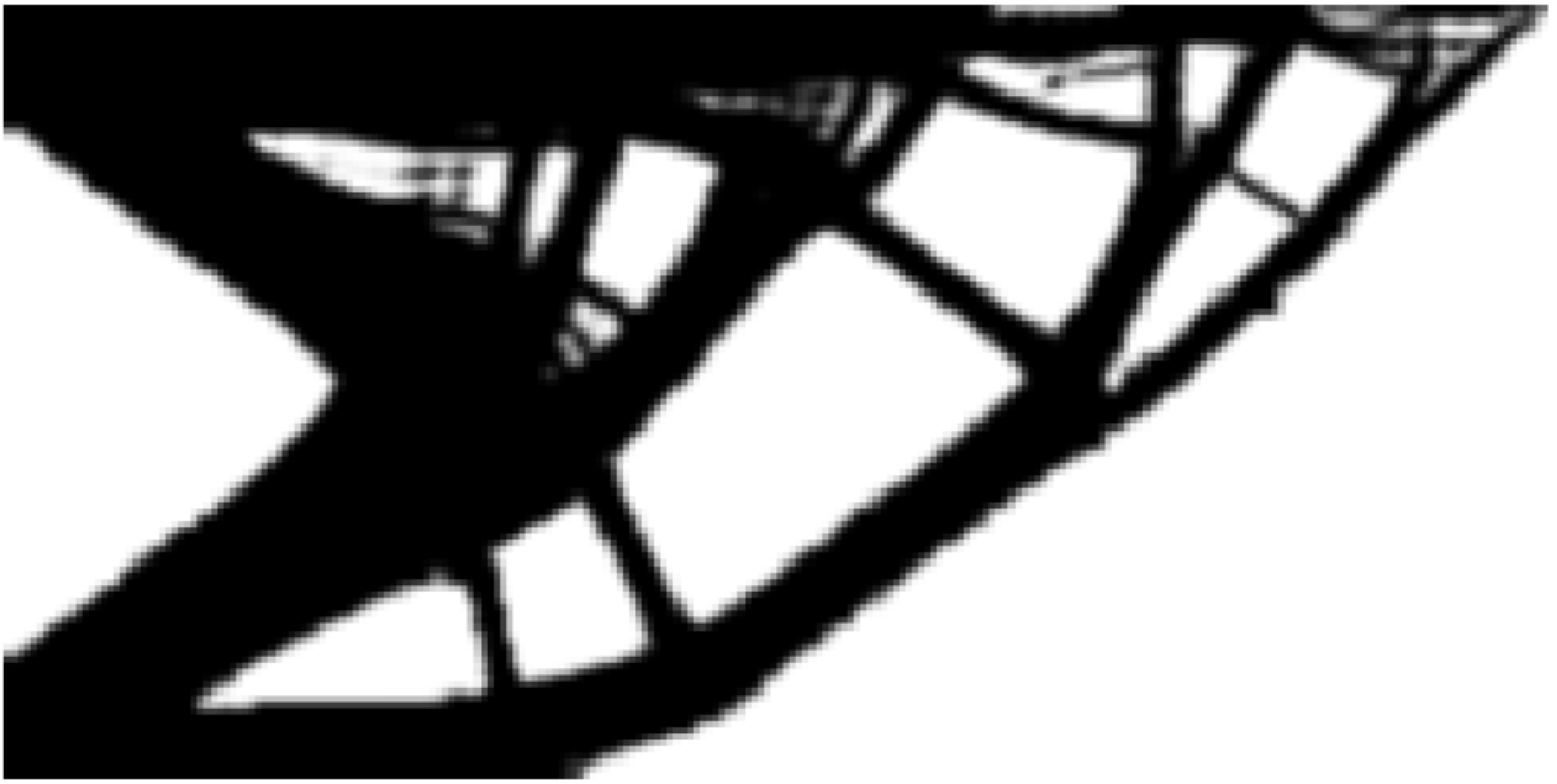}};
\end{tikzpicture}\hspace{0.5em}
\caption{Cycle: 2}
\end{subfigure}
\begin{subfigure}{1\textwidth}
\centering
\begin{tikzpicture}
    \node[anchor=south west,inner sep=0] at (0,0){
\includegraphics[scale=0.14]{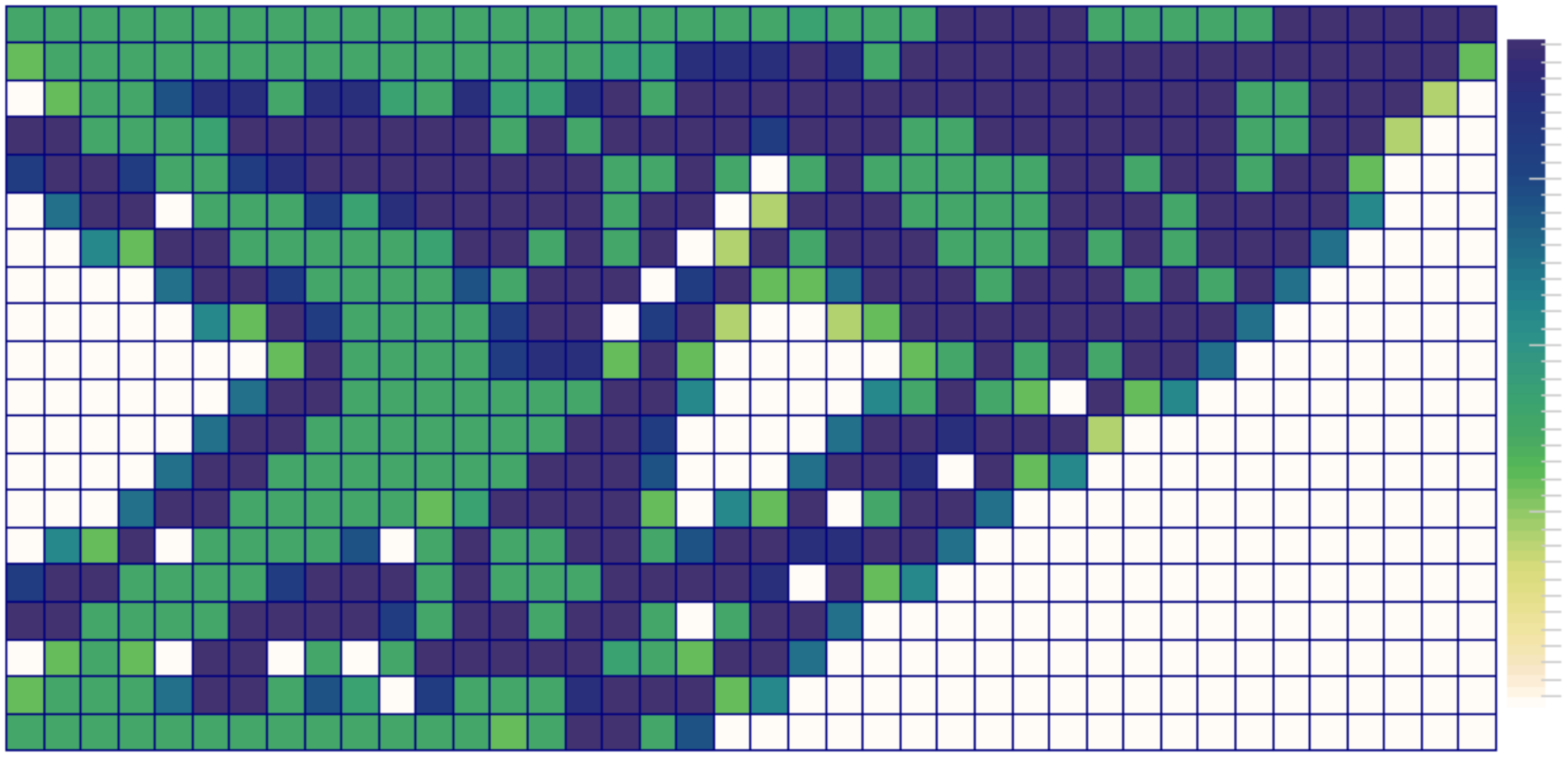}};
\node[text width=0cm] at (3.9, 1.77) 
    {\tiny \color{darkgray}{47}};
    \node[text width=0cm] at (3.9, 1.35) 
    {\tiny \color{darkgray}{34}};
    \node[text width=0cm] at (3.9, 0.92) 
    {\tiny \color{darkgray}{23}};
    \node[text width=0cm] at (3.9, 0.49) 
    {\tiny \color{darkgray}{11}};
\node[text width=0cm] at (3.9, 0.1) 
    {\tiny \color{darkgray}{1}};
\end{tikzpicture}\hspace{0.5em}
\begin{tikzpicture}
    \node[anchor=south west,inner sep=0] at (0,0){
\includegraphics[scale=0.14]{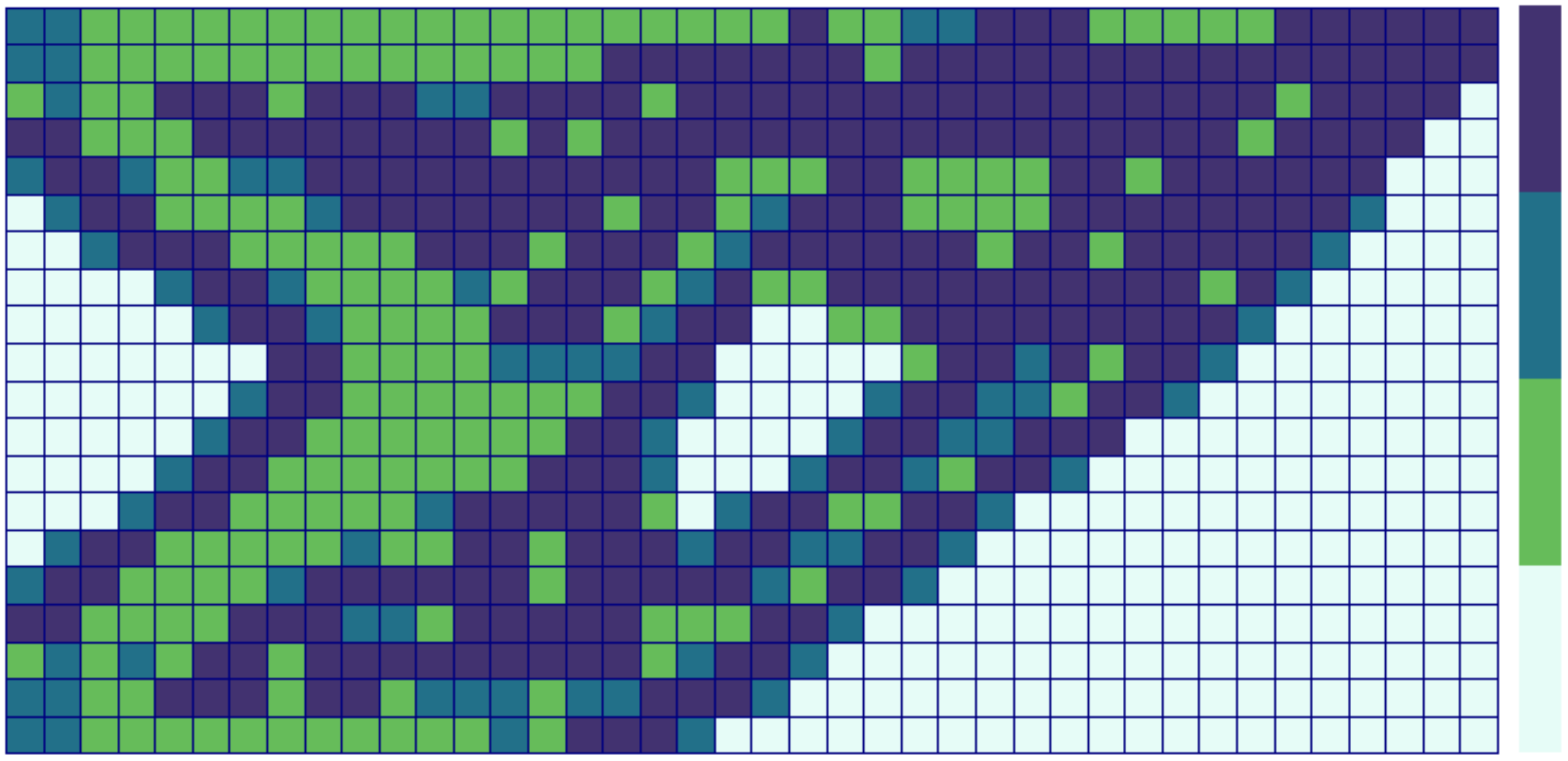}};
\node[text width=0cm] at (3.9, 1.77) 
    {\tiny \color{darkgray}{4}};
    \node[text width=0cm] at (3.9, 1.25) 
    {\tiny \color{darkgray}{3}};
    \node[text width=0cm] at (3.9, 0.72) 
    {\tiny \color{darkgray}{2}};
\node[text width=0cm] at (3.9, 0.15) 
    {\tiny \color{darkgray}{1}};
\end{tikzpicture}\hspace{0.5em}
\begin{tikzpicture}
    \node[anchor=south west,inner sep=0] at (0,0){
\includegraphics[scale=0.15]{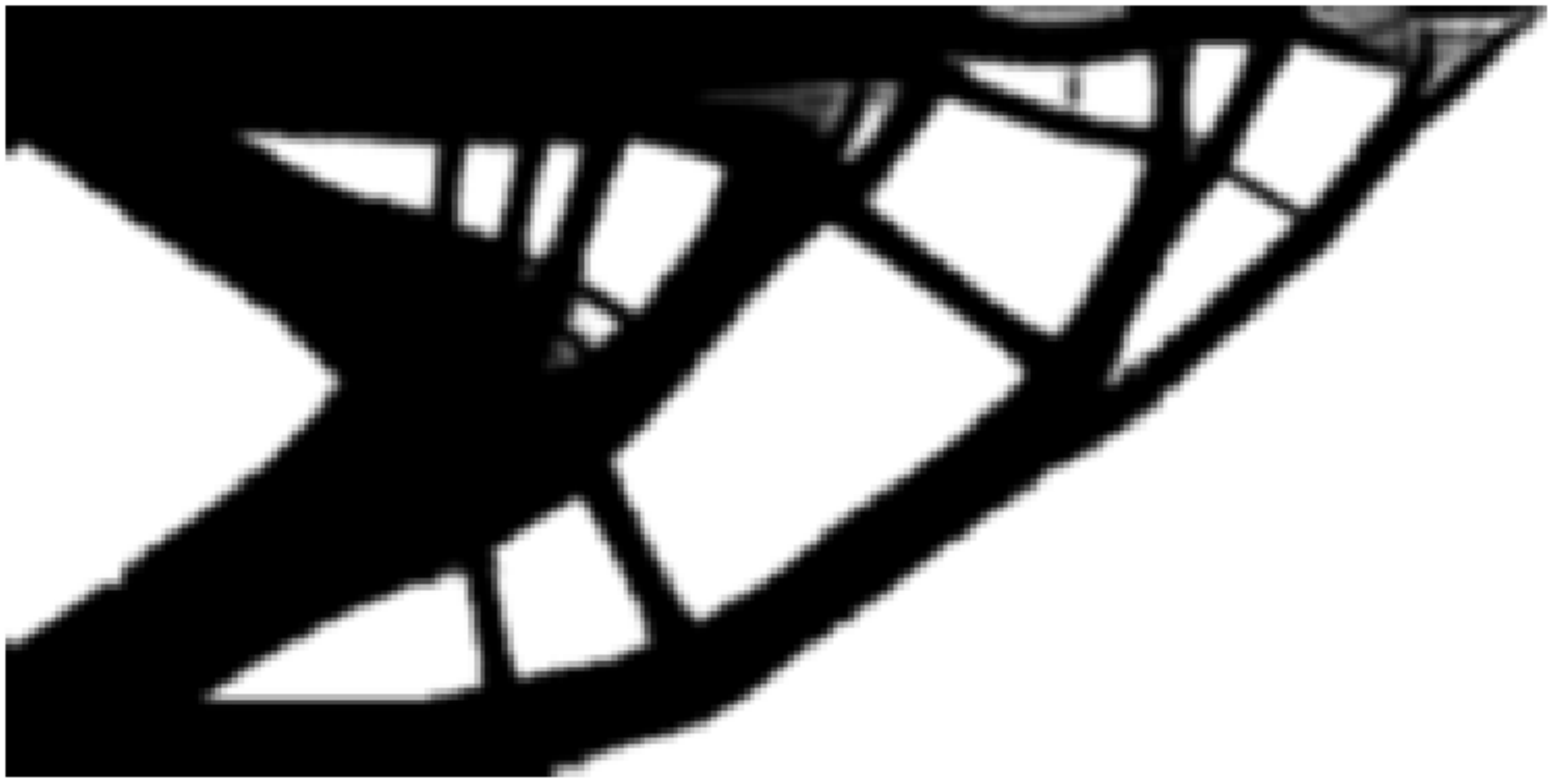}};
\end{tikzpicture}\hspace{0.5em}
\caption{Cycle: 3}
\end{subfigure}
\begin{subfigure}{1\textwidth}
\centering
\begin{tikzpicture}
    \node[anchor=south west,inner sep=0] at (0,0){
\includegraphics[scale=0.14]{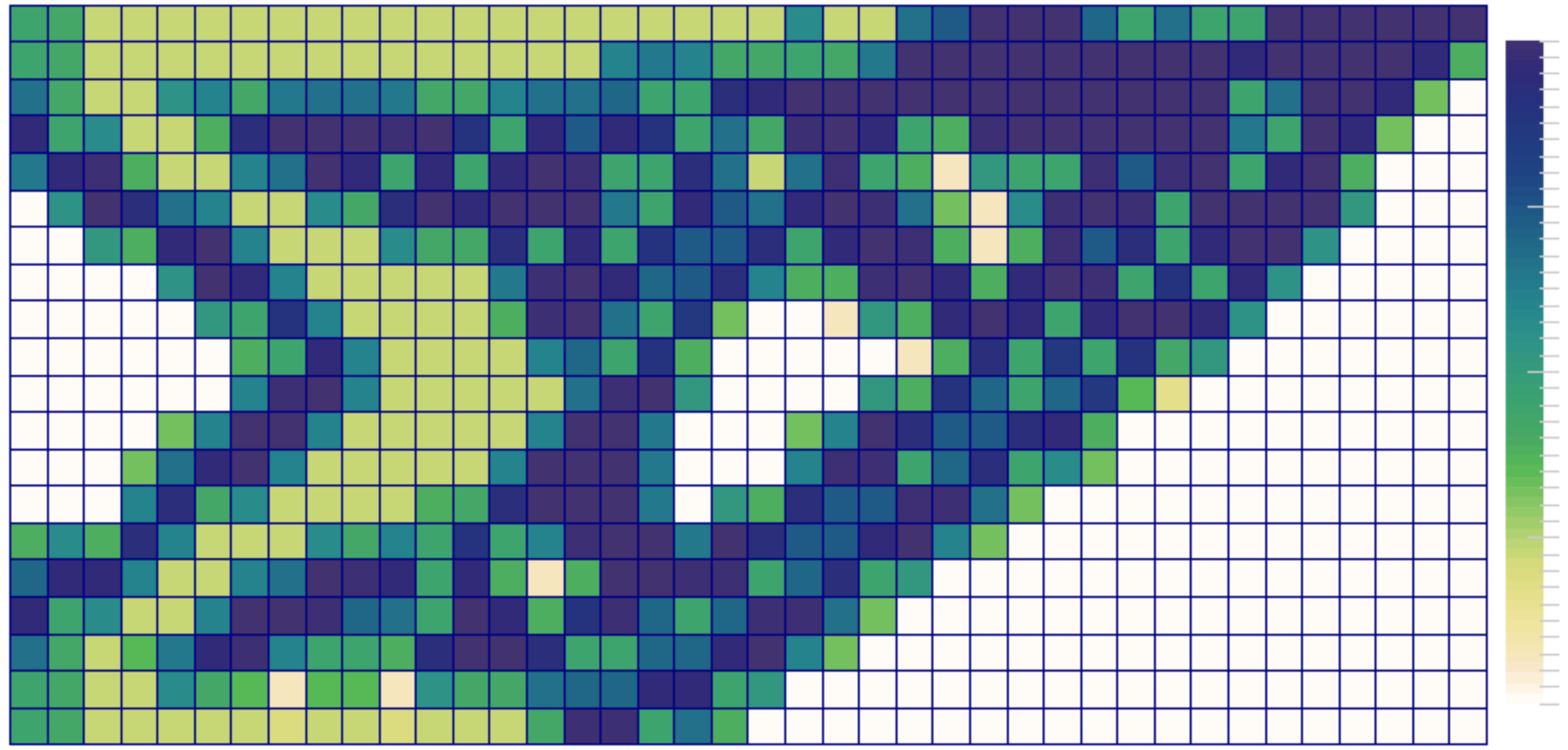}};
\node[text width=0cm] at (3.9, 1.77) 
    {\tiny \color{darkgray}{64}};
    \node[text width=0cm] at (3.9, 1.35) 
    {\tiny \color{darkgray}{48}};
    \node[text width=0cm] at (3.9, 0.92) 
    {\tiny \color{darkgray}{32}};
    \node[text width=0cm] at (3.9, 0.47) 
    {\tiny \color{darkgray}{16}};
\node[text width=0cm] at (3.9, 0.1) 
    {\tiny \color{darkgray}{1}};
\end{tikzpicture}\hspace{0.5em}
\begin{tikzpicture}
    \node[anchor=south west,inner sep=0] at (0,0){
\includegraphics[scale=0.14]{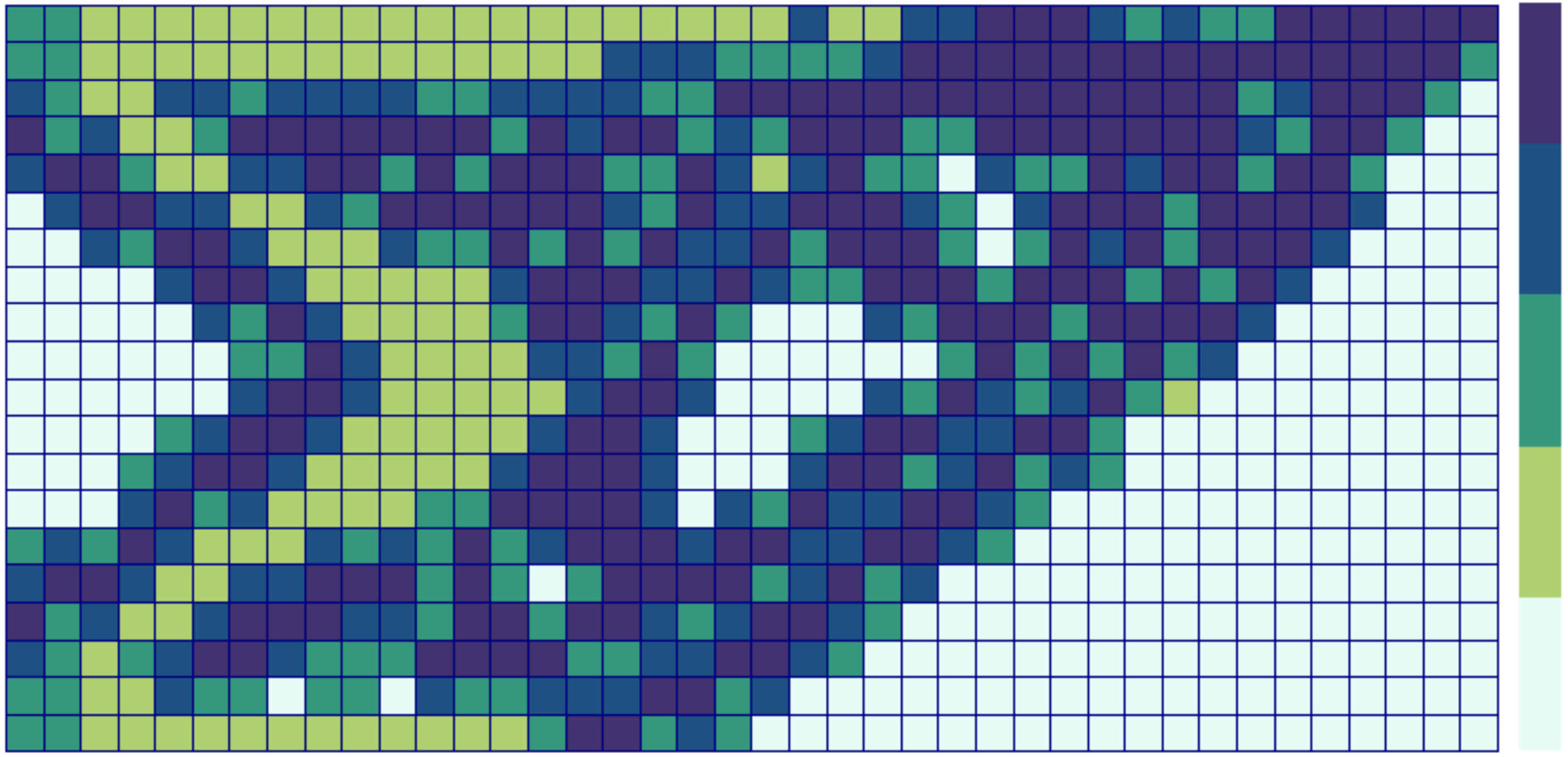}};
\node[text width=0cm] at (3.9, 1.77) 
    {\tiny \color{darkgray}{5}};
    \node[text width=0cm] at (3.9, 1.35) 
    {\tiny \color{darkgray}{4}};
    \node[text width=0cm] at (3.9, 0.92) 
    {\tiny \color{darkgray}{3}};
    \node[text width=0cm] at (3.9, 0.47) 
    {\tiny \color{darkgray}{2}};
\node[text width=0cm] at (3.9, 0.1) 
    {\tiny \color{darkgray}{1}};
\end{tikzpicture}\hspace{0.5em}
\begin{tikzpicture}
    \node[anchor=south west,inner sep=0] at (0,0){
\includegraphics[scale=0.15]{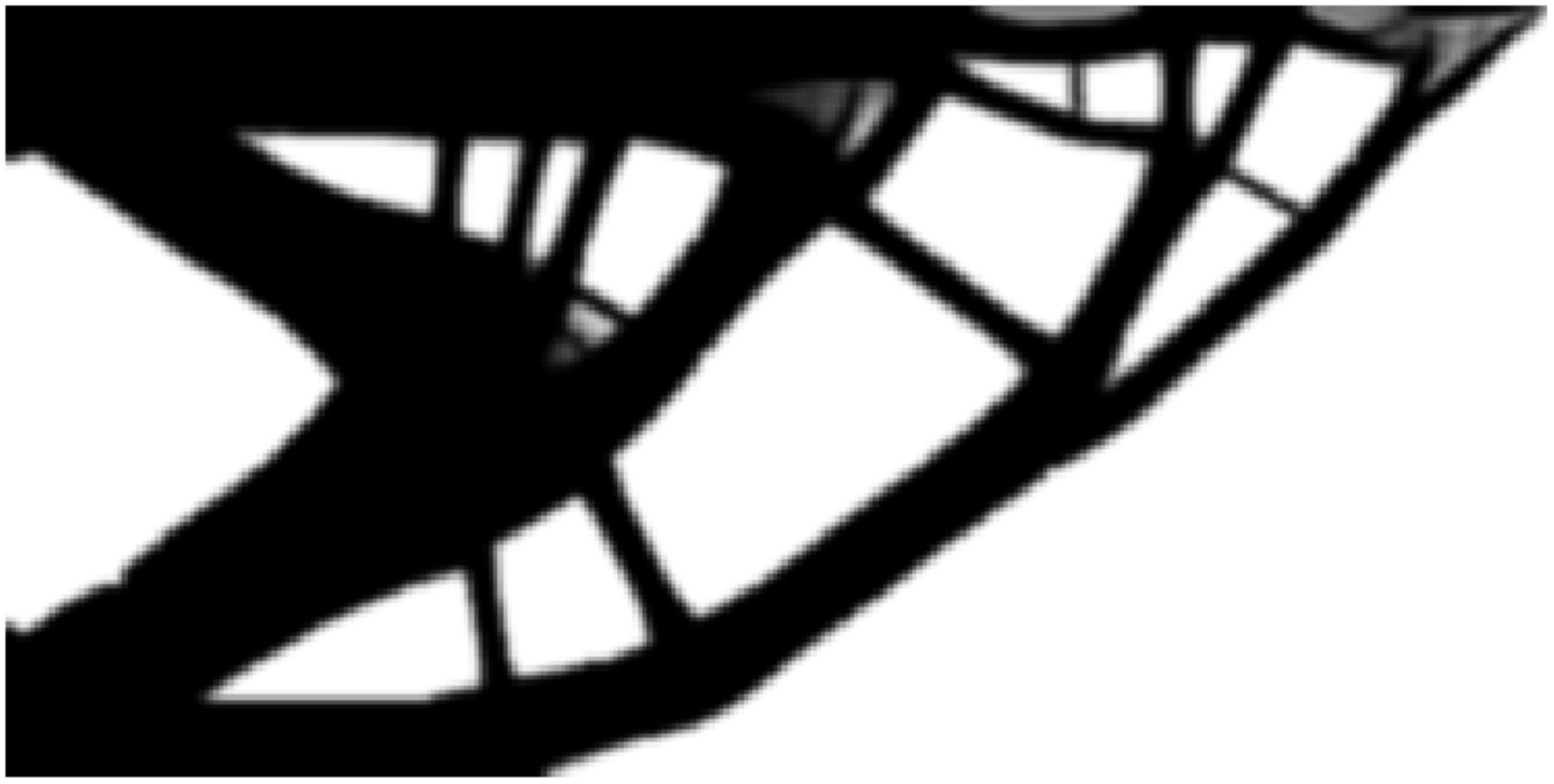}};
\end{tikzpicture}\hspace{0.5em}
\caption{Cycle: 4}
\end{subfigure}
\caption{Optimized designs (right), and the respective shape function orders (middle) and design field (left) obtained for 4 cycles of $dp$-adaptive MTO run for a cantilever beam subjected to distributed load (Fig. \ref{fig_cant_dist}). The domain has been discretized using  $40 \times 20$ quadrilateral finite elements ($r = 0.3h$). The initial mesh  is uniform and each element comprises shape functions of polynomial order 2 and 16 design points per element. The maximum allowed shape function order and number of design points are restricted to 5 and 64 per element, respectively. }
\label{fig_cant_dist_prog}
\end{figure*}

\indent To obtain an understanding on how the design evolves over 4 cycles of $dp$-adaptive refinement, see Fig. \ref{fig_cant_dist_prog}. Due to the low order of the shape function used in Cycle 1, QR-patterns are observed here. Similar to the previous cases, adaptive refinement in the affected regions helps to remove these artefacts. For Cycle 4, only 16 iterations are needed when using the $dp$-adaptive method, while the conventional MTO method uses 54 iterations in total. Also, the number of design points and DOFs used in the last cycle of the $dp$-adaptive MTO are lower than in the conventional MTO method. Together, these two factors make the $dp$-adaptive MTO method 4.6 times faster in this case.

\subsubsection{Force inverter compliant mechanism}
%
\begin{figure}
\centering
\begin{subfigure}{0.48\linewidth}
\centering
\includegraphics[scale=0.22]{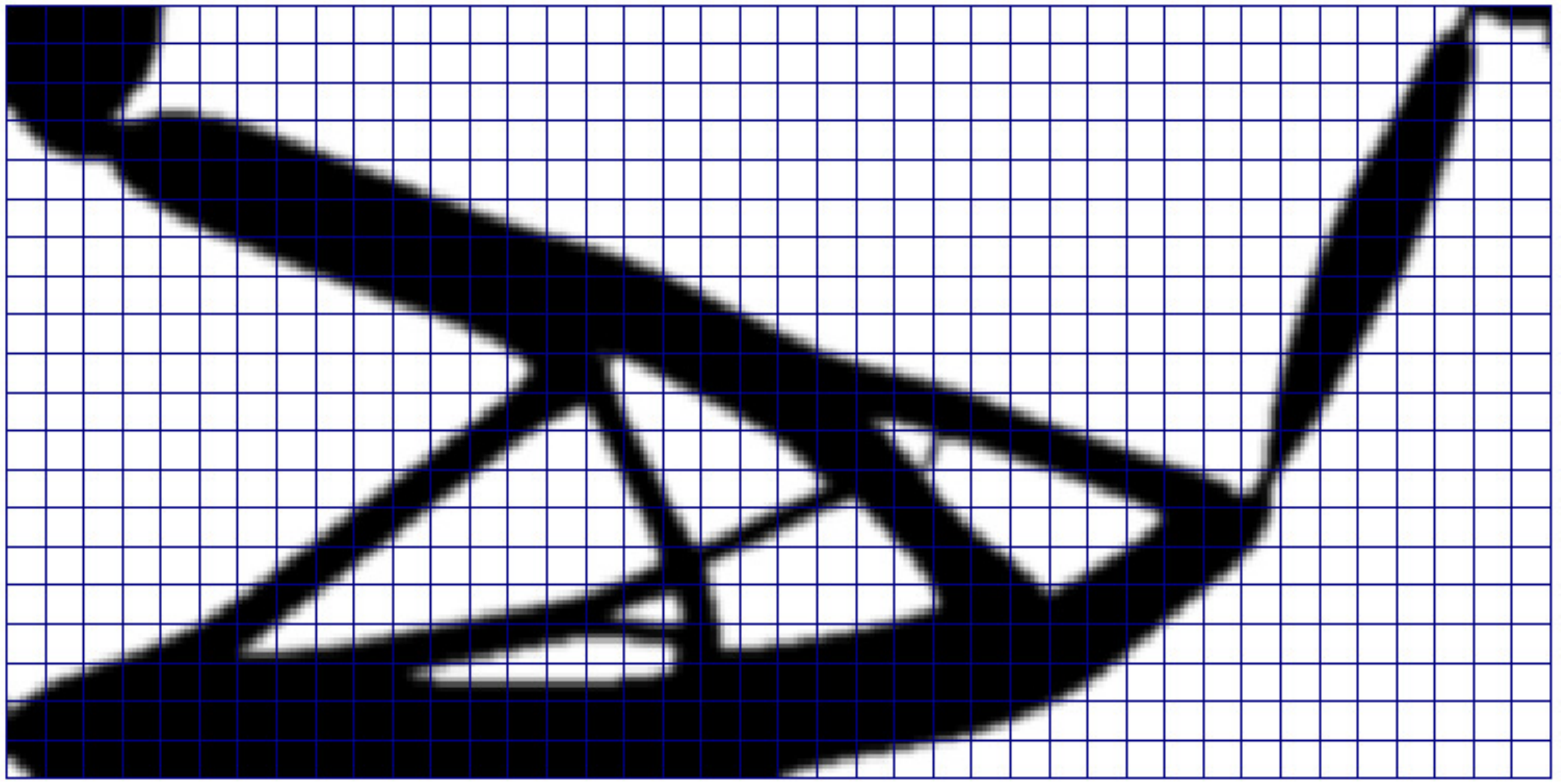}
\caption{MTO ($\mathcal{J}_0 = 2.224$m)}
\label{fig_comp_inv_2a}
\end{subfigure}
\begin{subfigure}{0.48\linewidth}
\centering
\includegraphics[scale=0.22]{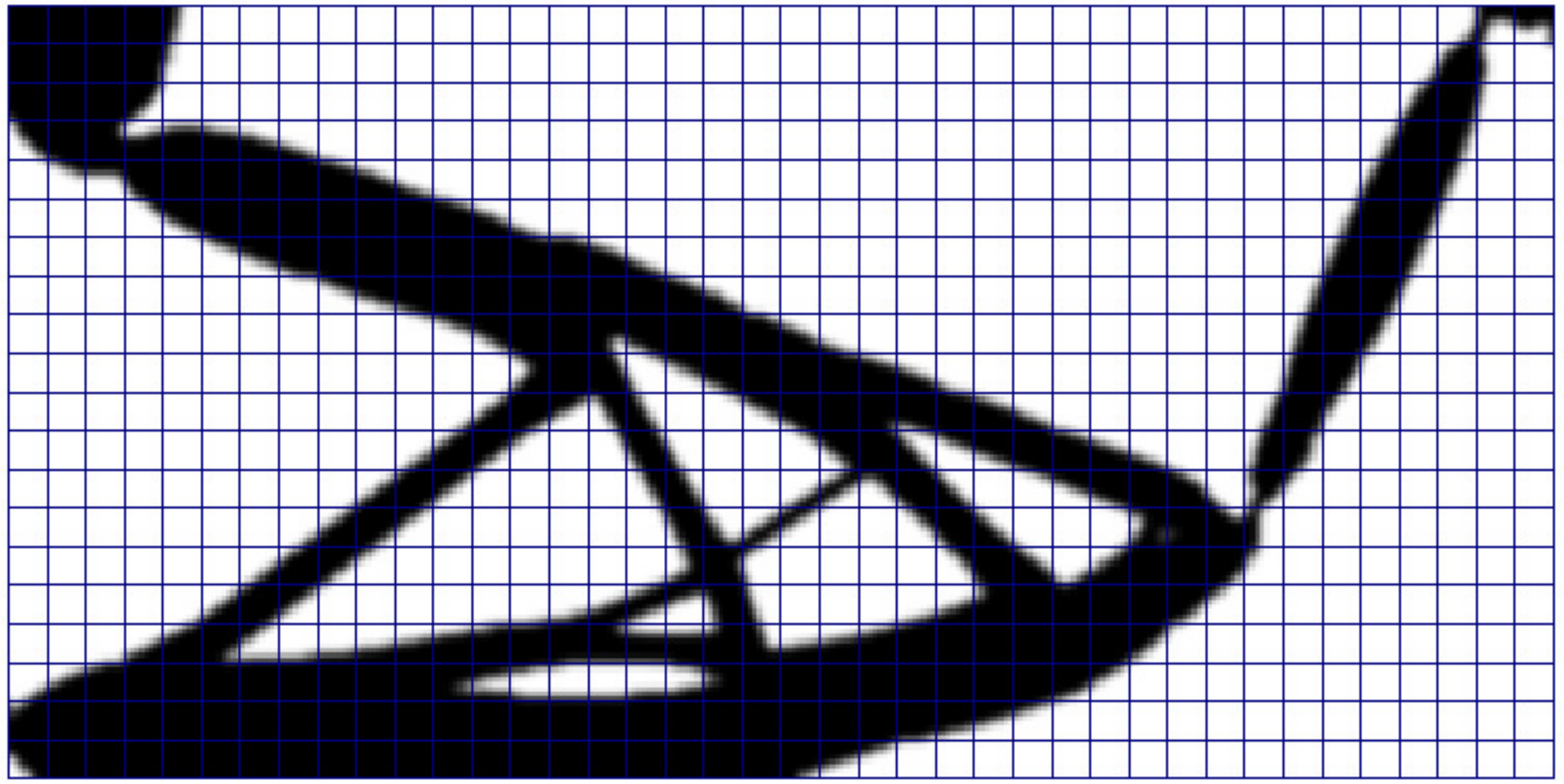}
\caption{$dp$-adaptive MTO ($\mathcal{J} = 2.258$m)}
\label{fig_comp_inv_2b}
\end{subfigure}
\caption{Optimized cantilever designs for the force inverter problem shown in Fig.\ref{fig_comp_inv}, obtained using a uniform MTO mesh (left) and $dp$-adaptive MTO approach (right). A speed-up of 6.2 folds is obtained using $dp$-adaptivity.}
\label{fig_comp_inv_2}
\end{figure}
\indent \vspace{0.5em} \\ To demonstrate the applicability of $dp$-adaptivity on topology optimization of compliant mechanisms, it is applied to the force inverter problem shown in Fig. \ref{fig_comp_inv}. The allowed volume fraction $V_0$ is set to 0.45 and the goal of the problem is to distribute the material in a way that the displacement $u_{out}$ is maximized. \mbox{Fig. \ref{fig_comp_inv_2}} shows the optimized designs obtained using conventional MTO (Fig. \ref{fig_comp_inv_2a}) and the $dp$-adaptive method (Fig. \ref{fig_comp_inv_2b}). As in the previous cases, the two designs are very similar. Details related to the MTO runs are reported in Table \ref{table_mto_runs}. It is observed that the objective ratio $\mathcal{J}/\mathcal{J}_0$ is 1.01. Since this is a maximization problem, a value of $\mathcal{J}/\mathcal{J}_0$ higher than 1 denotes that the design obtained using $dp$-adaptive MTO performs better. $\mathcal{J}/\mathcal{J}^*$ is equal to 1.0, which means that the solution is as accurate as the reference solution.

\begin{figure*}[!htb]
\centering
\begin{subfigure}{1\textwidth}
\centering
\begin{tikzpicture}
    \node[anchor=south west,inner sep=0] at (0,0){
\includegraphics[scale=0.14]{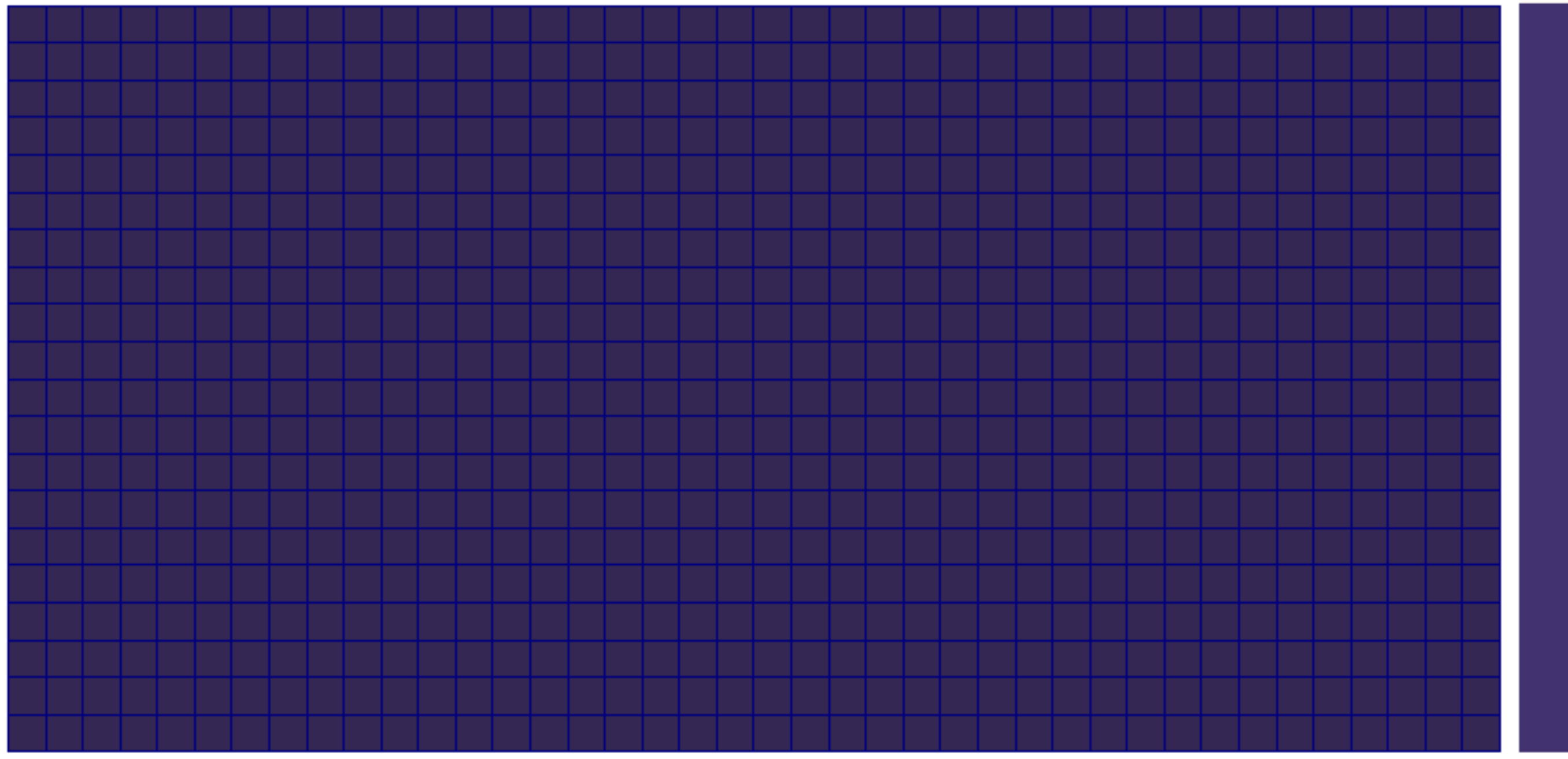}};
\node[text width=0cm] at (3.9, 1.77) 
    {\tiny \color{darkgray}{16}};
    \node[text width=0cm] at (3.9, 1.35) 
    {\tiny \color{darkgray}{}};
    \node[text width=0cm] at (3.9, 0.92) 
    {\tiny \color{darkgray}{16}};
    \node[text width=0cm] at (3.9, 0.49) 
    {\tiny \color{darkgray}{}};
\node[text width=0cm] at (3.9, 0.1) 
    {\tiny \color{darkgray}{16}};
    \node[text width=2cm] at (2.2, 2.3) 
    {\normalsize \color{darkgray}{Design field}};
\end{tikzpicture}\hspace{0.5em}
\begin{tikzpicture}
    \node[anchor=south west,inner sep=0] at (0,0){
\includegraphics[scale=0.14]{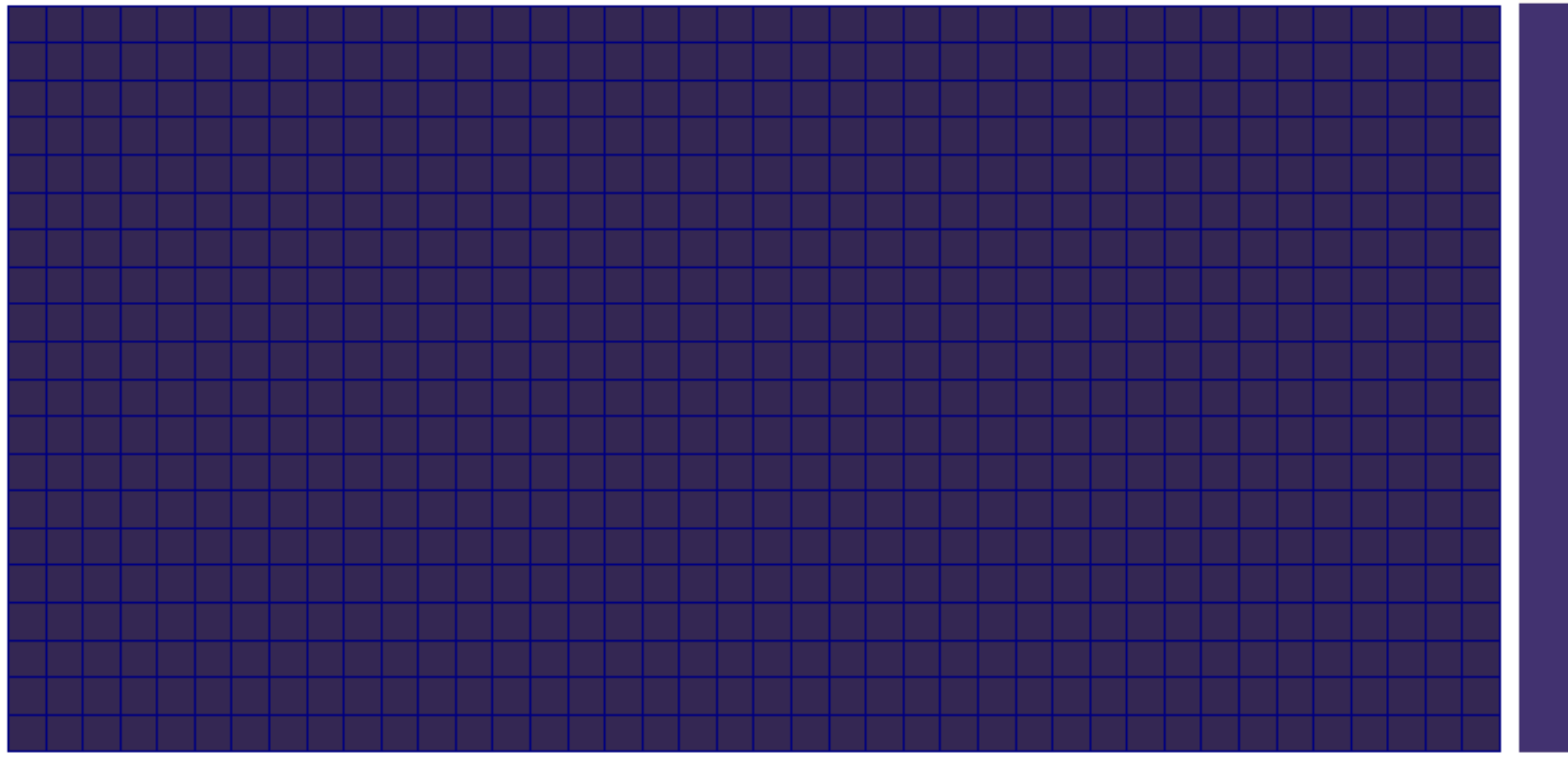}};
\node[text width=0cm] at (3.9, 1.77) 
    {\tiny \color{darkgray}{2}};
    \node[text width=0cm] at (3.9, 1.35) 
    {\tiny \color{darkgray}{}};
    \node[text width=0cm] at (3.9, 0.92) 
    {\tiny \color{darkgray}{2}};
    \node[text width=0cm] at (3.9, 0.49) 
    {\tiny \color{darkgray}{}};
\node[text width=0cm] at (3.9, 0.1) 
    {\tiny \color{darkgray}{2}};
    \node[text width=3.7cm] at (2.23, 2.3) 
    {\normalsize \color{darkgray}{Shape function order}};
\end{tikzpicture}\hspace{0.5em}
\begin{tikzpicture}
    \node[anchor=south west,inner sep=0] at (0,0){
\includegraphics[scale=0.15]{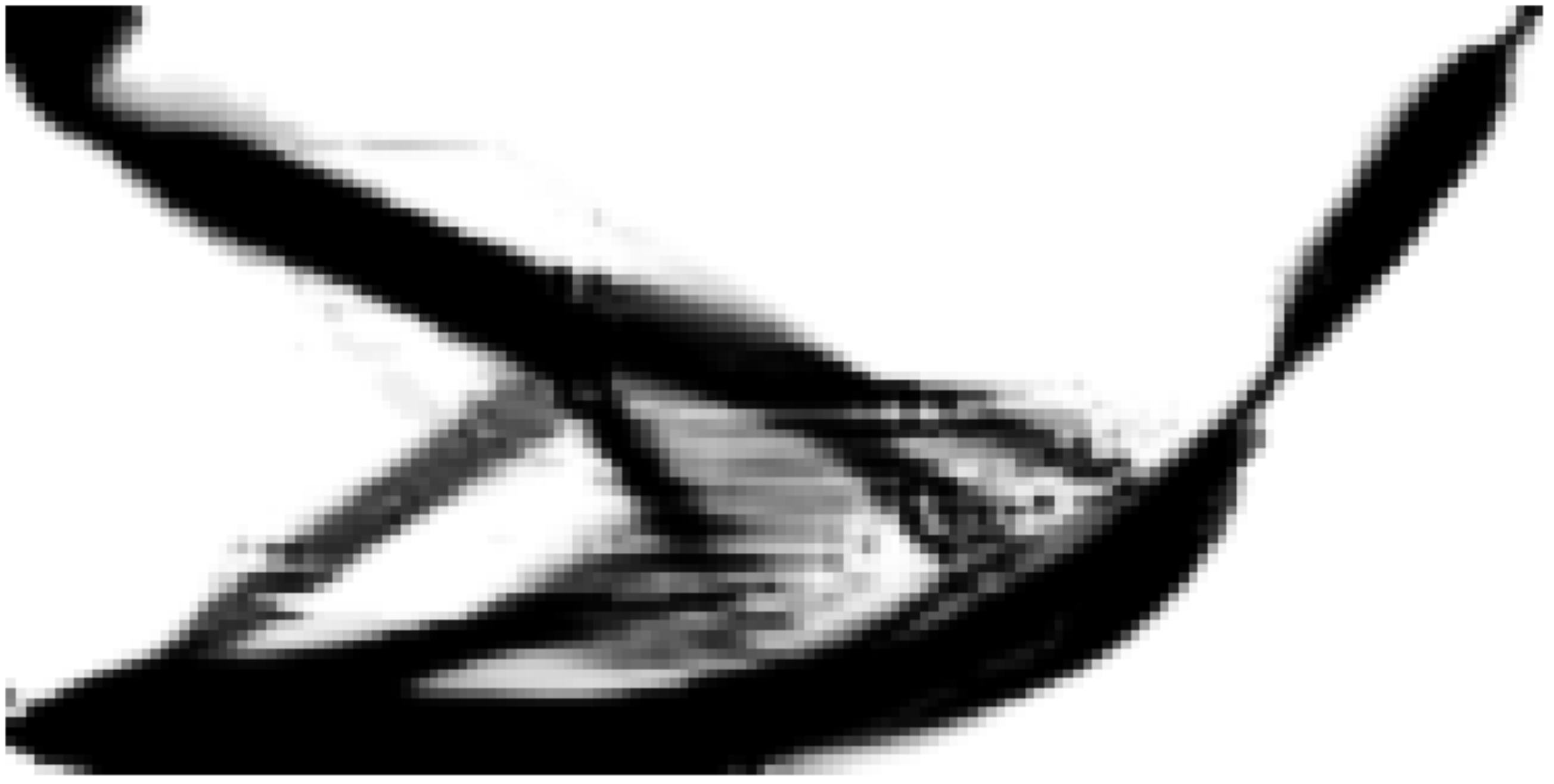}};
    \node[text width=3cm] at (2.23, 2.42) 
    {\normalsize \color{darkgray}{Optimized design}};
\end{tikzpicture}\hspace{0.5em}
\caption{Cycle: 1}
\end{subfigure}
\begin{subfigure}{\textwidth}
\centering
\begin{tikzpicture}
    \node[anchor=south west,inner sep=0] at (0,0){
\includegraphics[scale=0.14]{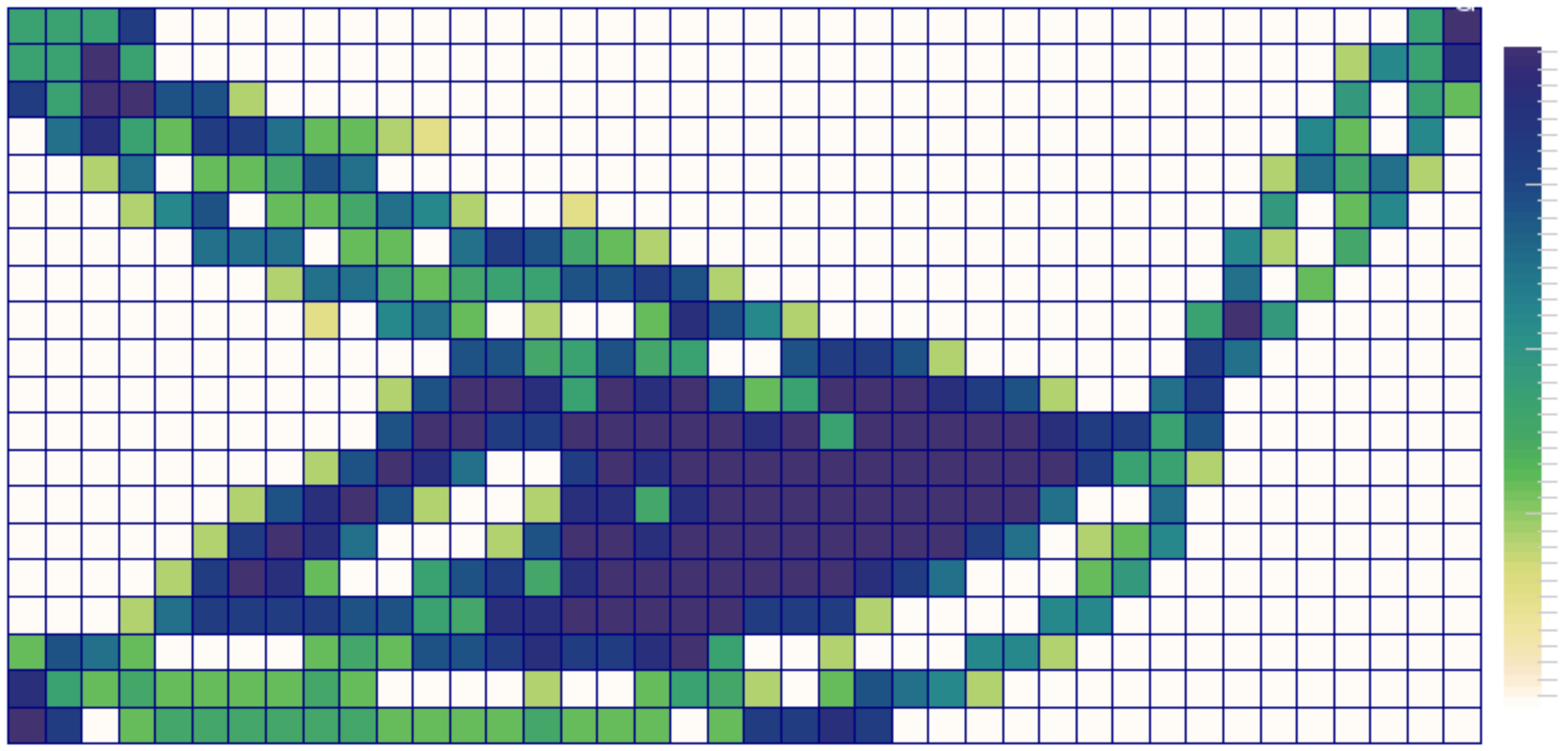}};
\node[text width=0cm] at (3.9, 1.77) 
    {\tiny \color{darkgray}{29}};
    \node[text width=0cm] at (3.9, 1.35) 
    {\tiny \color{darkgray}{24}};
    \node[text width=0cm] at (3.9, 0.92) 
    {\tiny \color{darkgray}{18}};
    \node[text width=0cm] at (3.9, 0.47) 
    {\tiny \color{darkgray}{12}};
\node[text width=0cm] at (3.9, 0.1) 
    {\tiny \color{darkgray}{5}};
\end{tikzpicture}\hspace{0.5em}
\begin{tikzpicture}
    \node[anchor=south west,inner sep=0] at (0,0){
\includegraphics[scale=0.14]{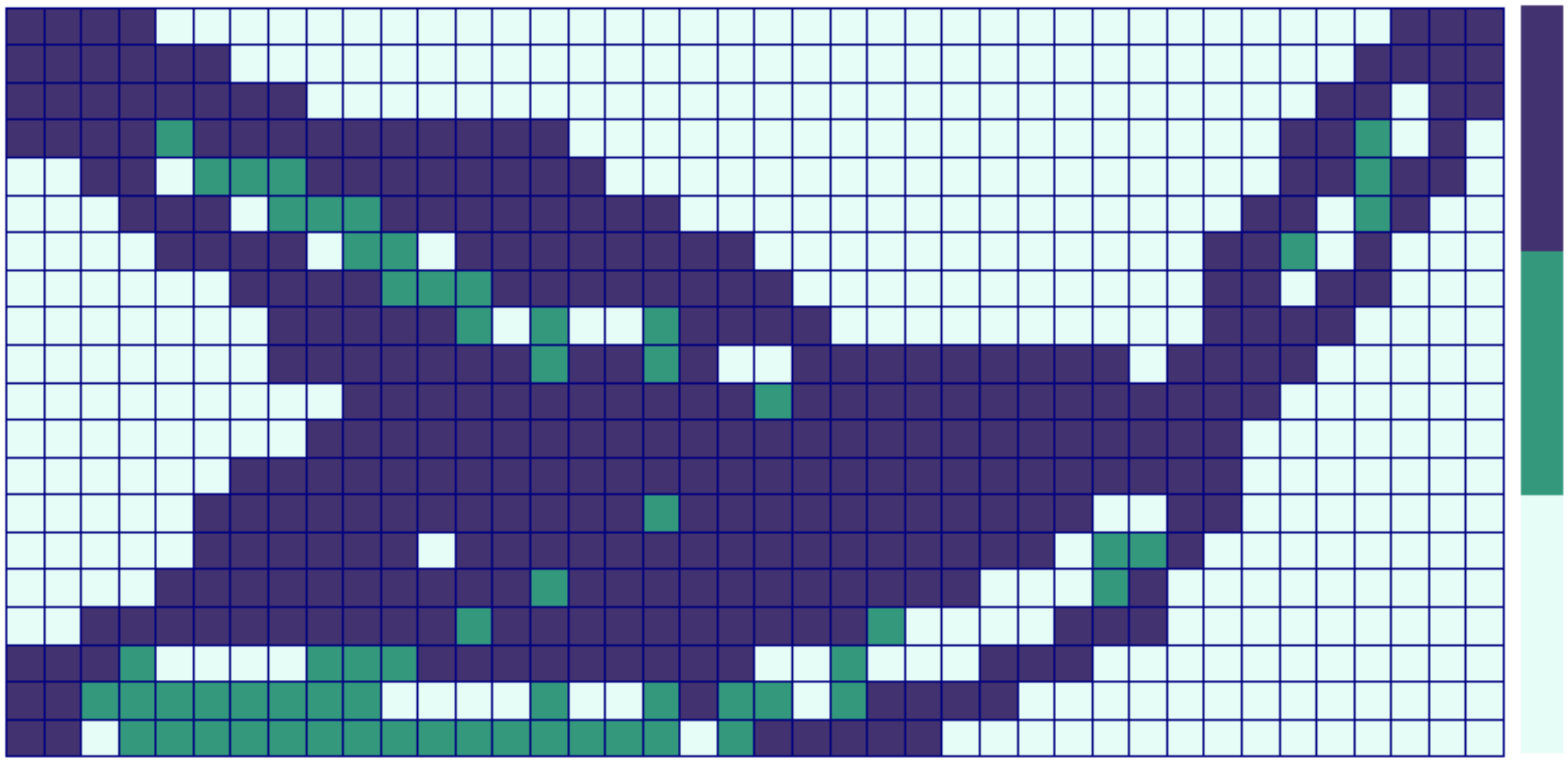}};
\node[text width=0cm] at (3.9, 1.77) 
    {\tiny \color{darkgray}{3}};
    \node[text width=0cm] at (3.9, 1.35) 
    {\tiny \color{darkgray}{}};
    \node[text width=0cm] at (3.9, 0.92) 
    {\tiny \color{darkgray}{2}};
    \node[text width=0cm] at (3.9, 0.47) 
    {\tiny \color{darkgray}{}};
\node[text width=0cm] at (3.9, 0.1) 
    {\tiny \color{darkgray}{1}};
\end{tikzpicture}\hspace{0.5em}
\begin{tikzpicture}
    \node[anchor=south west,inner sep=0] at (0,0){
\includegraphics[scale=0.15]{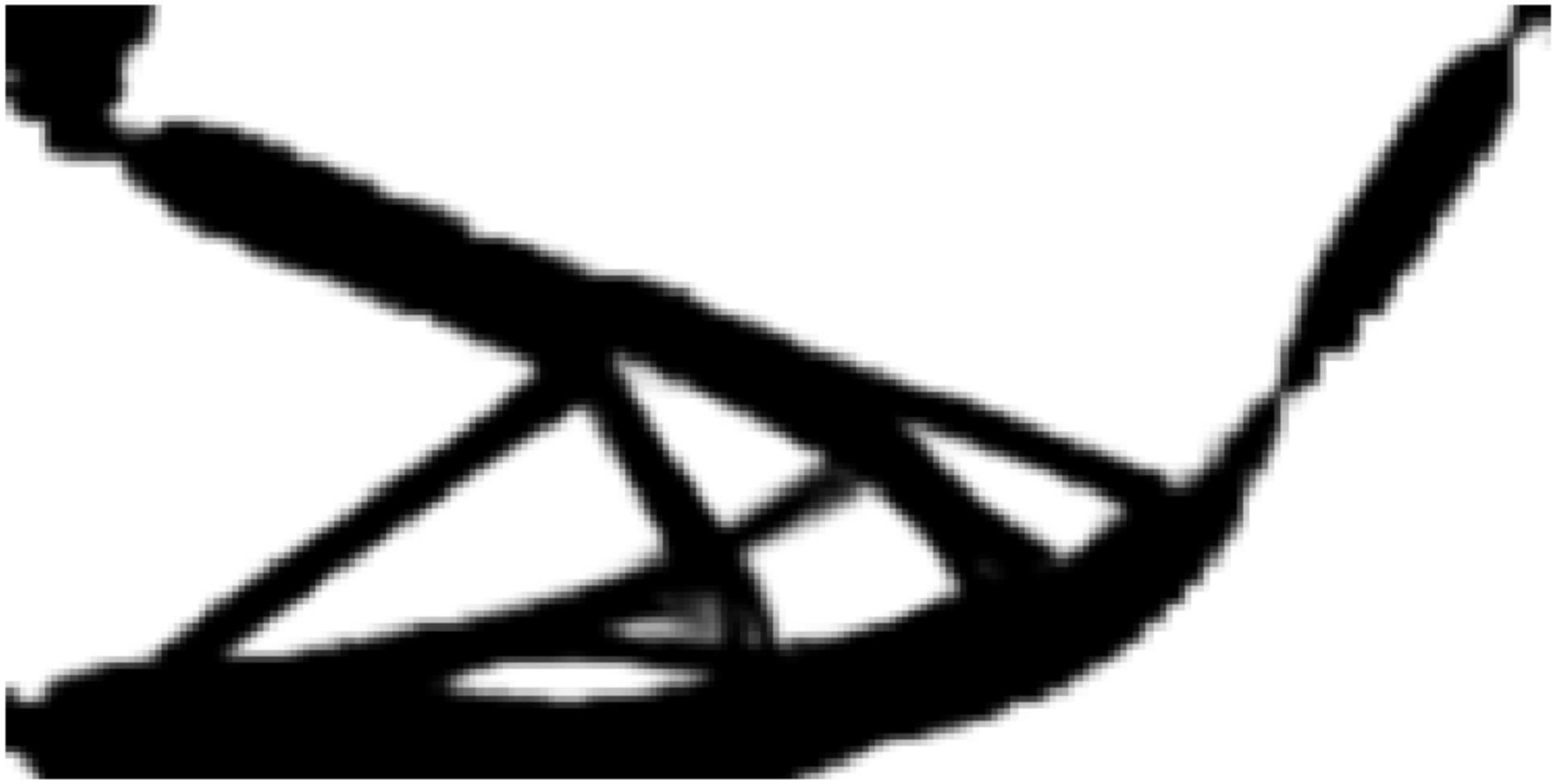}};
\end{tikzpicture}\hspace{0.5em}
\caption{Cycle: 2}
\end{subfigure}
\begin{subfigure}{1\textwidth}
\centering
\begin{tikzpicture}
    \node[anchor=south west,inner sep=0] at (0,0){
\includegraphics[scale=0.14]{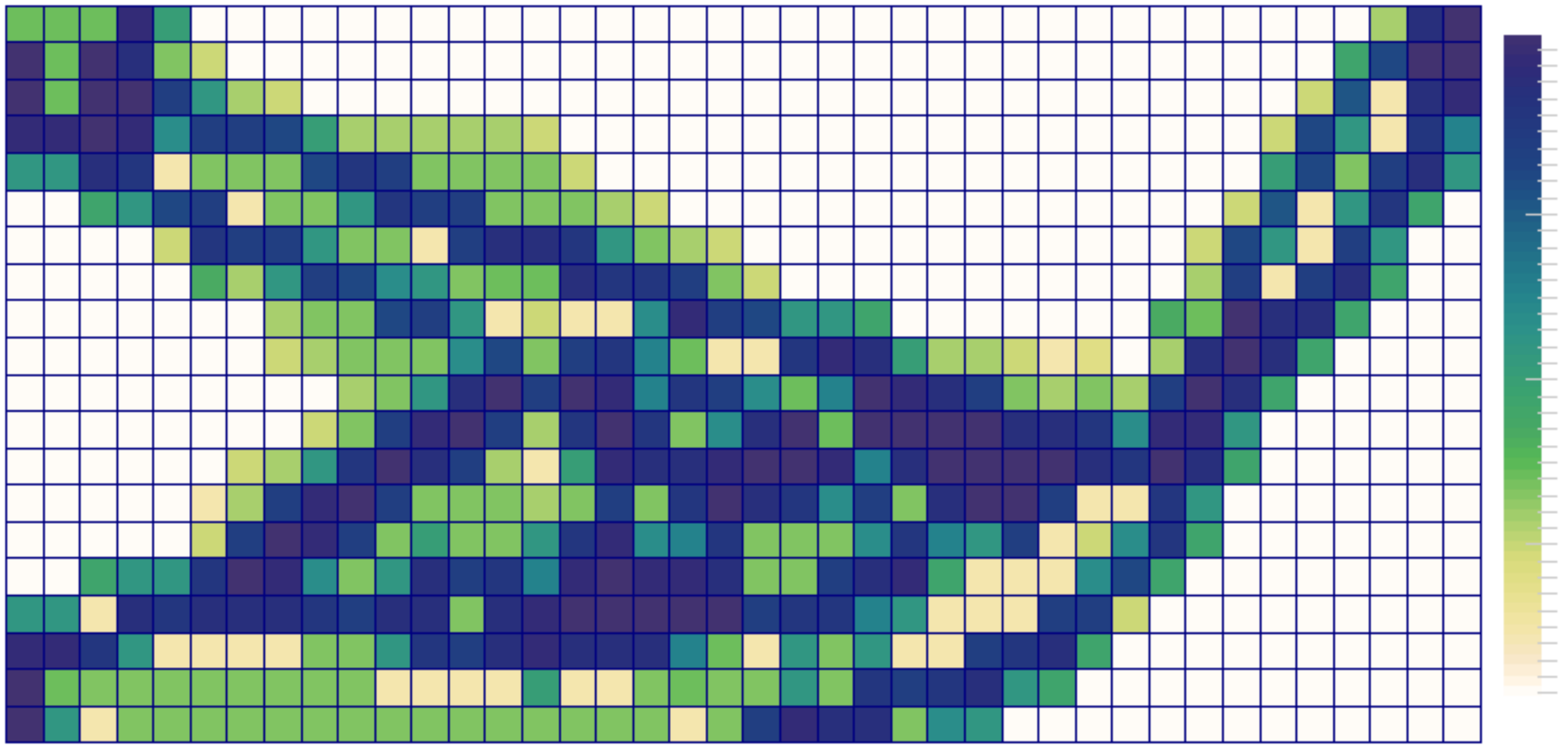}};
\node[text width=0cm] at (3.9, 1.77) 
    {\tiny \color{darkgray}{47}};
    \node[text width=0cm] at (3.9, 1.35) 
    {\tiny \color{darkgray}{34}};
    \node[text width=0cm] at (3.9, 0.92) 
    {\tiny \color{darkgray}{23}};
    \node[text width=0cm] at (3.9, 0.49) 
    {\tiny \color{darkgray}{11}};
\node[text width=0cm] at (3.9, 0.1) 
    {\tiny \color{darkgray}{1}};
\end{tikzpicture}\hspace{0.5em}
\begin{tikzpicture}
    \node[anchor=south west,inner sep=0] at (0,0){
\includegraphics[scale=0.14]{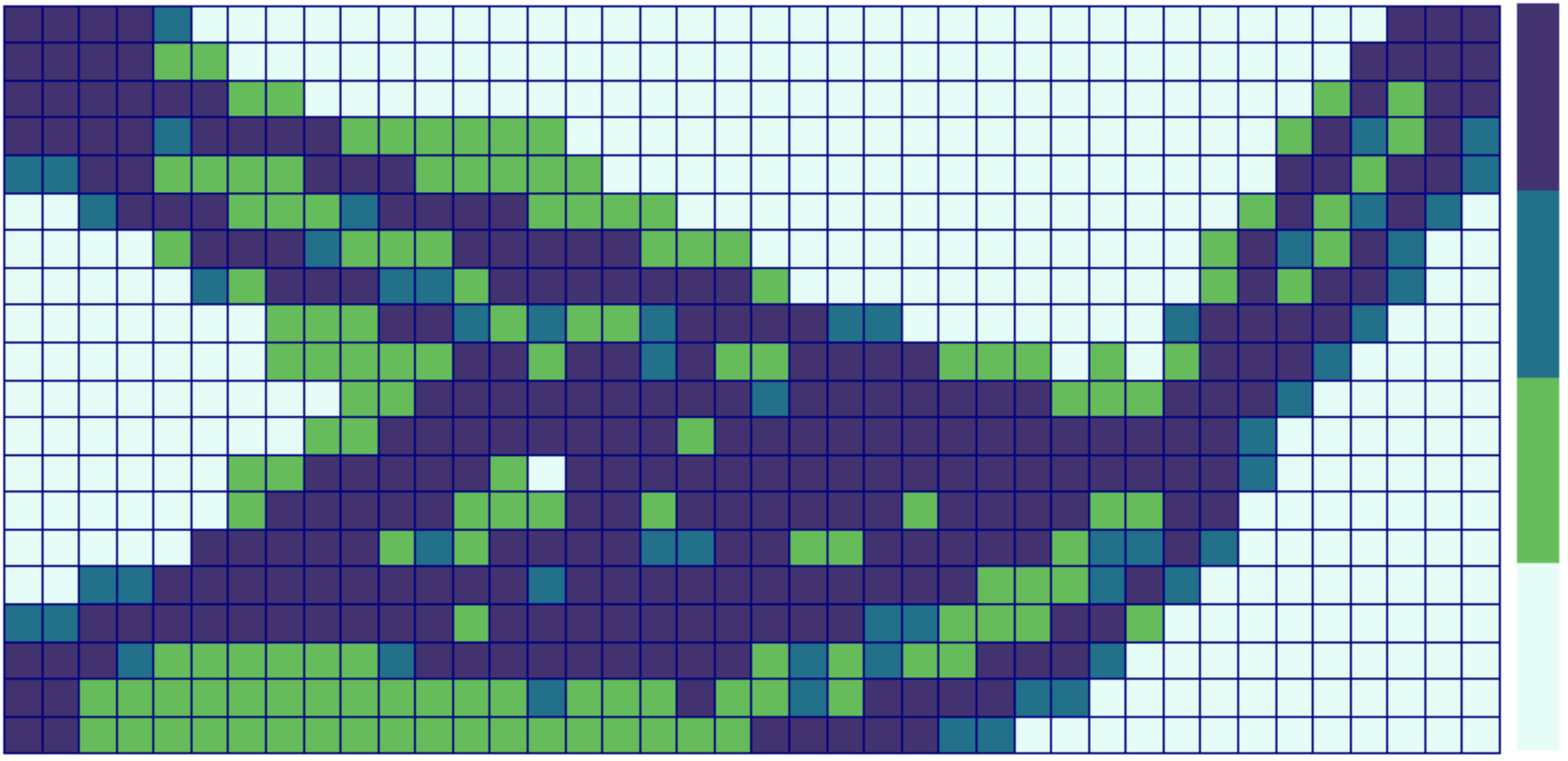}};
\node[text width=0cm] at (3.9, 1.77) 
    {\tiny \color{darkgray}{4}};
    \node[text width=0cm] at (3.9, 1.25) 
    {\tiny \color{darkgray}{3}};
    \node[text width=0cm] at (3.9, 0.72) 
    {\tiny \color{darkgray}{2}};
\node[text width=0cm] at (3.9, 0.15) 
    {\tiny \color{darkgray}{1}};
\end{tikzpicture}\hspace{0.5em}
\begin{tikzpicture}
    \node[anchor=south west,inner sep=0] at (0,0){
\includegraphics[scale=0.15]{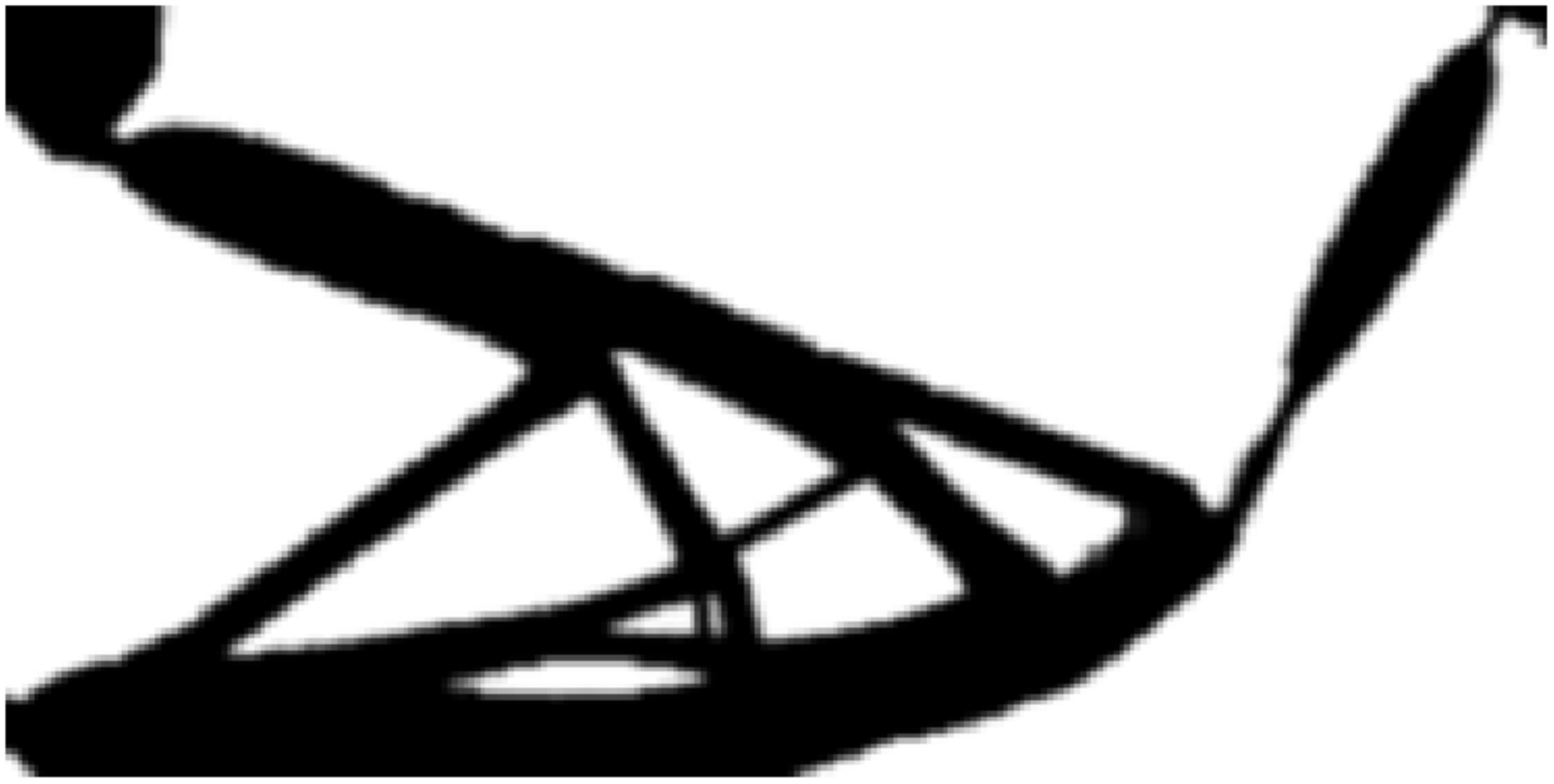}};
\end{tikzpicture}\hspace{0.5em}
\caption{Cycle: 3}
\end{subfigure}
\begin{subfigure}{1\textwidth}
\centering
\begin{tikzpicture}
    \node[anchor=south west,inner sep=0] at (0,0){
\includegraphics[scale=0.14]{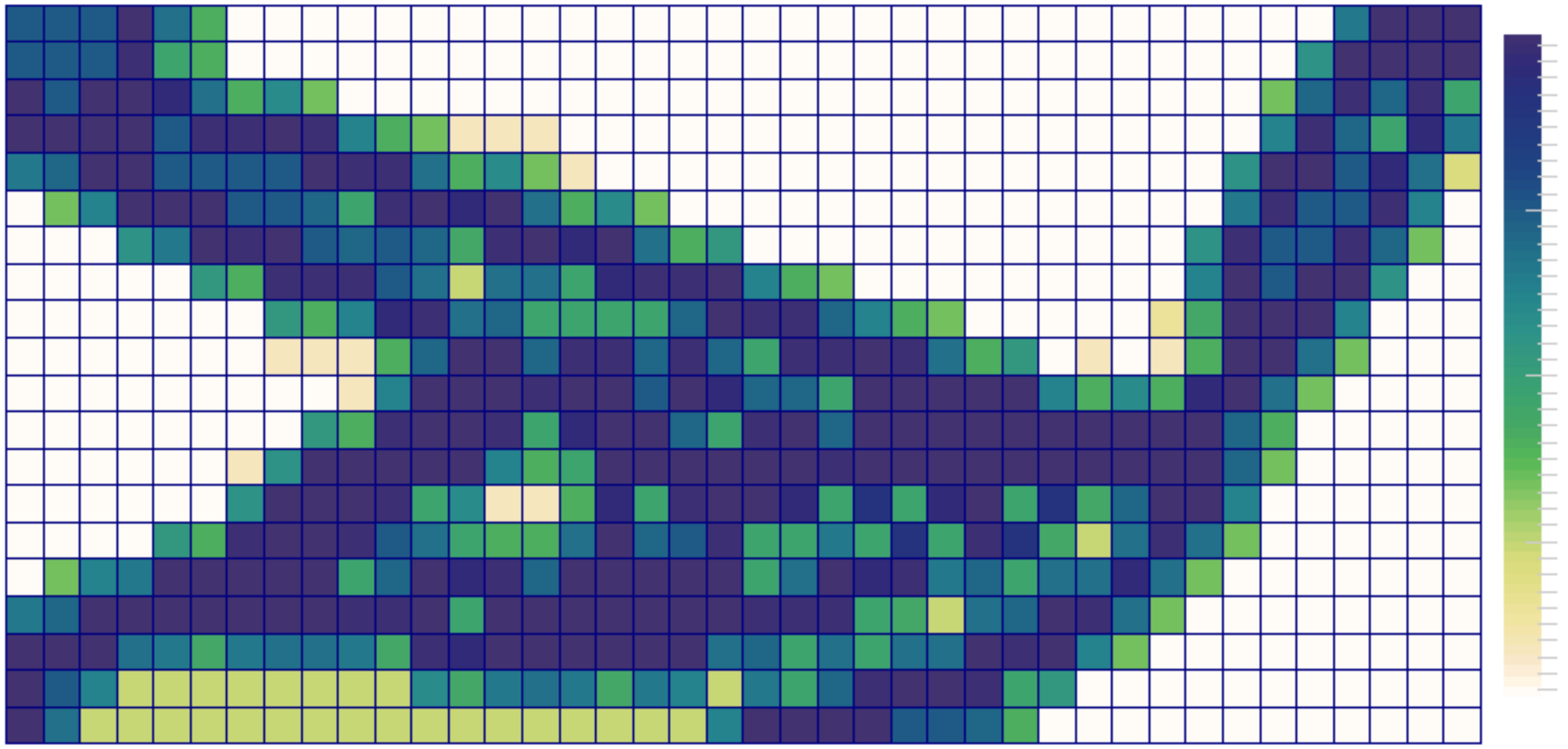}};
\node[text width=0cm] at (3.9, 1.77) 
    {\tiny \color{darkgray}{64}};
    \node[text width=0cm] at (3.9, 1.35) 
    {\tiny \color{darkgray}{48}};
    \node[text width=0cm] at (3.9, 0.92) 
    {\tiny \color{darkgray}{32}};
    \node[text width=0cm] at (3.9, 0.47) 
    {\tiny \color{darkgray}{16}};
\node[text width=0cm] at (3.9, 0.1) 
    {\tiny \color{darkgray}{1}};
\end{tikzpicture}\hspace{0.5em}
\begin{tikzpicture}
    \node[anchor=south west,inner sep=0] at (0,0){
\includegraphics[scale=0.14]{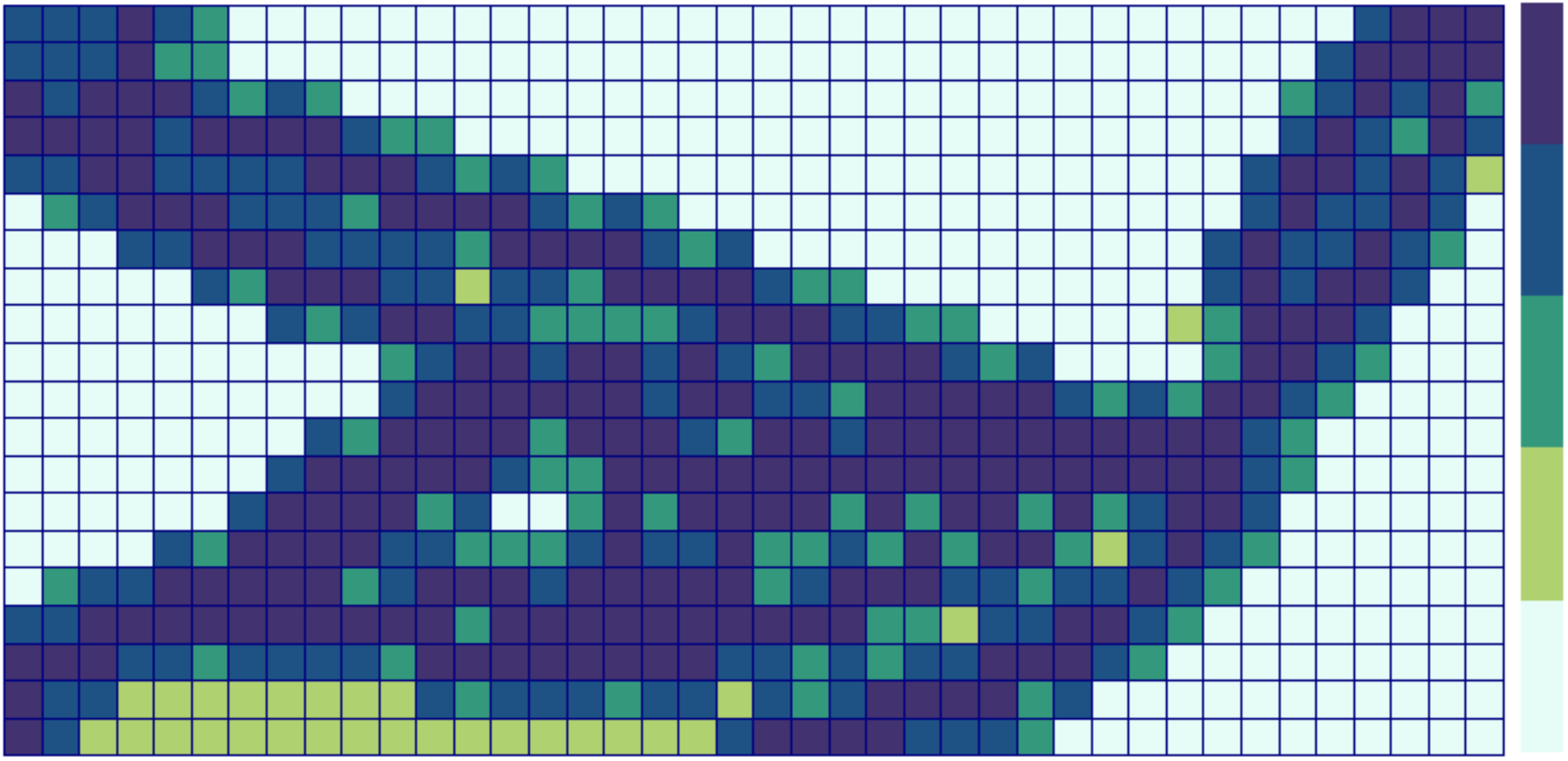}};
\node[text width=0cm] at (3.9, 1.77) 
    {\tiny \color{darkgray}{5}};
    \node[text width=0cm] at (3.9, 1.35) 
    {\tiny \color{darkgray}{4}};
    \node[text width=0cm] at (3.9, 0.92) 
    {\tiny \color{darkgray}{3}};
    \node[text width=0cm] at (3.9, 0.47) 
    {\tiny \color{darkgray}{2}};
\node[text width=0cm] at (3.9, 0.1) 
    {\tiny \color{darkgray}{1}};
\end{tikzpicture}\hspace{0.5em}
\begin{tikzpicture}
    \node[anchor=south west,inner sep=0] at (0,0){
\includegraphics[scale=0.15]{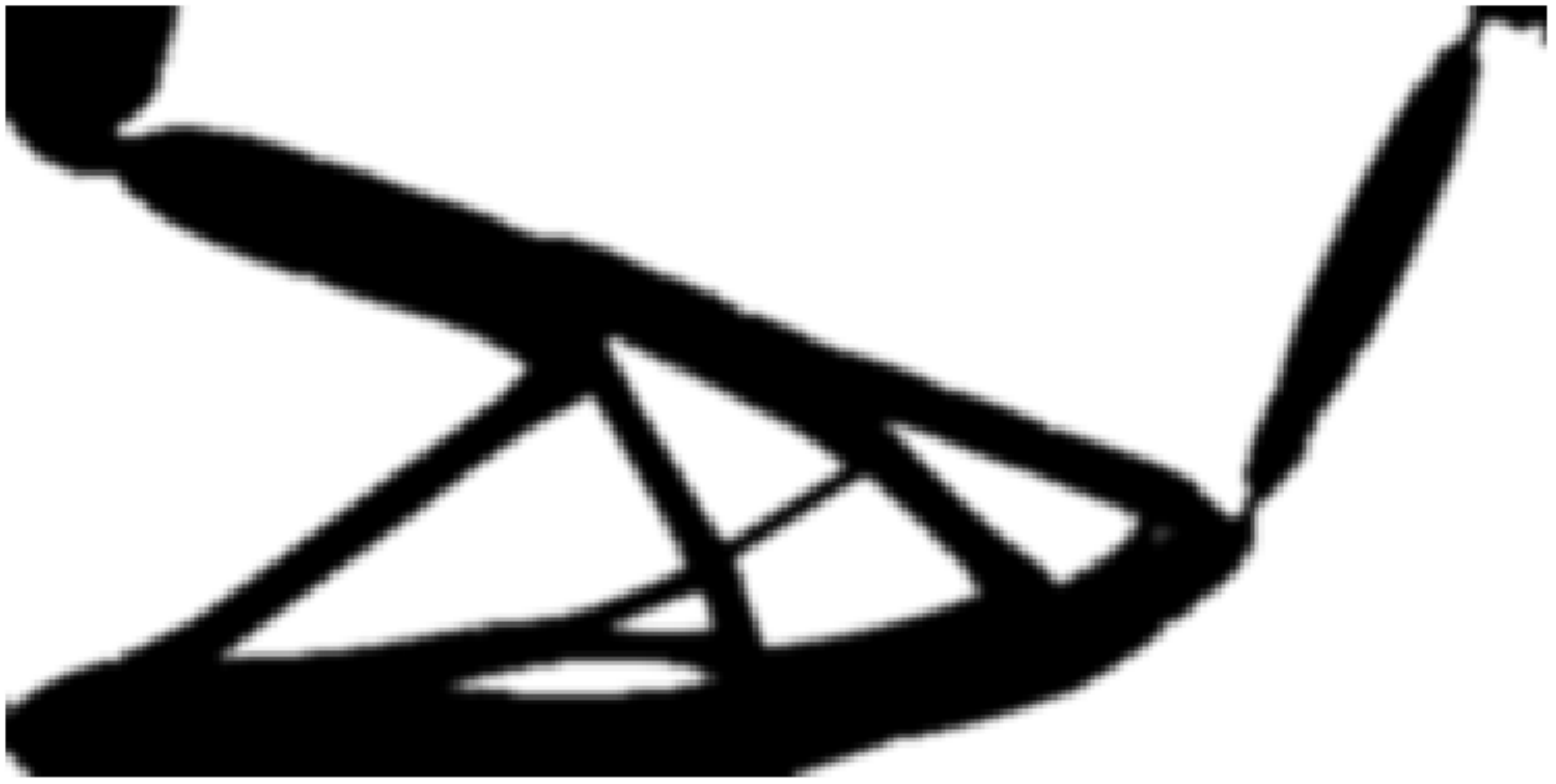}};
\end{tikzpicture}\hspace{0.5em}
\caption{Cycle: 4}
\end{subfigure}
\caption{Optimized designs (right), and the respective shape function orders (middle) and design field (left) obtained for 4 cycles of a $dp$-adaptive MTO run for the force inverter problem shown in Fig. \ref{fig_comp_inv}. The initial mesh  is uniform and each element comprises shape functions of polynomial order 2 and 16 design points per element. The maximum allowed order of the shape functions and number of design points are restricted to 5 and 64 per element, respectively.}
\label{fig_comp_inv_prog}
\end{figure*}

Fig. \ref{fig_comp_inv_prog} shows the distribution of design points and shape function orders, as well as the optimized designs for each cycle of $dp$-adaptivity. Similar to the other cases discussed in this paper, QR-patterns are observed in the results of the first cycle. Nevertheless, the overall material distribution after Cycle 1 already corresponds to the final solution. The QR-patterns eventually disappear in the subsequent cycles due to adaptive refinement of the domain. Refinement primarily occurs in regions where intermediate densities are prominent, and coarsening mainly occurs in the void and solid parts of the domain.

\section{Discussions}
\label{discuss}
The primary goal of using an MTO scheme is to obtain a high-resolution design at a relatively low computational cost. MTO decouples the design and analysis meshes in way that even for the choice of a coarse analysis mesh, a high-resolution density field can be obtained. The potential of MTO has already been demonstrated in \cite{Groen2016, Nguyen2017}. However, there are a few aspects of MTO (\emph{e.g.} computational cost, QR-patterns) where scope of improvement existed. The $dp$-adaptive approach presented in this paper addresses these aspects and further enhances the capability of the MTO method. \\
\begin{figure}
\centering
\begin{subfigure}{0.45\textwidth}
\centering
\begin{tikzpicture}
\node[inner sep=0pt](design1) at (0, 0)
{\includegraphics[width=4.91cm, height=4.91cm]{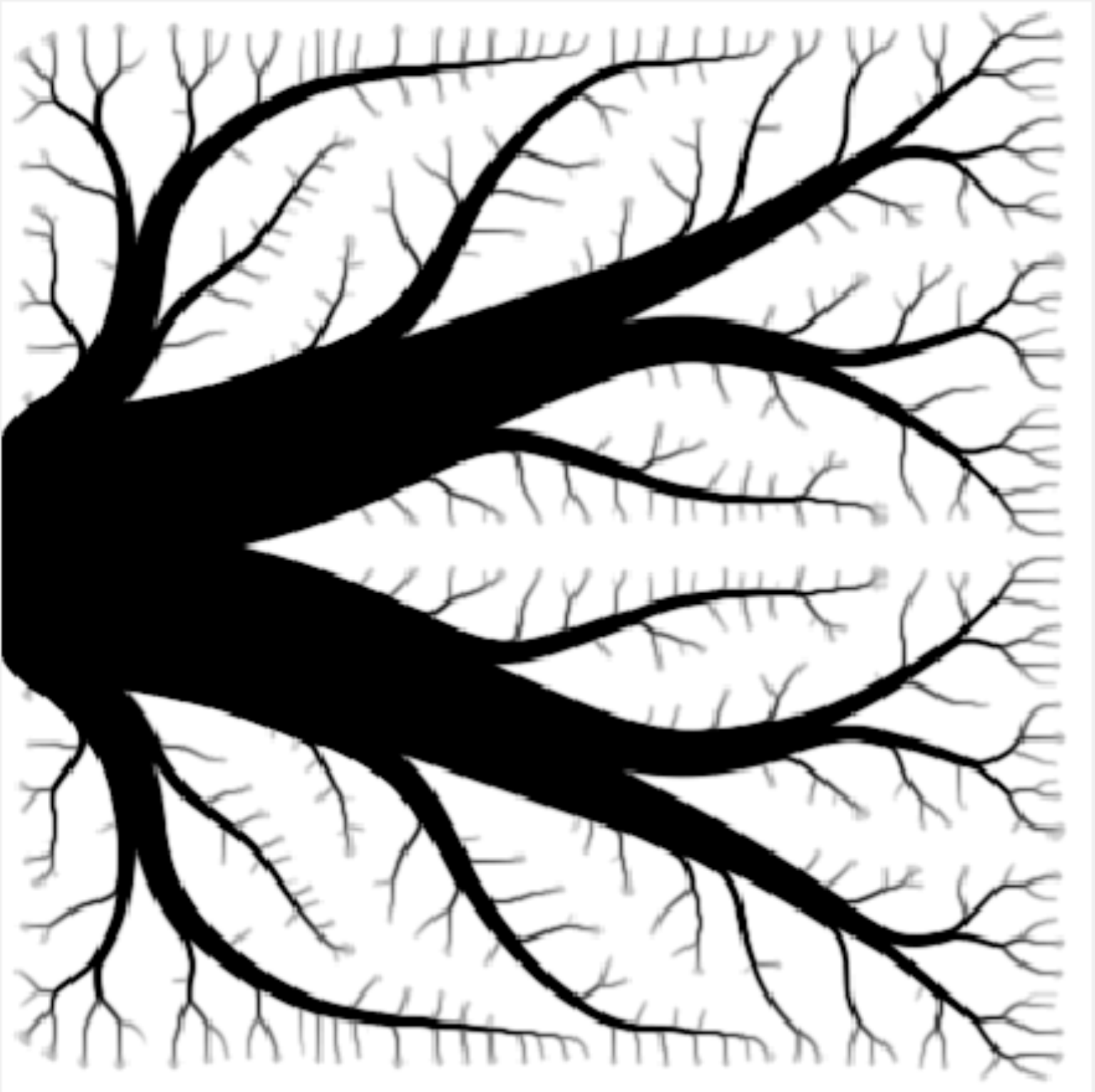}};
\draw [draw=black] (2.45,2.45) rectangle (-2.45,-2.45);
\end{tikzpicture}
\caption{Optimized design for a linear heat conduction problem}
\label{fig_elec1}
\end{subfigure}\hspace{1em}
\begin{subfigure}{0.45\textwidth}
\centering
\begin{tikzpicture}
\node[inner sep=0pt](design1) at (0, 0)
{\includegraphics[scale=0.2]{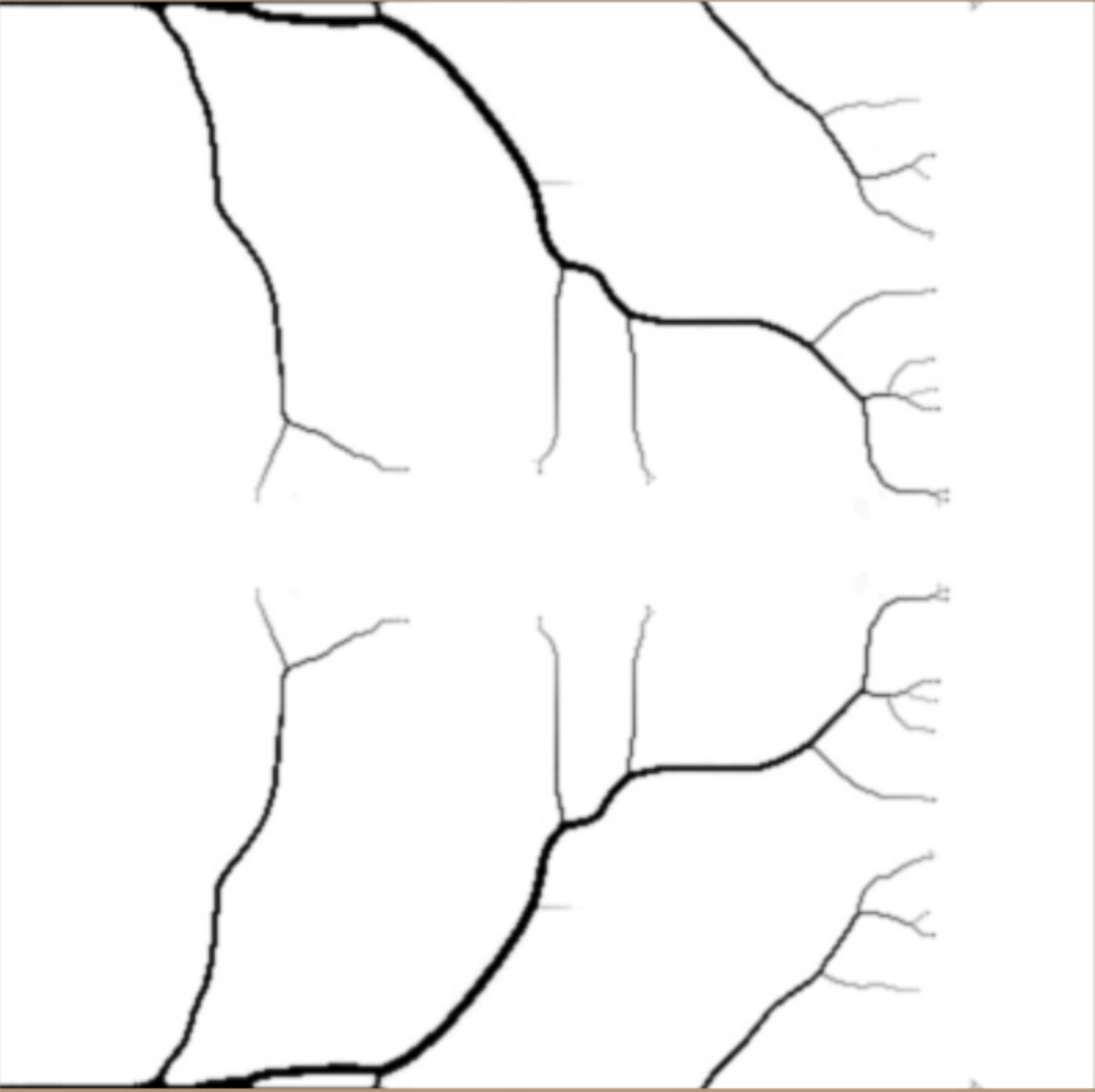}};
\draw [draw=black] (2.45,2.45) rectangle (-2.45,-2.45);
\end{tikzpicture}
\caption{Optimized metallization design for the front surface of a solar cell \cite{Gupta2015}}
\label{fig_elec2}
\end{subfigure}
\caption{Optimized designs obtained using the traditional TO approach on a mesh of $400 \times 400$ finite elements, with $R$ set to 1.5 elements. The two cases refer to (a) linear heat conduction problem with $V_0$ set to 0.3, and (b) nonlinear electrical conduction problem \cite{Gupta2015}. }
\end{figure}
\indent This paper has mainly been focused on presenting the rationale and detailed formulation of the method. To demonstrate the applicability of $dp$-adaptive MTO, 2D mechanical test problems have been considered in this study. Intended future work includes exploring the application of the proposed method on problems involving other physics as well as in 3D settings. In \cite{Groen2016}, it has been shown that MTO can bring a speed-up of up to 32 folds over the traditional TO scheme. The improvement in 3D is significantly higher than that observed in 2D. As $dp$-adaptive MTO reduces the DOFs compared to the conventional MTO method, it is certainly expected to pay off even more in 3D. To really understand the value of the $dp$-adaptive approach for 3D problems, this hypothesis needs to be tested, and this is a part of our future work. 

A preliminary investigation related to the application of $dp$-adaptive MTO on linear conduction (thermal/electrical) problems with loads distributed throughout the domain, revealed that this approach could bring only limited improvements in speed (less than twofolds) for this problem class. The primary reason is that for this type of problems, the optimized design comprises fine features, dendritic in nature, which spread all across the domain. For example, Fig. \ref{fig_elec1} shows an optimized design obtained for a linear thermal conduction problem using the traditional TO approach. A mesh of $400 \times 400$ elements was used and $R$ was set to 1.5 times the length of the element. The material volume fraction was set to 0.3. Details related to the definition of the problem can be found in \cite{Bendsoe2003}. It is seen that the optimized design has very few extended void areas, and most of the domain consists of fine material branches. Due to this, the majority of the domain gets refined at every adaptive cycle, which eventually reduces the relative advantage of \mbox{$dp$-adaptive} MTO method over its non-adaptive variant.

Fig. \ref{fig_elec2} shows an optimized solar cell front metallization design obtained using the traditional TO approach on a mesh of $400 \times 400$ finite elements and $R$ set to 1.5 times the element edge length \cite{Gupta2015}. This design has been obtained by solving a nonlinear electrical conduction problem, and only 4-5\% of the domain is filled with material. For this case, it is seen that significant parts of the domain consists of void regions, which can be easily modeled with low values of $d$ and $p$. Clearly, for such cases, the \mbox{$dp$-adaptive} approach can be used to significantly reduce the associated computational costs. From the two examples of conduction problems discussed here, it is clear that $dp$-adaptivity could certainly have a potential value for problems where designs feature extended void regions. 

To demonstrate the concept of $dp$-adaptivity, a composite indicator has been formulated in this paper. This indicator consists of an analysis error indicator, a density-based indicator and a QR-indicator. Although certain choices have been made for these indicators, the presented methodology itself is independent of these choices. Either of these indicators can be replaced with other alternatives that exist in the literature. For example, the Kelly estimator used as an analysis indicator in this work can be replaced with other analysis-based refinement indicators, \emph{e.g.}, goal-oriented error indicator \cite{Oden2001}. Such choices can provide a better control over the absolute error, accordingly helping to make a better choice of mesh resolution and solution accuracy. However, it is important that the tuning parameters associated with the chosen indicators are properly set so that issues related to excessive refinement are avoided. An addition to consider is a limit on, \emph{e.g.}, the increase in DOFs and/or design variables at a given adaptive cycle.

For the analysis indicator discussed in this paper, the top 10\% and bottom 5\% of the elements corresponding to $\boldsymbol\Gamma^a$ are chosen for refinement and coarsening, respectively. There is no particular motivation to choose these cut-offs. For problems where the design domain has prominent regions with large  jump across the element edges, it is recommended to allow more cells to be refined, so as to reduce the error in fewer cycles. For the density-based indicator, both $\alpha^d_r$ and $\alpha^d_c$ are set to 1.0 for the current study. This ensures that all the elements with $\boldsymbol\Gamma^d> 0$ are refined and all elements with $\boldsymbol\Gamma^d < 0$ are coarsened. The reason to set these parameters to 1.0 is that the stopping criterion chosen in this paper allows the design to converge sufficiently at every MTO cycle. Due to this, the intermediate densities are reduced. However, if fewer iterations are permitted per MTO cycle, it is advisable to set $\alpha^d_r$ and $\alpha^d_c$ to values less than 1, in order to avoid excessive refinement and coarsening. The tuning of all these meta-parameters forms an optimization problem in itself, and as adaptive design approaches become more sophisticated, setting such parameters can become highly nontrivial and time-consuming. For the present study, no extensive parameter tuning was performed, yet already significant performance gains are observed. We see opportunities for future research in further adaptive and intelligent tuning strategies of meta-parameters during the adaptive optimization itself, to take this burden away from the user. 

For the MTO method, $dp$-adaptivity serves as an add-on where the design distribution and shape function orders are adapted at every cycle of refinement based on a predefined criterion. However, there are additional aspects of MTO which can be adapted to gain further improvements in accuracy and associated computational cost. Among others, appropriately adapting the filter radius $R$ could lead to further improvements. In the context of adaptive $h$-refinement, the impact of adaptive filter radius has been explored in \cite{Gupta2016}. For MTO, this aspect has briefly been discussed \mbox{in \cite{Nguyen2012}}. However, the advantage of using an adaptive filter radius in an MTO setting remains an open question and needs to be explored. Also, based on our numerical tests, here we chose to use an adaptive stopping criterion, such that the cutoff value for minimum change in objective value between successive iterations is relaxed by a factor of 0.6 at every adaptive cycle. However, further investigations are needed to decide how this aspect can be adapted in the most efficient way. 

Additional directions associated with $dp$-adaptivity exist that could be investigated for further improvement of the methodology. For example, currently the number of design variables is set to the maximum allowed value based on the element-level upper bound described in \cite{Gupta2016a}. However, it is still an open question whether this is the most appropriate way to refine the design field. Moreover, for the problems presented in this paper, we observed that for the chosen setting, violating the system-level bounds (also derived in \cite{Gupta2016a}) did not have any detrimental impacts. Hence, we decided to not incorporate the system-level bounds in the method. However, for more complex problems, where the objective is very sensitive to small design changes, the system-level bounds might have to be enforced. 

To wrap up the discussions, there are several research aspects that can be explored in the context of adaptive MTO. This work lays the foundation for an adaptive MTO scheme that is mathematical reliable as well as computationally efficient. It is hoped that with further research along the directions outlined above, the proposed approach can be improved further.

\section{Conclusions}
\label{conclude}
Multiresolution topology optimization (MTO) methods decouple the analysis and design discretizations, such that high resolution design representations are permitted on relatively coarse analysis meshes. In this paper, the first adaptive variant of the MTO scheme, namely $dp$-adaptive MTO, has been presented. Through several 2D numerical examples, it has been demonstrated that the proposed method can obtain highly optimized designs at significantly lower computational cost than in conventional MTO, and high analysis accuracy. Moreover, undesired features such as intermediate densities and QR-patterns can be significantly reduced in the resulting designs, and a desired analysis accuracy can be enforced. A particularly interesting application of this $dp$-adaptive MTO method is for TO problems involving low material volume fractions. The speed-up over conventional MTO was found to increase with decreasing material volume fraction. It has been shown that for test cases with a 10\% maximum relative volume, 10-fold speed-up can be obtained over the conventional MTO scheme in 2D, when the $dp$-adaptive MTO method is used. For 3D problems, even higher speed-ups are expected.

Clearly, the proposed adaptive approach improves on the conventional MTO method by tacking some of the issues associated with it. For future work, we aim at exploring the application of $dp$-adaptive approach for problems involving different physics and three-dimensional problems. However, based on the results presented in this study, it can already be argued that the proposed approach could serve as an important methodology to obtain high resolution designs at an attractive computational cost.	
   
\section*{Acknowledgements}
This work is part of the Industrial Partnership Programme (IPP) Computational Sciences for Energy Research of the Foundation for Fundamental Research on Matter (FOM), which is part of the Netherlands Organisation for Scientific Research (NWO). This research programme is co-financed by Shell Global Solutions International B.V. The numerical examples presented in this paper are implemented using deal.II \cite{Bangerth2007}, an open-source C++ package for adaptive finite element analysis. We would like to thank the developers of this software. We would also like to thank Rajit Ranjan for his help.

\section*{Appendices}
\appendix
\section{$k$-means clustering}
\label{sec_kmeans}
$k$-means clustering is a cluster analysis technique popularly used in data mining \cite{Mackay2003}. It aims to partition $\psi$ observations into $k$ clusters such that the observations in each cluster tend to be close to each other. Note that although the problem is computationally difficult, there are various heuristic techniques that can quickly obtain a locally optimal solution. \\
\indent This technique can be used to choose locations of design points within a finite element (FE) and one of the primary advantages of this method is that it is easily applicable to various finite elements differing in geometry. Synonymous to the observations required in $k$-means clustering, a large number of uniformly distributed random points $\psi$ are chosen within the FE using Mersenne twister pseudorandom number generator \cite{Matsumoto1998}. Given that $k$ design points' locations need be to chosen in the FE, we choose $\psi = 1000k$. Next, an initial set of $k$ points is chosen in the FE using $k$-means++ cluster center initialization algorithm \cite{Arthur2007}. These points serve as the initial $k$ means for the $\psi$ observations.

 Let $m_1^{(1)}, m_1^{(2)}, \hdots, m_1^{(k)}$ denote the initial locations of $k$ design points, then the following two steps are iteratively performed to optimize these locations:
\begin{enumerate}
\item \textbf{Assignment step}: Each observation $x_p$ is assigned to exactly one out of $k$ clusters based on the shortest Euclidean distance. Thus, during the $t^{\text{th}}$ iteration, $x_p$ is assigned to the $i^{\text{th}}$ cluster, if
\begin{equation}
||x_p - m_i^{(t)}||^2 \leq ||x_p - m_j^{(t)}||^2 \enskip \forall \enskip 1 \leq j \leq k.
\end{equation}
\item \textbf{Update step}: The new centroids of each of the $k$ clusters then become the new locations of the design points. The centroids are calculated as follows:
\begin{equation}
m_i^{(t+1)} = \frac{1}{c_i}\sum_{p=1}^{c_i} x_p.
\end{equation}
\end{enumerate}
\begin{figure}
\centering
\begin{subfigure}{0.3\textwidth}
\centering
\includegraphics[height=3.5cm, width=4cm]{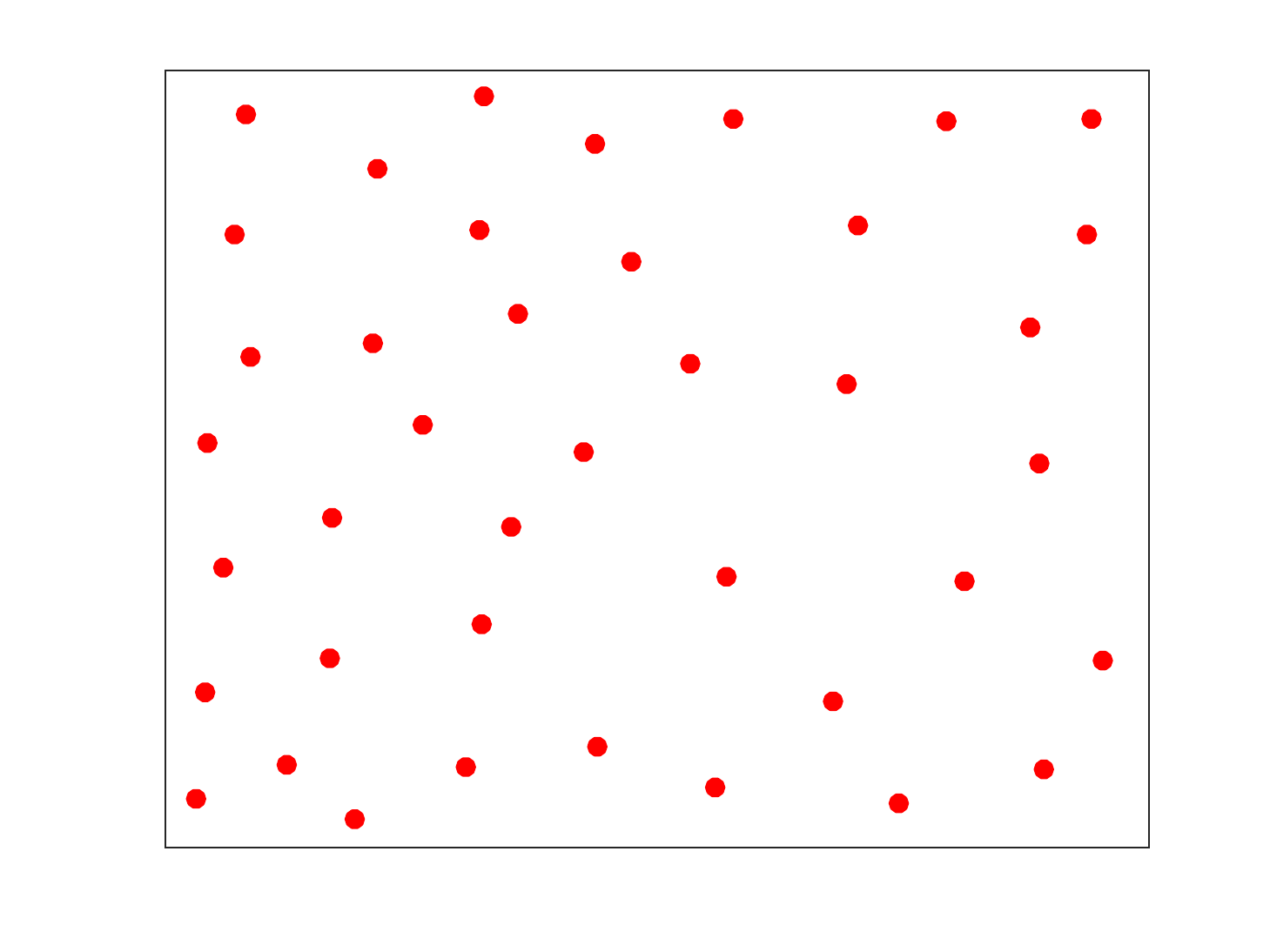}
\caption{initial design field}
\end{subfigure} 
\begin{subfigure}{0.3\textwidth}
\centering
\includegraphics[height=3.5cm, width=4cm]{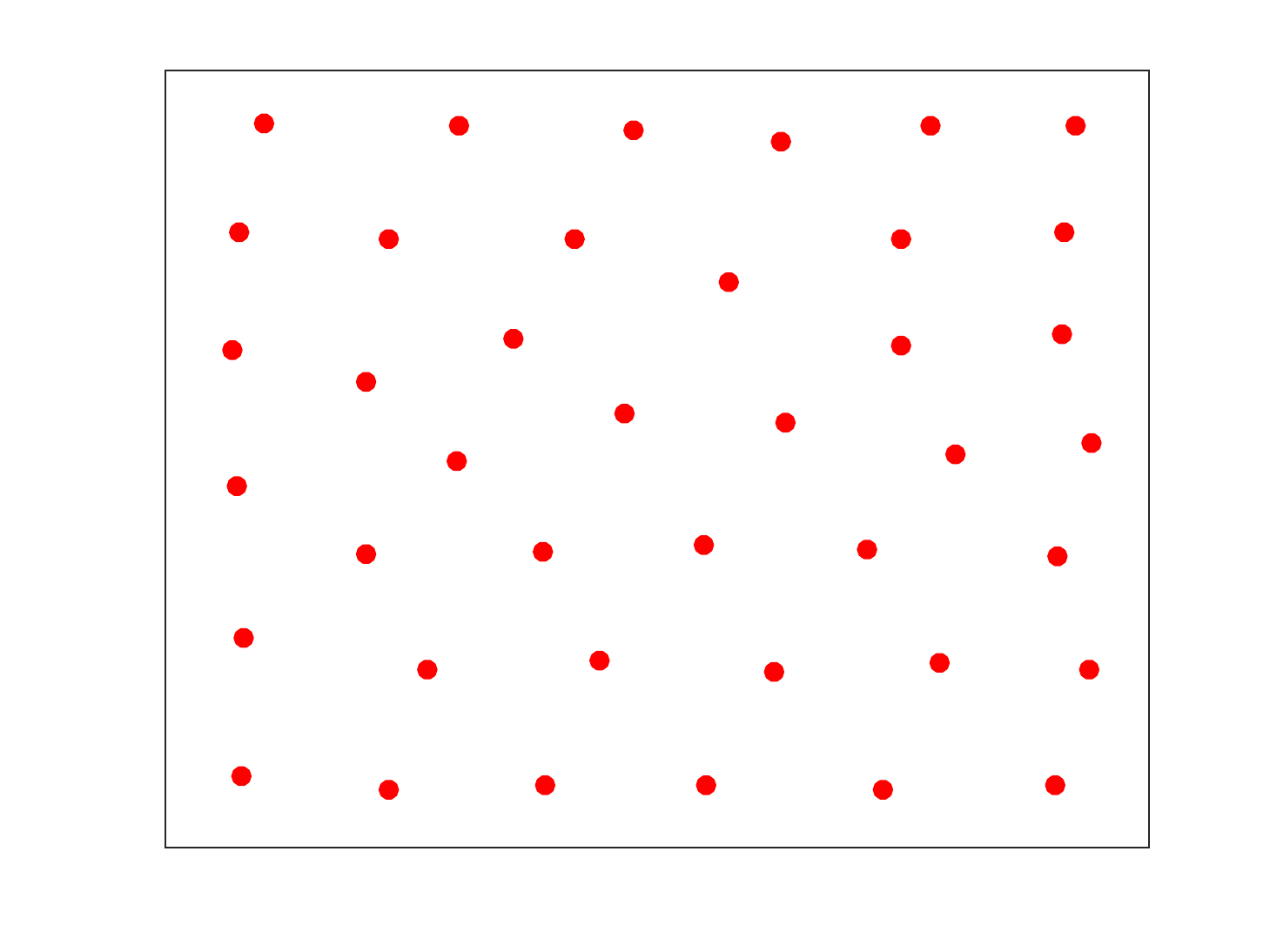}
\caption{optimized design field}
\end{subfigure}
\caption{Distribution of 40 design points in a Q-type finite element obtained using $k$-means clustering.}
\label{kmeans1}
\end{figure}
The two steps are repeated until locally optimal cluster-means are obtained. Note that for every number of design points, these distributions are generated once, and stored for use during optimization.\\
\indent  Fig. \ref{kmeans1} shows the initial and optimized distributions of 40 design points in a Q-type FE. The optimized design distribution has been obtained using the $k$-means clustering algorithm. Clearly, in the optimized design field, the design points are more uniformly distributed and away from the boundaries of the element. 

\section{Numerical integration scheme}
\label{num_int}
The element stiffness matrix $\mathbf{K}_e$ needs to be accurately integrated for every finite element. For the traditional TO using Q1 elements with elementwise constant densities, a $2 \times 2$ Gauss quadrature rule is sufficient. However, for more complex density fields and higher order shape functions, more advanced ways of integration are needed to obtain correct $\mathbf{K}_e$. One of the possibilities is to use higher order integration schemes. A drawback of this approach is that a solid-void boundary may not be correctly modeled. However, the associated error is very small, and with higher order integration schemes, numerically correct designs are obtained using MTO.

The density inside every voxel in the background mesh is constant. Thus, a composite integration scheme can also be used, where the voxel-contributions to the stiffness matrix are evaluated first, and these are then summed together to obtain the element stiffness matrix \cite{Groen2016}. Since density is assumed to be constant inside each voxel, the choice of integration scheme depends on the polynomial order of the shape functions only. The advantage of this scheme is that the solid-void boundaries are aligned with the edges of the voxels, due to which the stiffness matrix can be accurately integrated. 

The composite integration, in general, is superior over the traditional integration scheme which is based on higher order Gauss quadrature rule. However, since in TO the design changes during the course of optimization, significant amount of information related to the stiffness matrices needs to be precomputed to use it in an adaptive MTO formulation. To avoid this excessive storage issue and to reduce the additional computational costs related to assembling the stiffness matrix at each iteration of MTO, we prefer to use the traditional Gauss quadrature rule with higher number of integration points. 

\begin{table}[h!]
  \centering
  \caption{Choice of integration scheme for different combinations of design fields and polynomial shape functions for Q-type finite elements. Here, $d$ and $p$ denote the number of design points and polynomial order of the shape functions, respectively, $n_{\text{sup}}$ refers to the number of support points, and $\overline{\mathcal{P}}(d)$ and $\overline{\mathcal{P}}(\mathbf{K})$ denote the maximum possible polynomial order of the design field and stiffness matrix, respectively. }
  \label{tab:table1}
  \begin{tabular}{c|c|c|c|c|c}
    \hline
    $d$ & $\overline{\mathcal{P}}(d)$ & $p$ & $n_{\text{sup}}$ & $\overline{\mathcal{P}}(\mathbf{K})$ & Gauss quadrature rule\\
    \hline
    1 & 0 & 1 & 4 & 2 & $2 \times 2$\\
    4 & 2 & 1 & 4 & 4 & $3 \times 3$\\
    9 & 3 & 2 & 9 & 7 & $4 \times 4$\\
    16 & 5 & 3 & 16 & 11 &  $6 \times 6$\\
    25 & 6 & 3 & 16 & 12 &  $7 \times 7$\\
    36 & 7 & 4 & 25 & 15 & $8 \times 8$\\
    49 & 9 & 5 & 36 & 19 & $10 \times 10$\\
    64 & 10 & 5 & 36 & 20 & $11 \times 11$\\
  	\hline
    
    \hline
  \end{tabular}
\end{table}
\indent Table \ref{tab:table1} lists the minimum Gauss quadrature rule needed to accurately integrate the element stiffness matrix for several different density fields and polynomial shape functions. Here, only quadrilateral finite elements are considered. Based on the number of design points, a polynomial design field is constructed, and based on the shape functions, the order of element stiffness matrix is determined.  

\bibliography{template}

\end{document}